\documentclass[12pt]{article}

\usepackage{latexsym}
\usepackage{amssymb,amsfonts,amsmath}
\usepackage{graphicx} 
\usepackage{indentfirst}
\usepackage{bbm}
\usepackage{amssymb}
\usepackage{verbatim}
\usepackage{amsmath, amsthm,amssymb}
\usepackage{mathrsfs}
\usepackage{hyperref}
\usepackage{amsfonts}
\usepackage{dsfont}
\usepackage{cite}

\parskip=\medskipamount

\def \un{\underline}
\newcommand {\cD}{{\cal D}}
\newcommand {\cE}{{\cal E}}

\newcommand {\cH}{{\cal H}}

\newcommand {\cM}{{\cal M}}
\newcommand {\cN}{{\cal N}}

\newcommand {\cR}{{\cal R}}

\newcommand {\cT}{{\cal T}}

\def\a{\alpha}

\def\b{\beta}

\def\d{\delta}

\def\f{\phi}
\def\g{\gamma}

\def\j{\psi}

\def\l{\lambda}

\def\o{\omega}

\def\q{\theta}
\def\r{\rho}
\def\s{\sigma}

\def\x{\xi}
\def\z{\zeta}
\def\D{\Delta}
\def\F{\Phi}
\def\J{\Psi}
\def\L{\Lambda}

\def\U{\Upsilon}

\def\tr{{\rm tr}}
\def\rd{{\rm d}}
\def\ri{{\rm i}}

\newcommand{\ad}{{\dot{\alpha}}}                           
\newcommand{\bd}{{\dot{\beta}}}                            
\newcommand{\ve}{\varepsilon}                            

\newcommand{\pa}{\partial}                           
\newcommand{\hf}{\frac12}

\newcommand{\vf}{\varphi}

\newcommand{\be}{\begin{equation}}
\newcommand{\ee}{\end{equation}}
\newcommand{\bea}{\begin{eqnarray}}
\newcommand{\eea}{\end{eqnarray}}
\newcommand{\non}{\nonumber}
\newcommand{\1}{{\underline{1}}}

\newcommand{\dsR}{{\mathbb R}}

\newcommand{\bm}[1]{\mbox{\boldmath$#1$}}

\def\double #1{#1{\hbox{\kern-2pt $#1$}}}

\newcommand{\gd}{{\dot\g}}
\newcommand{\dd}{{\dot\d}}

\newcommand{\ts}{{\tilde{\s}}}

\newif\ifdtup

\newcommand{\bsubeq}{\begin{subequations}}
\newcommand{\esubeq}{\end{subequations}}

\numberwithin{equation}{section}

\newcommand{\sU}{\mathsf{U}}

\begin{document}

\begin{titlepage}
\begin{flushright}
February, 2019\\
\end{flushright}
\vspace{5mm}

\begin{center}
{\Large \bf Conformal geometry and  (super)conformal higher-spin gauge theories
}\\ 
\end{center}

\begin{center}

{\bf 
Sergei M. Kuzenko and Michael Ponds 
} \\
\vspace{5mm}

\footnotesize{
{\it Department of Physics M013, The University of Western Australia\\
35 Stirling Highway, Crawley W.A. 6009, Australia}}  
\vspace{2mm}
~\\
Email: \texttt{sergei.kuzenko@uwa.edu.au, michael.ponds@research.uwa.edu.au
}
\\
\vspace{2mm}

\end{center}

\begin{abstract}
\baselineskip=14pt
We develop a manifestly conformal approach to describe linearised 
(super)conformal higher-spin gauge theories 
in arbitrary conformally flat backgrounds in three and four spacetime dimensions. 
Closed-form expressions in terms of gauge prepotentials are given
for gauge-invariant higher-spin (super) Cotton 
and (super) Weyl tensors in three and four dimensions, respectively.
The higher-spin (super) Weyl tensors are shown to be conformal primary (super)fields
in arbitrary conformal (super)gravity backgrounds, however they are gauge invariant only if 
the background (super) Weyl tensor vanishes. The proposed higher-spin actions 
are (super) Weyl-invariant on arbitrary curved backgrounds, 
however the appropriate higher-spin gauge 
invariance holds only in the conformally flat case.
We also describe conformal models for generalised 
gauge fields that are used to describe partially massless dynamics in three and four dimensions. In particular, generalised higher-spin Cotton and Weyl tensors are introduced.
\end{abstract}

\vfill

\vfill
\end{titlepage}

\newpage
\renewcommand{\thefootnote}{\arabic{footnote}}
\setcounter{footnote}{0}

\tableofcontents{}
\vspace{1cm}
\bigskip\hrule

\allowdisplaybreaks


\section{Introduction}

The concept of conformal higher-spin (CHS) theory was introduced by 
Fradkin and Tseytlin \cite{FT} in 1985 as a generalisation of Maxwell's electrodynamics
and conformal gravity in four dimensions. 
Since then there has been much interest in CHS theories in diverse dimensions, see
\cite{FL-3D,PopeTownsend,FL,FL-4D,Tseytlin,Segal,Marnelius,Metsaev,Vasiliev2009}
for an incomplete list of works published within a quarter-century after \cite{FT}.
This interest has truly exploded in the last decade and, 
unfortunately, it is hardly possible to list all relevant publications 
(although comments on the literature will be given in the main body). 
Among the attractive features 
of CHS theories are the following: (i) maximal spin-$s$ gauge symmetry consistent with 
locality \cite{FT}; (ii) natural connection to the AdS/CFT correspondence \cite{Tseytlin};
(iii) Lagrangian formulation for a complete interacting bosonic CHS theory \cite{Segal};
and (iv) interesting quantum properties
\cite{Beccaria:2014jxa,BT2015,Beccaria:2016syk,Adamo:2018srx,Beccaria:2018rxp}.

Off-shell $\cN=1$ superconformal higher-spin (SCHS) multiplets in four dimensions
were briefly discussed, in the framework of supercurrent multiplets,
 by Howe, Stelle and Townsend  \cite{HST} in 1981,  a few years before 
Fradkin and Tseytlin \cite{FT} 
constructed the free CHS actions.
It was only in 2017 that the higher-spin gauge prepotentials
(describing superspin-$(s+\hf)$ multiplet, with $s=2,3,\dots$) introduced in \cite{HST} 
and more general off-shell gauge supermultiplets were finally used to construct 
free $\cN=1$ 
SCHS
actions \cite{KMT}.
Parallel studies in three dimensions ($3D$) describing  
 SCHS
 multiplets and the corresponding Chern-Simons actions 
 were conducted in  \cite{K16,KT,SK-MP} and \cite{KO,HKO} for the cases 
$\cN=1$ and $\cN=2$, respectively. These $3D$ and $4D$ off-shell constructions 
open the possibility to develop a manifestly supersymmetric setting for 
SCHS
theories first advocated 
by Fradkin and Linetsky 
 \cite{FL-3D,FL-4D} in the component approach.
 It also becomes feasible, as was briefly discussed in \cite{KMT},
 to formulate an interacting 
 SCHS
 theory 
 by developing a superfield analogue of
 the  bosonic 
 CHS
 theory in even dimensions constructed  
 in full generality by Segal \cite{Segal}
(as an extension of his earlier work \cite{Segal2001}),
  in agreement with Tseytlin's observations
  \cite{Tseytlin}.\footnote{For more recent derivations of the Segal theory see, e.g.,  \cite{BJM,Bonezzi} and references therein.}

An important feature of  the approach advocated in \cite{KMT} 
is that it provides a new avenue  to study the problem 
of consistent propagation of conformal higher-spin fields on curved backgrounds. 
It is believed that  a gauge-invariant action for  conformal fields of spin $s>2$
may be defined only  if the background metric is 
a solution of the equation of motion for conformal gravity, 
which means that  the Bach tensor is equal to zero. 
However, 
even the simplest $s=3$ case 
has not yet been studied in full generality
\cite{NT,GrigorievT,BeccariaT,Manvelyan}.
When dealing with $\cN=1$ SCHS theories in curved backgrounds, the gravitational field belongs to the so-called Weyl multiplet
\cite{KTvN1,KTvN2} which also  contains a conformal gravitino and a $\sU(1)$ gauge field. 
It appears that  consistent propagation of 
SCHS
multiplets in such a background may
be defined  if the corresponding super Bach tensor \cite{BK88,BK} 
vanishes\footnote{The terminology ``super Bach tensor'' was introduced in 
\cite{KMT}. In  linearised conformal supergravity the super Bach tensor was first computed by Ferrara and Zumino \cite{FZ}.}
and, therefore, the background Weyl multiplet is a solution to the equations of motion for conformal supergravity. So far explicit calculations have been carried out  
 only for the superconformal gravitino multiplet in a supergravity background
\cite{KMT}.

The $3D$ story is considerably simpler and more complete as far as the issue of 
consistent  propagation of higher-spin fields
on curved backgrounds is concerned. 
The equation of motion for conformal gravity requires the Cotton tensor to vanish 
\cite{DJT1,DJT2,vN}, and therefore curved spacetime is conformally flat. 
In off-shell $\cN$-extended conformal supergravity,  
with $1\leq \cN \leq 6$, 
the superfield Euler-Lagrange equation  states that the super Cotton tensor is equal to zero
\cite{BKNT-M1,BKNT-M2,KNT-M}, and therefore curved superspace is conformally flat. 
It was shown in \cite{SK-MP,HKO} that a gauge-invariant action exists for every conformal 
higher-spin (super)field on arbitrary conformally flat backgrounds for the cases $\cN=0,1,2$.
These results may be naturally extended (at least) to the $\cN=3$ case. 

This paper is a continuation of the research program initiated in \cite{KMT,SK-MP,HKO}.
Our main goal will be to develop a formalism 
with manifest {\it local} (super)conformal symmetry.  
This will allow us, in particular, to elaborate on several  
constructions that were only sketched in \cite{KMT,SK-MP,HKO}.

Two years ago, Ref. \cite{KMT} proposed
off-shell $4D$ $\cN=1$ superconformal higher-spin models
in arbitrary conformally flat supergravity backgrounds.
 Technical details of the corresponding formulation
were not spelled out in \cite{KMT} since the linearised higher-spin (super) Weyl 
tensors were explicitly given in terms of the gauge prepotentials only for the 
models describing conformal superspin values $s=1, \frac 32, \frac 52$.

More recently, off-shell actions were derived  for linearised $3D$ $\cN=0, 1$
(super)conformal higher-spin  gravity in general conformally flat 
(super)gravity backgrounds \cite{SK-MP}.
This construction was also extended 
to the $\cN=2$ superconformal case in \cite{HKO}.
As in the $4D$ analysis of \cite{KMT}, technical details of the 
$3D$ formulations were not given, since closed-form expressions 
for  the linearised higher-spin (super) Cotton
tensors in terms of the gauge prepotentials were not known. 

In this work we fill the technical gaps in the constructions of \cite{KMT,SK-MP,HKO}. In particular, we explicitly construct CHS models that are Weyl invariant in any curved $4D$ spacetime or any conformally flat $3D$ spacetime. In both dimensions the higher-spin gauge invariance of these models holds only in conformally flat spacetimes. Supersymmetric extensions of the models are also given. In addition, by extending the depth of the higher-spin gauge symmetry, we construct novel generalisations of the proposed CHS models whose Weyl and gauge invariance hold under the same conditions.

Of central importance to our approach are 
(i) the formulation of conformal gravity  as the gauge theory of the conformal group
 \cite{KTvN1}; and (ii) 
the off-shell formulations for conformal
supergravity in diverse dimensions known as conformal superspace
\cite{ButterN=1,ButterN=2,BKNT-M1,BKNT-M2,BKNT-M5D,BKNT}, 
an approach pioneered by Butter in the $4D$ case \cite{ButterN=1,ButterN=2}.
Since superfield techniques are not well known within 
the higher-spin community, and also since the conformal superspace approach 
is still familiar only to a limited number of superspace practitioners, 
the details of our approach and its application to CHS theory will be presented from a non-supersymmetric point of view. 
Therefore, the majority of this paper
will be devoted to non-supersymmetric CHS models and their supersymmetric counterparts will be presented at the end with the technical details being simply sketched.

This paper is organised as follows. In section \ref{section2} we review the formulation 
of conformal gravity in $D>2$ dimensions as the gauge theory of the conformal group.
Section \ref{section3} is devoted to $3D$ CHS theories in curved backgrounds. 
Section \ref{section4} discusses $4D$ CHS theories in curved backgrounds. 
Supersymmetric extensions are studied in sections \ref{section5} and \ref{section6}.
Concluding comments are given in section \ref{section7}. The main body of the paper is accompanied by six technical appendices. Appendix \ref{appendixA} and \ref{appendixB} present some of the various conventions that we adopt.  Proofs for several properties of the higher-spin generalised Cotton tensors are provided in Appendix \ref{AppendixC}. Appendix \ref{appendixE} examines the issue of integration by parts in the conformal space. Appendices \ref{appendix-gravitino} and \ref{appendix-graviton} discuss the construction of conformal spin $s=3/2$ and $s=2$ models that are gauge invariant in any $4D$ Bach-flat spacetime.

Before turning to the main body of this paper, several comments are in order 
regarding the existence of 
different ways to describe conformal higher-spin fields. 
They differ only in the sector 
of purely gauge degrees of freedom (compensators) that can be eliminated 
algebraically by applying local symmetry transformations without derivatives. 
The original Fradkin-Tseytlin model \cite{FT} for a conformal field of integer spin 
$s>1$ is described in terms of a symmetric rank-$s$ tensor field
${\bm h}_{a_1 \dots a_s} = {\bm h}_{(a_1 \dots a_s)} \equiv {\bm h}_{a(s)}$ with the gauge transformation law 
\bea
\d {\bm h}_{a_1\dots a_s} = \pa_{(a_1} \x_{a_2 \dots a_s)} 
+ \eta_{(a_1 a_2} \l_{a_3 \dots a_s)}~, \qquad 
\eta^{bc} \x_{bc a_1 \dots a_{s-3}} =0~,
\label{1.1}
\eea 
 where both  gauge parameters $\x_{a( s-1) } $ and  
 $\l_{a(s-2)}$ are symmetric, and  $\x_{a( s-1)} $ is in addition 
 traceless.\footnote{The gauge transformation law \eqref{1.1} is often 
 generalised  by removing the condition  $\eta^{bc} \x_{bc a (s-3)} =0$  imposed on the parameter $\x_{a(s-1)}$. However the resulting transformation law is equivalent to \eqref{1.1} with a modified algebraic 
 parameter $\l_{a(s-2)}$.
 }
 It is natural to interpret 
 the gauge symmetries generated by   $\x_{a( s-1)} $ and $\l_{a(s-2)}$ 
 for $s>2$ as linearised higher-spin gauge and 
 ``generalised Weyl'' transformations, respectively.
The $\l$-gauge freedom in \eqref{1.1}
may be used to make the gauge field ${\bm h}_{a(s)}$
traceless
by requiring
\bea
  {\bm h}_{a(s)} =h_{a(s)}~, \qquad 
  \eta^{bc} {h}_{bc a (s-2) } =0~. 
 \eea 
  If one switches to the two-component spinor notation and introduces
  \bea
  h_{a(s)} ~\to ~h_{\a_1 \dots \a_s \ad_1 \dots \ad_s}
  := (\s^{a_1})_{\a_1 \ad_1 } \dots (\s^{a_s})_{\a_s \ad_s} 
  {h}_{a_1 \dots a_s}~,
  \eea
then the field $h_{\a_1 \dots \a_s \ad_1 \dots \ad_s}$ proves to be symmetric in its undotted indices and, 
separately, in its dotted indices, $h_{\a_1 \dots \a_s \ad_1 \dots \ad_s}
=h_{( \a_1 \dots \a_s)( \ad_1 \dots \ad_s )} \equiv h_{\a(s) \ad(s)}$.
In accordance with \eqref{1.1}, the gauge transformation of $h_{\a(s)\ad(s) }$ is 
\bea
\d h_{\a_1 \dots \a_{s} \ad_1 \dots \ad_{s} } 
&=& \pa_{(\a_1 (\ad_1} \x_{\a_2\dots \a_{s}) \ad_2 \dots \ad_{s})}~.
\label{1.4}
\eea
It is natural 
to think of  $h_{a(s)}$ (or equivalently  $h_{\a(s)\ad(s) }$) as the  genuine conformal spin-$s$ 
gauge field, due to several reasons. Firstly, one
can consistently  define $h_{\a(s)\ad(s) }$ to be a conformal primary field, see section \ref{section4}. Secondly, the other degrees of freedom 
contained in ${\bm h}_{a(s)}$ are purely gauge ones, 
and as such they may become essential only at the nonlinear level.
Finally,  the nonlinear  conformal higher-spin theory of
 \cite{Segal} is formulated in terms of the fields $h_{a(s)}$, with $s=0,1,2, \dots$,
 in the $4D$ case.
  
In principle, one may instead use Fronsdal's doubly traceless
spin-$s$ gauge field 
\cite{Fronsdal1,Fronsdal2}
\bea
{\mathfrak h}_{a_1 \dots a_s} =h_{a_1\dots a_s} 
+ \eta_{(a_1 a_2} \vf_{a_3 \dots a_s) }~, \qquad
\eta^{bc} \vf_{bc a(s-4)} =0~,
\eea
to describe conformal spin-$s$ dynamics. In such an approach $\vf_{a(s-2)}$ is a compensator.
The gauge transformation law of ${\mathfrak h}_{a(s)}$ is given by 
\bea
\d {\mathfrak h}_{a(s)} = \pa_{(a_1} \x_{a_2 \dots a_s)} 
+ \eta_{(a_1 a_2} \tilde{\l}_{a_3 \dots a_s)}~, \qquad 
\eta^{bc} \x_{bc a(s-3)} =0~,\quad
\eta^{bc} \tilde{\l}_{bc a(s-4)} =0~.
\label{1.6}
\eea 
It is clear that the compensator $\vf_{a(s-2)}$  may be gauged away by applying 
a  $\tilde \l$-transformation, 
and then we are back to the formulation in terms of $h_{a(s)}$.

Another description of conformal spin-$s$ dynamics is obtained by employing 
 Vasiliev's frame field \cite{Vasiliev,Vasiliev87}
\bea
{\bm e}_{m, \, a_1 \dots a_{s-1}} = {\bm e}_{m, \, (a_1 \dots a_{s-1})} ~,
\qquad \eta^{bc} {\bm e}_{m, \, bc a (s-3)} =0~.
\eea 
In addition to a higher-spin $\x$-transformation, $\d {\bm e}_{m, a(s-1) } = \pa_m \x_{a(s-1)}$,
there are two additional local symmetries in this setting. These are 
 generalised Lorentz and Weyl transformations,  which do not involve 
 derivatives and allow one to gauge away two compensating degrees of freedom 
 contained in ${\bm e}_{m, a(s-1) }$ by imposing the gauge condition 
 that ${\bm e}_{m, a(s-1) }$ is completely symmetric, 
 ${\bm e}_{m, a(s-1) } = h_{m a(s-1)}$.

Not all  of the field realisations discussed above 
originate in the $4D$ $\cN=1$ superconformal setting.
We recall that the conformal  superspin-$(s+\hf)$ prepotential \cite{HST,KMT}
$H_{\a(s)\ad(s)} := H_{\a_1 \dots \a_{s} \ad_1 \dots \ad_{s}} (\q,\bar \q) $ is a real superfield, which is symmetric in its undotted indices and, independently, in its dotted indices.
The gauge transformation  law of  $H_{\a(s)\ad(s)} $ is 
\bea
 \d H_{\a_1 \dots \a_{s} \ad_1 \dots \ad_{s}} 
 = \bar D_{(\ad_1} \L_{\a_1 \dots \a_{s} \ad_2 \dots \ad_{s} )} 
- D_{(\a_1} \bar{\L}_{\a_2 \dots \a_{s})\ad_1 \dots \ad_{s}} \ ,
\label{1.8}
\eea
with  the gauge parameter $\L_{\a(s) \ad(s-1)}$ being  unconstrained.
For the $s=1$ case this transformation law  corresponds to linearised 
conformal supergravity \cite{FZ}. The gauge freedom makes it possible to choose a
Wess-Zumino gauge 
\bea
H_{\a_1 \dots \a_s \ad_1 \dots \ad_s} (\q, \bar \q) 
&=& \q^\b \bar \q^\bd {\bm e}_{\b, \a_1 \dots \a_s \bd,  \ad_1 \dots \ad_s } 
+\bar \q^2 \q^\b {\bm \j}_{\b,  \a_1 \dots \a_s \ad_1 \dots \ad_s} - \q^2 \bar \q^\bd \bar {\bm \j}_{ \a_1 \dots \a_s \bd , \ad_1 \dots \ad_s } \non \\
&&
+\q^2 {\bar \q}^2 h_{\a_1 \dots \a_s  \ad_1 \dots \ad_s} ~,
\label{1.9}
\eea
where the bosonic fields ${\bm e}_{\b, \a(s) \bd, \ad(s)}
= (\s^m)_{\b\bd} {\bm e}_{m, \, \a (s) \ad(s)} $ 
and $h_{\a(s) \ad(s)}$ are real. In  the Wess-Zumino gauge \eqref{1.9}, 
we stay  with a restricted set of local 
transformations \eqref{1.8}. It is not difficult to check that the transformation law of 
${\bm e}_{m, \, \a (s) \ad(s)} $ coincides with that of the spin-$(s+1)$ frame field 
 \cite{Vasiliev,Vasiliev87}. The gauge transformation of $h_{\a(s) \ad(s)}$ coincides
 with \eqref{1.4}. The fermionic field ${\bm \j}_{\b,  \a (s)  \ad (s)}$ and its conjugate 
 in \eqref{1.9} describe the conformal spin-$(s+\hf)$ gauge field. 
 This field realisation coincides neither with the Fradkin-Tseytlin 
 conformal spin-$(s+\hf)$ field \cite{FT} nor with Vasiliev's fermionic frame
field \cite{Vasiliev,Vasiliev87}.
 
 The residual  gauge freedom \eqref{1.8}, which preserves 
 the Wess-Zumino gauge \eqref{1.9}, contains algebraic local transformations that 
 can be used to eliminate the compensators 
such that $H_{\a(s) \ad(s)}$ takes the form \cite{KMT}
\bea
H_{\a_1 \dots \a_s \ad_1 \dots \ad_s} (\q, \bar \q) 
&=& \q^\b \bar \q^\bd h_{(\b \a_1 \dots \a_s) (\bd \ad_1 \dots \ad_s) } 
+\bar \q^2 \q^\b \j_{(\b \a_1 \dots \a_s) \ad_1 \dots \ad_s} \non \\
&&- \q^2 \bar \q^\bd \bar \j_{ \a_1 \dots \a_s (\bd \ad_1 \dots \ad_s) } 
+\q^2 {\bar \q}^2 h_{\a_1 \dots \a_s  \ad_1 \dots \ad_s} ~.
\eea
The gauge transformation of $\j_{\a(s+1) \ad(s)}$ is 
\bea
\d \j_{\a_1 \dots \a_{s+1} \ad_1 \dots \ad_{s} } 
&=& \pa_{(\a_1 (\ad_1} \r_{\a_2\dots \a_{s+1}) \ad_2 \dots \ad_{s})}~.
\eea
It is natural to think of  field $\j_{\a(s+1) \ad(s)}$ and its conjugate  
as the genuine conformal spin-$(s+\hf)$ gauge field.


\section{Conformal geometry}\label{section2}
 
Conformal (super)gravity as the gauge theory of the (super)conformal group
was constructed long ago \cite{KTvN1,KTvN2,BdeRdeW,vN}, 
see  \cite{FT,FVP} for pedagogical reviews. 
In this section we give a brief review of the formulation for 
conformal gravity in $D>2$ dimensions following \cite{BKNT-M1}. 
 This setting is known to be ideal for extensions to 
 the superspace formulations for conformal supergravity 
 in diverse dimensions \cite{ButterN=1,ButterN=2,BKNT-M2,BKNT-M5D}.
It also turns out to be useful in the framework of higher-spin (super)conformal dynamics, as will be shown below.


\subsection{Gauging the conformal algebra} \label{subsection2.1}

The conformal algebra in $D>2$ dimensions, $\mathfrak{so}(D,2)$, consists of the translation $(P_a)$, Lorentz $(M_{ab})$, special conformal $(K_a)$ and dilatation $(\mathbb{D})$ generators. The non-vanishing commutators are given by
\begin{subequations} \label{confal}
\begin{align}
&[M_{ab},M_{cd}]=2\eta_{c[a}M_{b]d}-2\eta_{d[a}M_{b]c}~, \phantom{inserting blank space inserting}\\
&[M_{ab},P_c]=2\eta_{c[a}P_{b]}~, \qquad \qquad \qquad \qquad ~ [\mathbb{D},P_a]=P_a~,\\
&[M_{ab},K_c]=2\eta_{c[a}K_{b]}~, \qquad \qquad \qquad \qquad [\mathbb{D},K_a]=-K_a~,\\
&[K_a,P_b]=2\eta_{ab}\mathbb{D}+2M_{ab}~.
\end{align}
\end{subequations}
The generators $M_{ab},K_a$ and $\mathbb{D}$ span a subalgebra 
of $\mathfrak{so}(D,2)$
and are collectively referred to as $X_{\un{a}}$. In contrast, we denote the generators of the full algebra by $X_{\tilde{a}}$. Then, the commutation relations  \eqref{confal} may be rewritten as follows\footnote{We adopt the convention whereby a factor of 1/2 is inserted when summing over pairs of antisymmetric indices. For example, $f_{\un{a}\un{b}}{}^{\un{c}}X_{\un{c}}=f_{\un{a}\un{b}}{}^{K_c}K_{c}+\frac{1}{2}f_{\un{a}\un{b}}{}^{M_{cd}}M_{cd}+\dots$.}
\begin{subequations}\label{confal2}
\begin{align}
&[X_{\un{a}},X_{\un{b}}]=-f_{\un{a}\un{b}}{}^{\un{c}}X_{\un{c}}~,\\
&[X_{\un{a}},P_b]=-f_{\un{a}b}{}^{\un{c}}X_{\un{c}}-f_{\un{a}b}{}^{c}P_c~ \label{confal2b}
\end{align}
\end{subequations}
where $f_{\tilde{a}\tilde{b}}{}^{\tilde{c}}$ are
the structure constants whose non-vanishing components  are: 
\begin{subequations} \label{strcon1}
\begin{align}
f_{M_{ab},M_{cd}}{}^{M_{fg}}&=4\eta_{a[c}\delta_{d]}^{[f}\delta_{b}^{g]}-4\eta_{b[c}\delta_{d]}^{[f}\delta_{a}^{g]}~,\\
f_{M_{ab},P_c}{}^{P_d}&=-2\eta_{c[a}\delta_{b]}^d~, \qquad \qquad \qquad \qquad ~ f_{\mathbb{D}, P_a}{}^{P_b}=-\delta_a^b~, \\
f_{M_{ab},K_c}{}^{K_d}&=-2\eta_{c[a}\delta_{b]}^d~, \qquad \qquad \qquad \qquad f_{\mathbb{D}, K_a}{}^{K_b}=\delta_a^b~,\\
f_{K_a,P_b}{}^{\mathbb{D}}&=-2\eta_{ab}~,\qquad \qquad \qquad \qquad  ~~ f_{K_a,P_b}{}^{M_{cd}}=-4\delta_{a}^{[c}\delta_{b}^{d]}~.
\end{align}
\end{subequations} 
The structure constants satisfy the Jacobi identities
\begin{align} \label{jacobi}
f_{[\tilde{a}\tilde{b}}{}^{\tilde{d}}f_{\tilde{c}]\tilde{d}}{}^{\tilde{e}}=0~.
\end{align}

Let $\mathcal{M}^D$
be a $D$-dimensional spacetime 
 parameterised 
 by local coordinates $x^m$, where $m=0,1,\dots,D-1$. 
  To gauge the conformal algebra, we associate with each generator $X_{\un{a}}$ a connection one-form, $\omega^{\un{a}}=\text{d}x^m\omega_{m}{}^{\un{a}}$, and with $P_a$ the vielbein $e^a=\text{d}x^me_m{}^{a}$. We denote by $\mathcal{H}$ the gauge group generated by $X_{\un{a}}$ and postulate that 
  $e^a$ and $\omega^{\un{a}}$ transform
  under $\mathcal{H}$
as
\begin{subequations}\label{contrans}
\begin{align}
\delta_{\mathcal{H}}e^{a}&=e^{b}\Lambda^{\un{c}}f_{\un{c}b}{}^{a}~,
\label{contrans-a}
\\
\delta_{\mathcal{H}}\omega^{\un{a}}&=\text{d}\Lambda^{\un{a}}+e^b\Lambda^{\un{c}}f_{\un{c}b}{}^{\un{a}}+\omega^{\un{b}}\Lambda^{\un{c}}f_{\un{c}\un{b}}{}^{\un{a}}~,
\label{contrans-b}
\end{align}
\end{subequations}
with gauge parameter $\Lambda^{\un{a}}$. 

Given a field $\Phi$ (with its indices suppressed),
we say that $\F$  is $\cH$-covariant 
if it transforms under the action of $\cH$
with no derivative on the parameter, 
$\delta_{\mathcal{H}}\Phi=\Lambda^{\un{a}}X_{\un{a}}\Phi$. 
In addition, if $\Phi$ satisfies
\begin{align}
K_{a}\Phi=0~,\qquad \mathbb{D}\Phi=\Delta\Phi~,
\end{align}
it is called a primary field of dimension (or Weyl weight) $\D$.

It is clear that $\pa_m \F$ is no longer $\cH$-covariant.
We are therefore led to introduce a covariant derivative according to
\begin{align}\label{2.5}
\nabla_m=\partial_m-\omega_m{}^{\un{a}}X_{\un{a}}~.
\end{align} 
 It follows from \eqref{contrans}
 that $\nabla_a\Phi=e_a{}^{m}\nabla_m\Phi$ transforms covariantly,
 \begin{align}
 \delta_{\mathcal{H}}(\nabla_a\Phi)=\Lambda^{\un{b}}\nabla_aX_{\un{b}}\Phi-\Lambda^{\un{b}}f_{\un{b}a}{}^{c}\nabla_c\Phi-\Lambda^{\un{b}}f_{\un{b}a}{}^{\un{c}}X_{\un{c}}\Phi~. \label{2.6}
 \end{align}
From eq \eqref{2.6} we can deduce the commutation relations of $X_{\un{a}}$ with $\nabla_a$,
 \begin{align}\label{confal3}
 [X_{\un{a}},\nabla_b]=-f_{\un{a}b}{}^{\un{c}}X_{\un{c}}-f_{\un{a}b}{}^{c}\nabla_c~.
 \end{align}
Comparing this with \eqref{confal2b} we see that $X_{\un{a}}$ satisfies the same commutation relations with $\nabla_{{b}}$
as it does with $P_b$. However, unlike the translation generators $P_a$, 
the commutator of two covariant derivatives is not zero but is given by 
\begin{align} 
[\nabla_a,\nabla_b]=-\mathcal{T}_{ab}{}^{c}\nabla_c-\mathcal{R}_{ab}{}^{\un{c}}X_{\un{c}}~.\label{nablacom1}
\end{align}
In eq \eqref{nablacom1}, $\mathcal{T}_{ab}{}^{c}$ and $\mathcal{R}_{ab}{}^{\un{c}}$ are the torsion and curvature tensors respectively,
\begin{subequations} \label{fieldstr}
\begin{align}
\mathcal{T}_{ab}{}^{c}&=-\mathscr{C}_{ab}{}^{c}+2\omega_{[a}{}^{\un{d}}f_{b]\un{d}}{}^{c}~,\\
\mathcal{R}_{ab}{}^{\un{c}}&=-\mathscr{C}_{ab}{}^{c}\omega_{c}{}^{\un{c}}+2\omega_{[a}{}^{\un{d}}f_{b]\un{d}}{}^{\un{c}}+\omega_{[a}{}^{\un{e}}\omega_{b]}{}^{\un{d}}f_{\un{de}}{}^{\un{c}}+2e_{[a}\omega_{b]}{}^{\un{c}} ~,
\end{align}
\end{subequations}
where $e_a=e_a{}^{m}\partial_m$ is the inverse vielbein and the anholonomy coefficients, 
$\mathscr{C}_{ab}{}^{c}$, are given by
\begin{align}
\mathscr{C}_{ab}{}^{c}=(e_ae_b{}^{m}-e_be_a{}^{m})e_m{}^{c}~.
\end{align}
The definitions \eqref{2.5} and \eqref{nablacom1} differ
from
those used in some  previous publications. See appendix \ref{appendixB} for a dictionary to convert between these conventions.

Using the transformation rules \eqref{contrans} and the Jacobi identities \eqref{jacobi}, we find that the torsion and curvature tensors \eqref{fieldstr} transform 
covariantly under $\mathcal{H}$ according to
\begin{subequations}\label{2.11}
\begin{align}
\delta_{\mathcal{H}}\mathcal{T}_{ab}{}^{c}&=\mathcal{T}_{ab}{}^{d}\Lambda^{\un{e}}f_{\un{e}d}{}^{c}
-2\Lambda^{\un{a}}f_{\un{a}[a}{}^{d}\mathcal{T}_{b]d}{}^c~,\\
\delta_{\mathcal{H}}\mathcal{R}_{ab}{}^{\un{c}}&=\mathcal{R}_{ab}{}^{\un{e}}\Lambda^{\un{d}}f_{\un{de}}{}^{\un{c}}+2\Lambda^{\un{d}}f_{\un{d}[a}{}^{e}\mathcal{R}_{b]e}{}^{\un{c}}+\mathcal{T}_{ab}{}^{e}\Lambda^{\un{f}}f_{\un{f}e}{}^{\un{c}}~.
\end{align}
\end{subequations}
In this formulation infinitesimal general coordinate transformations, generated by a local parameter $\xi^a$, are not covariant with respect to $\mathcal{H}$. To remedy this, they must be supplemented by an additional $\mathcal{H}$-transformation with gauge parameter $\Lambda^{\un{a}}=\xi^a\omega_a{}^{\un{a}}~$,
\begin{align}
\delta_{\text{cgct}}(\xi^a)=\delta_{\text{gct}}(\xi^a)-\delta_{\mathcal{H}}(\xi^a\omega_a{}^{\un{a}})~.
\end{align}
 It follows that such transformations act on fields $\Phi$ (with all indices Lorentz) as $\delta_{\text{cgct}}\Phi=\xi^a\nabla_a\Phi$. The conformal gravity gauge group, denoted by $\mathcal{G}$, is then generated by the set of operators $(\nabla_a,X_{\un{a}})$ under which $\Phi$ transforms as
 \begin{align}
 \delta_{\mathcal{G}}\Phi=\mathcal{K}\Phi,\qquad \mathcal{K}=\xi^b\nabla_b+\Lambda^{\un{b}}X_{\un{b}}~.
 \end{align}
 Finally, the gauge transformation of $\nabla_a$ under $\mathcal{G}$ proves to obey the relation
 \begin{align}
 \delta_{\mathcal{G}}\nabla_a=[\mathcal{K},\nabla_a]
 \end{align}
 provided we interpret
 \begin{align}
 \nabla_a\xi^b:=e_a\xi^b+\omega_a{}^{\un{c}}\xi^df_{d\un{c}}{}^{b}~,\qquad \nabla_{a}\Lambda^{\un{b}}:=e_a\Lambda^{\un{b}}+\omega_a{}^{\un{c}}\xi^df_{d\un{c}}{}^{\un b}+\omega_{a}{}^{\un{c}}\Lambda^{\un{d}}f_{\un{dc}}{}^{\un{b}}~.
 \end{align}
 
Through this procedure the entire conformal algebra has been gauged in such a way that the generators $X_{\un{a}}$ act on $\nabla_a$ in the same way as they do on $P_a$.


\subsection{Conformal gravity}

 The covariant derivatives given by eq. \eqref{2.5} are
\begin{align}
\nabla_a=e_a-\frac{1}{2}\hat{\o}_{a}{}^{bc}M_{bc}-\mathfrak{f}_{a}{}^{b}K_b-\mathfrak{b}_a\mathbb{D}~,
\label{2.177}
\end{align}
where $\hat{\o}_{a}{}^{bc}$, $\mathfrak{f}_a{}^{b}$ and $\mathfrak{b}_a$ are the Lorentz, special conformal and dilatation connections respectively. They satisfy the commutation relations
\begin{align}
[\nabla_a,\nabla_b]=-\mathcal{T}_{ab}{}^{c}\nabla_c-\frac{1}{2}\mathcal{R}(M)_{ab}{}^{cd}M_{cd}-\mathcal{R}(K)_{ab}{}^{c}K_c-\mathcal{R}(\mathbb{D})_{ab}\mathbb{D}
\end{align}
where the torsion and curvatures are
\begin{subequations}
 \begin{align}
 \mathcal{T}_{ab}{}^c&=-\mathscr{C}_{ab}{}^{c}+2\hat{\o}_{[ab]}{}^c+2\mathfrak{b}_{[a}\delta_{b]}{}^{c}~,\\
 \mathcal{R}(M)_{ab}{}^{cd}&=\hat{R}_{ab}{}^{cd}+8\mathfrak{f}_{[a}{}^{[c}\delta_{b]}{}^{d]}~,\label{LorCur}\\ 
 \mathcal{R}(K)_{ab}{}^{c}&=-\mathscr{C}_{ab}{}^d\mathfrak{f}_d{}^{c}-2\hat{\o}_{[a}{}^{cd}\mathfrak{f}_{b]d}-2\mathfrak{b}_{[a}\mathfrak{f}_{b]}{}^{c}+2e_{[a}\mathfrak{f}_{b]}{}^{c}~, \label{2.19cc}\\
 \mathcal{R}(\mathbb{D})_{ab}&= -\mathscr{C}_{ab}{}^{c}\mathfrak{b}_c+4\mathfrak{f}_{[ab]}+2e_{[a}\mathfrak{b}_{b]}~,\\
  \hat{R}_{ab}{}^{cd}&=-\mathscr{C}_{ab}{}^{f}\hat{\o}_{f}{}^{cd}+2e_{[a}\hat{\o}_{b]}{}^{cd}-2\hat{\o}_{[a}{}^{cf}\hat{\o}_{b]f}{}^{d}~. \label{RiemannB}
 \end{align}
 \end{subequations}
Here $\hat{R}_{abcd}$ is the Riemann tensor 
corresponding to the spin connection  $\hat{\o}_{abc}$. 

To ensure that the vielbein is the only independent field in the theory
modulo purely gauge degrees of freedom, 
we have to impose covariant constraints.
 These constraints are
 \begin{subequations}\label{constraints}
 \begin{align}
 \mathcal{T}_{ab}{}^c=0~,\\
 \eta^{bd}\mathcal{R}(M)_{abcd}=0~.
 \end{align}
 \end{subequations}
Indeed, the conditions \eqref{constraints} are preserved by $\mathcal{H}$-transformations, which may be verified through \eqref{2.11}. The first constraint determines the spin connection in terms of the vielbein and the dilatation connection, 
  \begin{align}\label{spincon}
  \hat{\omega}_{abc}&=\o_{abc}-2\eta_{a[b}\mathfrak{b}_{c]}~,
  \end{align}
 where $\o_{abc} \equiv \o_{abc} (e)
 =\frac{1}{2}\big(\mathscr{C}_{abc}-\mathscr{C}_{acb}-\mathscr{C}_{bca}\big)$ 
 is the standard torsion-free Lorentz connection. 
 Similarly, the second constraint fixes the special conformal connection to be
 \begin{align}
 \mathfrak{f}_{ab}=-\frac{1}{2(D-2)}\hat{R}_{ab}+\frac{1}{4(D-1)(D-2)}\eta_{ab}\hat{R}~, \label{2.16}
 \end{align}
 where $\hat{R}_{ab}=\eta^{cd}\hat{R}_{acbd}$ is the (non-symmetric) Ricci tensor and $\hat{R}=\eta^{ab}\hat{R}_{ab}$ is the scalar curvature. 
 
 Rather than imposing an extra constraint to fix $\mathfrak{b}_a$ in terms of the vielbein, we observe that under a $K$-gauge transformation, $\mathfrak{b}_a$ transforms as
 \begin{align}\label{2.17}
 \delta_{K}\mathfrak{b}_a=-2\Lambda(K)_a~.
 \end{align}
 It follows that we may impose the gauge condition 
 \begin{align}\label{degauge}
 \mathfrak{b}_a=0~.
 \end{align}
After this choice, only the vielbein remains as an independent field. The gauge \eqref{degauge} breaks the special conformal symmetry. For our purposes, it is desirable to keep this symmetry intact throughout calculations and impose \eqref{degauge} only at the end when we wish to extract physically meaningful results. This process is referred to as `degauging'.
 
Making use of \eqref{constraints} and the Jacobi identity
\begin{align}
0=\big[\nabla_a,[\nabla_b,\nabla_c]\big]+\big[\nabla_b,[\nabla_c,\nabla_a]\big]+\big[\nabla_c,[\nabla_a,\nabla_b]\big]~,
\end{align}  
we find that the dilatation field strength vanishes,
\begin{align}
\mathcal{R}(\mathbb{D})_{ab}&=0~,
\end{align}
along with the following Bianchi identities
\begin{subequations}
\begin{align}
\mathcal{R}(K)_{[abc]}&=0~,\\
\mathcal{R}(M)_{[abc]d}&=0~,\\
\nabla_{[a}\mathcal{R}(K)_{bc]d}&=0~,\\
\nabla_{[a}\mathcal{R}(M)_{bc]}{}^{de}-4\mathcal{R}(K)_{[ab}{}^{[d}\delta_{c]}{}^{e]}&=0~.\label{2.87}
\end{align}
\end{subequations} 

 Making use of  \eqref{RiemannB} and \eqref{spincon}
 allows us to decompose $\hat{R}_{abcd}$ 
  into those terms which depend solely on the vielbein
 and those involving the dilatation connection,
\begin{subequations} \label{2.99}
 \bea
 \hat{R}_{abcd}&=&R_{abcd}-4e_{[a}\eta_{b][c}\mathfrak{b}_{d]}
 -4\eta_{[c[a}\omega_{b]d]}{}^{g}\mathfrak{b}_g
 +4\mathfrak{b}_{[c}\eta_{d][a}\mathfrak{b}_{b]}
 +2\eta_{c[a}\eta_{b]d}\mathfrak{b}^f\mathfrak{b}_f~,
\label{2.999}\\
 \hat{R}_{ab}&=&R_{ab}
+ (D-2) \Big\{ e_a\mathfrak{b}_b - \o_{ab}{}^{c}\mathfrak{b}_c - \mathfrak{b}_a\mathfrak{b}_b
 \Big\}
\non\\
&&
\phantom{R_{ab}}
+\eta_{ab} \Big\{ e_c\mathfrak{b}^c -\o^c{}_{cd}\mathfrak{b}^d
+  (D-2)  \mathfrak{b}_c  \mathfrak{b}^c\Big\}
~,\\
\hat{R}&=&R+2 (D-1) \Big\{ e_a\mathfrak{b}^a-\o^a{}_{ab}\mathfrak{b}^b
+\hf (D-2)\mathfrak{b}^a\mathfrak{b}_a \Big\}
~.
 \eea
 \end{subequations}
Here $R_{abcd}$ is the Riemann tensor associated with  
the spin connection $\o_{abc}$,
\begin{align}
R_{ab}{}^{cd}=2e_{[a}\omega_{b]}{}^{cd}-2\o_{[ab]}{}^{f}\omega_{f}{}^{cd}-2\omega_{[a}{}^{cf}\omega_{b]f}{}^{d}~,
\end{align}
and $R_{ab}$ and $R$  stand for  the corresponding (symmetric) Ricci tensor 
and scalar curvature, respectively. 
Inserting the relations \eqref{2.99} into the solution 
to the conformal gravity constraint \eqref{2.16} yields
 \begin{align}\label{2.96}
 \mathfrak{f}_{ab}=-\frac{1}{2}P_{ab}+\frac{1}{2}\mathfrak{b}_a\mathfrak{b}_b-\frac{1}{4}\eta_{ab}\mathfrak{b}^c
\mathfrak{b}_c+\frac{1}{2}\o_{ab}{}^{c}\mathfrak{b}_c-\frac{1}{2}e_{a}\mathfrak{b}_b~,
 \end{align} 
 where $P_{ab}$ is the  Schouten tensor,
 \begin{align}\label{2.97}
 P_{ab}=\frac{1}{(D-2)}\bigg(R_{ab}-\frac{1}{2(D-1)}\eta_{ab}R\bigg)~.
 \end{align}
 Using eqs. \eqref{2.999} and \eqref{2.96} allows us to show that 
 the dependence on the dilatation connection drops out of \eqref{LorCur}
 and we obtain
 \begin{align}
\mathcal{R}(M)_{abcd}=C_{abcd}~. \label{2.98}
 \end{align}
Here $C_{abcd}$ is the Weyl tensor,
\begin{align}\label{Weyl}
C_{abcd}=R_{abcd}-\frac{2}{(D-2)}\bigg(R_{a[c}\eta_{d]b}-R_{b[c}\eta_{d]a}\bigg)+\frac{2}{(D-1)(D-2)}\eta_{a[c}\eta_{d]a}R~,
\end{align}
which is a primary field of dimension +2,
\begin{align}
K_eC_{abcd}=0~, \qquad \mathbb{D}C_{abcd}=2C_{abcd}~.
\end{align}

For further analysis of the constraints, 
it is necessary to consider separately the choices $D=3$ and $D>3$. 
In both cases we make use of the Lorentz covariant derivative defined by
\begin{align}\label{2.100}
 \hat{\mathcal{D}}_{a}&=e_a-\frac{1}{2}\hat{\o}_{a}{}^{bc}M_{bc}=\mathcal{D}_a+\mathfrak{b}^cM_{ac}
 \end{align}
 where $\mathcal{D}_a=e_a-\frac{1}{2}\o_{a}{}^{bc}M_{bc}$ is the torsion-free 
 Lorentz covariant derivative.

 We note that whenever the gauge \eqref{degauge} is chosen, all hatted objects coincide with their non-hatted counterparts.  In particular
 \begin{align}
 \hat{\mathcal{D}}_a\big|_{\mathfrak{b}_a=0}=\mathcal{D}_a~,\qquad \hat{R}_{abcd}\big|_{\mathfrak{b}_a=0}=R_{abcd}~,
 \end{align}
 and in this gauge we may therefore abandon the hat notation without any ambiguity. Furthermore, in this case it is clear that the conformal covariant derivative takes the form  
\be
\nabla_a=\mathcal{D}_a+\frac 12 P_{a}{}^{b}K_b~. \label{2.377}
\ee 
Therefore, in any spacetime with vanishing Schouten tensor, the degauging process is trivial.  
  
In the  $D>3$ case, it follows from \eqref{2.87} and \eqref{2.98} that the special conformal curvature is given by 
 \begin{align}\label{2.60}
 \mathcal{R}(K)_{abc}=\frac{1}{2(D-3)}\nabla^dC_{abcd}~.
 \end{align}
 As a result, the algebra of conformal covariant derivatives is
\begin{align}\label{2.61}
[\nabla_a,\nabla_b]=-\frac{1}{2}C_{abcd}M^{cd}-\frac{1}{2(D-3)}\nabla^dC_{abcd}K^c~.
\end{align}
It is determined by a single primary tensor field, 
the Weyl tensor.

The expressions \eqref{2.19cc} and \eqref{2.60} are two equivalent representations for the special conformal curvature. Upon imposing the gauge \eqref{degauge} these relations 
lead
to the well-known Bianchi identity 
\be 
\mathcal{D}^dC_{abcd}=-2(D-3)\mathcal{D}_{[a}P_{b]c}~.
\ee 

From \eqref{2.61} it follows that if the spacetime under consideration is conformally flat, 
then the conformal covariant derivatives commute,
\begin{align}\label{2.62}
C_{abcd}=0 \quad \implies \quad [\nabla_a,\nabla_b]=0~.
\end{align}
This observation will be important for our subsequent analysis.


\subsection{Conformal gravity in three dimensions}

The Weyl tensor vanishes identically in three dimensions.
As a result, the Lorentz curvature \eqref{2.98} also vanishes and the algebra of conformal covariant derivatives takes the form 
\begin{align}\label{2.21}
[\nabla_a,\nabla_b]=-\mathcal{R}(K)_{ab}{}^cK_c~.
\end{align}
Therefore, all information about conformal geometry 
is encoded in 
a single primary field,
 $\mathcal{R}(K)_{abc}$, which proves to be proportional to the Cotton tensor, 
 as we now show.

The Lorentz covariant derivative \eqref{2.100} allows us to represent \eqref{2.19cc} as
\begin{align}\label{2.23}
 \mathcal{R}(K)_{abc}&= 2\mathcal{D}_{[a}\mathfrak{f}_{b]c}-2\mathfrak{b}_{[a}\mathfrak{f}_{b]c}+2\mathfrak{b}_c\mathfrak{f}_{[ab]}+2\eta_{c[a}\mathfrak{f}_{b]d}\mathfrak{b}^d~.
 \end{align}
  Using \eqref{2.96}, one may show that dependence on $\mathfrak{b}_a$ in eq. \eqref{2.23} drops out such that 
 \begin{align}\label{2.25}
 \mathcal{R}(K)_{abc}=-\frac{1}{2}W_{abc}~,\qquad W_{abc}=2\mathcal{D}_{[a}P_{b]c}~.
 \end{align}
Here  $W_{abc}$ is the Cotton tensor, which is 
 a primary field of dimension $+3$,
\begin{align}
K_dW_{abc}=0~,\qquad \mathbb{D}W_{abc}=3W_{abc}~.
\end{align} 
 It is useful to introduce the dual of the Cotton tensor,
 \begin{align}
 W_{ab}=\frac{1}{2}\ve_{acd}W^{cd}{}_{b}~, \qquad W_{abc}=-\ve_{abd}W^d{}_{c}~,
 \end{align}
 which is symmetric and traceless,
 \begin{align}
 W_{ab}=W_{ba}~,\qquad W^b{}_{b}=0~.
 \end{align}
 On account of the Bianchi identity $\mathcal{D}^aR_{ab}=\frac{1}{2}\mathcal{D}_bR~$, it is also 
 conserved, 
 \begin{align}
 \mathcal{D}^aW_{ab}=0~.
 \end{align}
 
The Cotton tensor 
contains  all information about
   the conformal geometry of $D=3$ spacetime, 
 and it vanishes if and only if spacetime is conformally flat. 
 As follows from   \eqref{2.21}, 
  the commutator of conformal covariant derivatives vanishes
  in the conformally flat case,
 \begin{align} \label{2.33}
 W_{abc}=0\quad \implies \quad [\nabla_a,\nabla_b]=0~.
 \end{align}

 In three dimensions, the Einstein-Hilbert action is known 
 to propagate no local degrees of freedom. However, non-trivial 
 dynamics emerge in 
 topologically massive gravity \cite{DJT1,DJT2} which is obtained by 
 combining 
 the Einstein-Hilbert action with a Lorentz Chern-Simons term.
 The latter proves to coincide with the action for $D=3$ conformal gravity\footnote{An
 alternative approach to conformal gravity in three dimensions was developed in \cite{HW}.}
   \cite{vN} 
\begin{align}\label{CG}
S_{\text{CG}}= \frac 16
 \int\Sigma_{\text{CS}}~,
\end{align} 
where  
the three-form 
 \begin{align}\label{csform}
  \Sigma_{\text{CS}}=\mathcal{R}^{\tilde{b}}\wedge \o^{\tilde{a}}\Gamma_{\tilde{a}\tilde{b}}+\frac{1}{6}\o^{\tilde{c}}\wedge\o^{\tilde{b}}\wedge\o^{\tilde{a}}f_{\tilde{a}\tilde{b}\tilde{c}}
 \end{align}
 varies 
 under an infinitesimal $\mathcal{H}$-transformation by  an exact form,
\begin{align}\label{exact}
\delta_{\mathcal{H}}\Sigma_{\text{CS}}=\text{d}\big(\text{d}\o^{\tilde{b}}\Lambda^{\tilde{a}}\Gamma_{\tilde{a}\tilde{b}}\big)~,\qquad \Lambda^{\tilde{a}}=\big(0,~\Lambda^{\un{a}}\big)~.
\end{align} 
Here and in \eqref{csform}, $\Gamma_{\tilde{a}\tilde{b}}=f_{\tilde{a}\tilde{d}}{}^{\tilde{c}}f_{\tilde{b}\tilde{c}}{}^{\tilde{d}}$ is the symmetric non-degenerate Cartan-Killing 
metric on $\mathfrak{so}(3,2)$ and $f_{\tilde{a}\tilde{b}\tilde{c}}=f_{\tilde{a}\tilde{b}}{}^{\tilde{d}}\Gamma_{\tilde{d}\tilde{c}}$ are the totally antisymmetric structure constants, see appendix \ref{appendixA}. 
We have also used a unified notation \cite{BKNT-M2}  
whereby the connection one-forms are written as $\o^{\tilde{a}}=(e^a,\o^{\un{a}})$ and the curvature two-forms as $\mathcal{R}^{\tilde{a}}=\frac{1}{2}e^c\wedge e^b \mathcal{R}_{bc}{}^{\tilde{a}}=(\mathcal{T}^a,\mathcal{R}^{\un{a}})$. 
It should be remarked that we have adopted the super-form conventions for differential forms, see, e.g.,  \cite{WB} for the details.

The action for conformal gravity \eqref{CG}
can be rewritten in  
 the form
 \begin{align}\label{CG1}
 S_{\text{CG}}=\frac 14
 \int\text{d}^3x\,e\,\ve^{abc}\bigg\{\hat{R}_{ab}{}^{fg}\hat{\o}_{cfg}
 -\frac{2}{3}\hat{\o}_{ad}{}^{e}\hat{\o}_{be}{}^{f}\hat{\o}_{cf}{}^{d}+8\mathfrak{f}_{ab}\mathfrak{b}_c\bigg\}~,\qquad e:=\text{det}(e_m{}^{a})~.
 \end{align}
 Since \eqref{CG} is inert under $K$-transformations up to a total derivative, 
 the dependence on $\mathfrak{b}_a$ once again drops out\footnote{This may be shown explicitly using the relations \eqref{spincon}, \eqref{2.999} and \eqref{2.96}.} and the action simplifies to
 \begin{align}\label{CG2}
 S_{\text{CG}}=\frac14
 \int\text{d}^3x\,e\,\ve^{abc}\bigg\{R_{ab}{}^{fg}\o_{cfg}-\frac{2}{3}\o_{ad}{}^{e}\o_{be}{}^{f}\o_{cf}{}^{d}\bigg\}~.
 \end{align}
Equivalently, one may arrive at equation \eqref{CG2} from \eqref{CG1} by making use of the special conformal symmetry to impose the gauge \eqref{degauge}. 
  
  The equation of motion derived from  the action \eqref{CG2} is 
 \begin{align}
 W_{ab}=0~. \label{2.80}
 \end{align}
 Such a conformally flat background has to be used in order to linearise the conformal 
 gravity action \eqref{CG2}. Before doing that, let us work out how geometric objects change 
 under an infinitesimal deformation of the vielbein, 
 \begin{subequations}
 \begin{align}
 \delta e_{a}{}^{m}=h_{a}{}^{b}e_{b}{}^{m}~, \qquad \delta e_m{}^{a}=-e_m{}^{b}h_b{}^{a}~.
 \end{align}
for some second-rank tensor $h_{ab}$.
Since the antisymmetric 
and trace parts of $h_{ab}$ correspond  to  Lorentz and Weyl transformations,
respectively, and we know the behaviour of the geometric objects under such transformations, 
it suffices to
choose $h_{ab}$ to be symmetric and traceless,
\begin{align}
h_{ab}=h_{ba}~,\qquad h^a{}_{a}=0~.
\end{align}
 \end{subequations}
We represent the corresponding change that the covariant derivative suffers as
 \begin{subequations} \label{deform1}
 \begin{align}
 \d \mathcal{D}_a =
 h_a{}^{b}\mathcal{D}_b-\frac{1}{2}\Xi_a{}^{bc}M_{bc}~,
 \end{align}
 where $\mathcal{D}_a$ is the background torsion-free Lorentz covariant derivative 
 and $\Xi_{abc}$ is a  deformation of the spin connection. 
 The latter is determined by
  imposing the torsion-free condition on 
 $\mathcal{D}'_a = \mathcal{D}_a+ \d \mathcal{D}_a  $, and the result is
 \begin{align}
\Xi_{abc}=-2\mathcal{D}_{[b}h_{c]a}~.
 \end{align}
 \end{subequations}
Making use of \eqref{deform1} leads to the well-known relations
\begin{subequations} \label{deform2}
 \begin{align}
 \delta R_{abcd}&=-2h^f{}_{[a}R_{b]fcd}-2\mathcal{D}_a\mathcal{D}_{[c}h_{d]b}+2\mathcal{D}_b\mathcal{D}_{[c}h_{d]a}~,\\
 \delta R_{ab}&=2h^f{}_{(a}R_{b)f}-\square h_{ab}+2\mathcal{D}^f\mathcal{D}_{(a}h_{b)f}~,\\
\delta R&= 2h^{ab}R_{ab}+2\mathcal{D}^a\mathcal{D}^bh_{ab}~,
 \end{align}
 \end{subequations}
 where $\square=\mathcal{D}^a\mathcal{D}_a$. 
 The relations  \eqref{deform1} and  \eqref{deform2} allow
 one to read off the deformation $\d W_{ab}$
 of the Cotton tensor. For our subsequent consideration in section \ref{section3}
 it is suitable to give the expression for $\d W_{ab}$ in the spinor 
 notation. A summary  of our spinor conventions is given in 
 appendix \ref{appendixA} to which the reader is referred for the technical details.
 
 Associated with  the traceless  Ricci tensor, $R_{ab}-\frac{1}{3}\eta_{ab}R$, and  
 the Cotton tensor, $W_{ab}$,
 are symmetric rank-four spinors defined by 
 \begin{subequations}
 \bea
  R_{\a\b\g\d}&=&\big(\gamma^a\big)_{\a\b}\big(\gamma^b\big)_{\g\d}\bigg(R_{ab}-\frac{1}{3}\eta_{ab}R\bigg)=R_{(\a\b\g\d)}~, \\
 W_{\a\b\g\d}&=&(\g^a)_{\a\b}(\g^b)_{\g\d}W_{ab}=W_{(\a\b\g\d)}~.
 \eea
 \end{subequations}
 The latter can be represented in the form
 \bea 
 W_{\a\b\g\d}=\mathcal{D}^{\sigma}{}_{(\a}R_{\b\g\d)\sigma}~,
 \eea
 where 
 $\mathcal{D}_{\a\b}=(\g^a)_{\a\b}\mathcal{D}_a$. 
 The infinitesimal deformation defined by \eqref{deform1} and \eqref{deform2} 
 may be shown to lead to 
 \begin{subequations}
 \begin{align}
 \delta R_{\a(4)}&=-\mathcal{D}_{(\a_1}{}^{\b_1}\mathcal{D}_{\a_2}{}^{\b_2}h_{\a_3\a_4)\b(2)}+\frac{1}{2}R^{\b(2)}{}_{(\a_1\a_2}h_{\a_3\a_4)\b(2)}+\frac{1}{6}Rh_{\a(4)}~,\\
 \delta W_{\a(4)}&=\frac{1}{2}W^{\b(2)}{}_{(\a_1\a_2}h_{\a_3\a_4)\b(2)}-\frac{1}{2}\mathcal{D}_{(\a_1}{}^{\b_1}\mathcal{D}_{\a_2}{}^{\b_2}\mathcal{D}_{\a_3}{}^{\b_3} h_{\a_4)\b(3)}-\frac{1}{2}\Box\mathcal{D}_{(\a_1}{}^{\b_1} h_{\a_2\a_3\a_4)\b_1}
\notag\\
&+\big(\mathcal{D}_{(\a_1}{}^{\b_1}R_{\a_2\a_3}{}^{\b_2\b_3}\big) 
 h_{\a_4)\b(3)}+\frac{1}{12}\big(\mathcal{D}_{(\a_1}{}^{\b_1}R\big) h_{\a_2\a_3\a_4)\b_1}
-\frac{1}{12}R\mathcal{D}_{(\a_1}{}^{\b_1} h_{\a_2\a_3\a_4)\b_1}\notag\\
&+2R^{\b_1\b_2}{}_{(\a_1\a_2}\mathcal{D}_{\a_3}{}^{\b_3} h_{\a_4)\b(3)}-\frac{3}{4}R^{\b_1}{}_{\d(\a_1\a_2}\mathcal{D}^{\d\b_2}  h_{\a_3\a_4)\b(2)}~, \label{lincot}
 \end{align}
 \end{subequations}
 where $h_{\a\b\g\d}=(\g^a)_{\a\b}(\g^b)_{\g\d}h_{ab}=h_{(\a\b\g\d)}$. 
 
The conformal gravity action \eqref{CG2} may be linearised around a background spacetime that is a solution 
to the equation of motion \eqref{2.80}.
 The result is 
 \begin{align}\label{CG3}
 S_{\text{CG,\,linearised}}=  - \frac 14\int \text{d}^3x\, e \, h^{\a(4)} \mathfrak{C}_{\a(4)} ~,
 \end{align}
 where $\mathfrak{C}_{\a(4)} =- \delta W_{\a(4)}$  and $\d W_{\a(4)} $ is obtained from \eqref{lincot}, 
 by setting  $W_{\a(4)} =0$.
The linearised action proves to be conformal
(assuming  $ {h}_{\a(4)}$ to be a primary field of dimension 0)  
as well as  it is invariant under the gauge transformations
\bea
\d_\x {h}_{\a(4) } =\cD_{(\a_1 \a_2 } \x_{\a_3 \a_4) }~,
\eea
where the gauge parameter $\x_{\a \b} $ is a primary field of dimension $-1$.

 
 \section{Conformal higher-spin models in three dimensions} \label{section3}

 In order to describe  higher-spin models,
 it is useful to convert to spinor notation. Then
 the commutator of two  covariant derivatives takes the form  
\begin{subequations}\label{confal4}
\bea
[\nabla_{\a\b},\nabla_{\g\d}]=\frac{1}{4}\ve_{\g(\a}W_{\b)\d}{}^{\r(2)}K_{\r(2)}
+\frac{1}{4}\ve_{\d(\a}W_{\b)\g}{}^{\r(2)}K_{\r(2)}
~,
\eea
and the commutation relations of the other generators of the conformal group 
with  the covariant derivatives are as follows:
\bea
\big[M_{\a\b},\nabla_{\g\d}\big]&=&\ve_{\g(\a}\nabla_{\b)\d}+\ve_{\d(\a}\nabla_{\b)\g}~,\\
\big[\mathbb{D},\nabla_{\a\b} \big]&=&\nabla_{\a\b}~,\\
\big[K_{\a\b},\nabla_{\g\d} \big]&=&4\ve_{\g(\a}\ve_{\b)\d}\mathbb{D}
-2\ve_{\g(\a}M_{\b)\d}-2\ve_{\d(\a}M_{\b)\g}
~.
\eea
\end{subequations}
The remaining non-vanishing commutators  
between the generators 
are given in appendix \ref{appendixA}. 
\subsection{CHS prepotentials and field strengths}  
We introduce conformal higher-spin gauge fields  by extending the discussion in \cite{SK-MP}.   
Consider a real totally symmetric rank-$n$ spinor
field $h_{\a(n)}:=h_{\a_1 \dots \a_n} = h_{(\a_1 \dots \a_n)}$ which is primary and of 
dimension
$(2-n/2)$,
 \begin{align}
 K_{\b(2)}h_{\a(n)}=0~,\qquad \mathbb{D}h_{\a(n)}=\left(2-\frac{n}{2}\right)h_{\a(n)}~.
 \end{align}
 Its dimension  
 is uniquely fixed by requiring $h_{\a(n)}$ 
 to be defined modulo gauge transformations of the form
\bea
\d_\x {h}_{\a(n) } =\nabla_{(\a_1 \a_2 } \x_{\a_3 \dots \a_n) }~,
\label{33}
\eea
with the real gauge parameter $\x_{\a(n-2)}$ being also primary.
We say that $h_{\a(n)}$ is  a conformal 
spin-$\frac n2$ gauge field. 

Starting from $h_{\a(n)}$ one may construct  a 
descendant $\mathfrak{C}_{\a(n)} (h)$, known as the higher-spin Cotton tensor, with the following properties:
\begin{enumerate}
\item $\mathfrak{C}_{\a(n)}$ is of the form $\mathcal{A}h_{\a(n)}$, where $\mathcal{A}$ is a linear differential operator involving the covariant derivatives, the Cotton tensor $W_{\a(4)}$, and its covariant derivatives.
\item $\mathfrak{C}_{\a(n)}$ is a primary field of dimension $(1+n/2)~,$
\bea
K_{\b(2)}\mathfrak{C}_{\a(n)}=0~,\qquad \mathbb{D}\mathfrak{C}_{\a(n)}=\left(1+\frac{n}{2} \right)\mathfrak{C}_{\a(n)}~. \label{3.444}
\eea
\end{enumerate} 

Here  we give the most general expressions for $\mathfrak{C}_{\a(n)}$
for $n=2,3,4,5$. They are:
\begin{subequations}\label{58.9}
\begin{align}
\mathfrak{C}_{\a(2)}&=\frac{1}{2^{~}}\bigg(2\nabla_{(\a_1}{}^{\b}h_{\a_2)\b}\bigg)~,\\
\mathfrak{C}_{\a(3)}&=\frac{1}{2^2}\bigg(3\nabla_{(\a_1}{}^{\b_1}\nabla_{\a_2}{}^{\b_2}h_{\a_3)\b(2)}+\Box_ch_{\a(3)} \bigg)~,\\
\mathfrak{C}_{\a(4)}&=\frac{1}{2^3}\bigg(4\nabla_{(\a_1}{}^{\b_1}\nabla_{\a_2}{}^{\b_2}\nabla_{\a_3}{}^{\b_3}h_{\a_4)\b(3)}+4\Box_c\nabla_{(\a_1}{}^{\b}h_{\a_2\a_3\a_4)\b}+a_0W_{(\a_1\a_2}{}^{\b(2)}h_{\a_3\a_4)\b(2)} \bigg)~,\label{3.6c} \\
\mathfrak{C}_{\a(5)}&=\frac{1}{2^4}\bigg(5\nabla_{(\a_1}{}^{\b_1}\nabla_{\a_2}{}^{\b_2}\nabla_{\a_3}{}^{\b_3}\nabla_{\a_4}{}^{\b_4}h_{\a_5)\b(4)} +10\Box_c\nabla_{(\a_1}{}^{\b_1}\nabla_{\a_2}{}^{\b_2}h_{\a_3\a_4\a_5)\b(2)}+(\Box_c)^2h_{\a(5)}\notag\\
&\phantom{BLAN}+\big(\frac{745}{16}-\frac{1}{2}a_1+\frac{5}{2}a_2+3a_3\big)W^{\b_1\b_2}{}_{(\a_1\a_2}\nabla_{\a_3}{}^{\b_3}h_{\a_4\a_5)\b(3)}\notag\\
&\phantom{BLAN}+\big(\frac{564}{48}-\frac{1}{2}\a_1+\frac{1}{2}a_2+a_3\big)W_{\d(\a_1\a_2}{}^{\b_1}\nabla^{\d\b_2}h_{\a_3\a_4\a_5)\b(2)}\notag\\
&\phantom{BLAN}+\big(-40+\frac{1}{2}a_1-\frac{1}{2}a_2-a_3\big)W^{\b(3)}{}_{(\a_1}\nabla_{\a_2\a_3}h_{\a_4\a_5)\b(3)}\notag\\
&\phantom{BLAN}+a_1\nabla^{\d\b_1}W^{\b_2}{}_{\d(\a_1\a_2}h_{\a_3\a_4\a_5)\b(2)}+a_2\nabla_{(\a_1}{}^{\b_1}W_{\a_2\a_3}{}^{\b_2\b_3}h_{\a_4\a_5)\b(3)}\notag\\
&\phantom{BLAN}+a_3\nabla_{(\a_1\a_2}W_{\a_3}{}^{\b(3)}h_{\a_4\a_5)\b(3)}\bigg)~,
\end{align}
\end{subequations}
 where $\Box_c=\nabla^a\nabla_a$ is the conformal d'Alembertian and the $a_i$ are arbitrary constants. 
 
 In a general curved background, for  $n\geq 4$ the requirements outlined above do not determine $\mathfrak{C}_{\a(n)}$ uniquely, since we can always add appropriate terms involving $W_{\a(4)}$. However, in the  $n=4$ case, one may  fix the constant to $a_0=4$ by explicitly linearising $W_{\a(4)}$ around an arbitrary background as in \eqref{lincot}.  
 
The higher-spin Cotton tensor is generally not gauge invariant and the aforementioned ambiguity associated with its definition cannot rectify this. From the expressions \eqref{58.9} it is evident that as $n$ increases this approach will become exceedingly difficult and the ambiguity will worsen. However, an attractive feature of this formulation occurs when the spacetime under consideration is conformally flat,
 \bea
 W_{\a(4)} =0~,
 \eea
  and therefore the conformal covariant derivatives commute \eqref{2.33}. Consequently, this ambiguity is eliminated and as we will now show, the unique expression, up to an overall normalisation, for the spin-$\frac{n}{2}$ Cotton tensor is 
 \begin{align}\label{3.7}
 \mathfrak{C}_{\a(n)}=\frac{1}{2^{n-1}}\sum_{j=0}^{\lceil n/2\rceil -1}\binom{n}{2j+1}(\Box_c)^j\nabla_{(\a_1}{}^{\b_1}\dots\nabla_{\a_{n-2j-1}}{}^{\b_{n-2j-1}}h_{\a_{n-2j}\dots\a_n)\b_1\dots\b_{n-2j-1}}~.
 \end{align} 
Here $\lceil n/2\rceil$ denotes the ceiling function and is equal to $s$ for $n = 2s$ and $s + 1$ for $n = 2s + 1$, with $s \geq 0$ an integer. In the flat limit, \eqref{3.7} reduces to the one derived 
in \cite{K16}.
For even values of $n$,  $n=2,4,\dots$, the flat-space version of 
\eqref{3.7} is equivalent to the one originally obtained by Pope and Townsend
\cite{PopeTownsend}.

It is clear that \eqref{3.7} is of the form $\mathfrak{C}_{\a(n)}=\mathcal{A}h_{\a(n)}$ and has Weyl weight equal to $(1+n/2)$. It remains to show that it is primary.
Since $\mathfrak{C}_{\a(n)}$ is a covariant field, it suffices to show that under a $K$-transformation we have $\delta_{K}\mathfrak{C}_{\a(n)}=-\frac{1}{2}\Lambda(K)^{\g(2)}K_{\g(2)}\mathfrak{C}_{\a(n)}=0$. 

 Using the algebra \eqref{confal4} it is possible to show, by induction on $j$, that the following two identities hold true
\begin{subequations}\label{40.40}
\begin{align}
&\Lambda(K)^{\g(2)}\big[K_{\g(2)},(\Box_c)^j\big]=\Lambda(K)^{\g(2)}\bigg\{2j(\Box_c)^{j-1}\nabla_{\g(2)}(2\mathbb{D}+2j-3)\notag\\
&\phantom{INSERT BLANK SPACE HER}-4j(\Box_c)^{j-1}\nabla_{\g_1}{}^{\d}M_{\g_2\d} \bigg\}~,\label{40.40a}\\
&\Lambda(K)^{\g(2)}\big[K_{\g(2)},\nabla_{(\a_1}{}^{\b_1}\dots\nabla_{\a_j}{}^{\b_j}\big]\z_{\a_{j+1}\dots\a_m)\b_1\dots\b_{j}}\phantom{LOTS OF BLANK SPACE HERE} \notag\\
&=\Lambda(K)^{\g(2)}\bigg\{4j\d_{\g_1}{}^{\b_1}\ve_{\g_2(\a_1}\nabla_{\a_2}{}^{\b_2}\dots\nabla_{\a_j}{}^{\b_j}(\mathbb{D}+j-1)\notag\\
&\phantom{SPACE}-2j\nabla_{(\a_1}{}^{\b_1}\dots\nabla_{\a_{j-1}}{}^{\b_{j-1}}\ve_{\a_j|\g_1}M_{\g_2|}{}^{\b_j}-2j\nabla_{(\a_1}{}^{\b_1}\dots\nabla_{\a_{j-1}}{}^{\b_{j-1}}\d_{|\g_1}{}^{\b_j}M_{\g_2|\a_j}\notag\\
&\phantom{SPACE}-j(j-1)\nabla_{\g(2)}\d_{(\a_1}{}^{\b_1}\d_{\a_2}{}^{\b_2}\nabla_{\a_3}{}^{\b_3}\dots\nabla_{\a_{j}}{}^{\b_{j}}\bigg\}\z_{\a_{j+1}\dots\a_m)\b_1\dots\b_{j}}~,\label{40.40b}
\end{align}
\end{subequations}
where $\zeta_{\a(m)}$ is an arbitrary primary field. Therefore, under a special conformal transformation we have
\begin{align}
&-2^n\delta_K\mathfrak{C}_{\a(n)}=\Lambda(K)^{\g(2)}\sum_{j=0}^{\lceil n/2\rceil -1}\binom{n}{2j+1}\bigg\{\big[K_{\g(2)},(\Box_c)^j\big]\nabla_{(\a_1}{}^{\b_1}\dots\nabla_{\a_{n-2j-1}}{}^{\b_{n-2j-1}}\phantom{BLANKSP}\notag\\ 
&\phantom{BLANK}+(\Box_c)^j\big[K_{\g(2)},\nabla_{(\a_1}{}^{\b_1}\dots\nabla_{\a_{n-2j-1}}{}^{\b_{n-2j-1}}\big]\bigg\}h_{\a_{n-2j}\dots\a_n)\b_1\dots\b_{n-2j-1}} \notag\\
&=\Lambda(K)^{\g(2)}\sum_{j=0}^{\lceil n/2\rceil -1}\binom{n}{2j+1}\notag\\
&\phantom{=}\times\bigg\{2j(2j+1)(\Box_c)^{j-1}\bigg[\nabla_{\g(2)}\nabla_{(\a_1}{}^{\b_1}\dots\nabla_{\a_{n-2j-1}}{}^{\b_{n-2j-1}}h_{\a_{n-2j}\dots\a_n)\b_1\dots\b_{n-2j-1}} \notag\\
&\phantom{BLA}-2\nabla_{\g_1(\a_1}\nabla_{\a_2}{}^{\b_2}\dots\nabla_{\a_{n-2j}}{}^{\b_{n-2j}}h_{\a_{n-2j+1}\dots\a_n)\b_2\dots\b_{n-2j}\g_2}\bigg] -(n-2j-1)(n-2j-2)\notag\\
&\phantom{BLA}\times(\Box_c)^j\bigg[\nabla_{\g(2)}\nabla_{(\a_1}{}^{\b_1}\dots\nabla_{\a_{n-2j-3}}{}^{\b_{n-2j-3}}h_{\a_{n-2j-2}\dots\a_n)\b_1\dots\b_{n-2j-3}} \notag\\
&\phantom{BLA}-2\nabla_{\g_1(\a_1}\nabla_{\a_2}{}^{\b_2}\dots\nabla_{\a_{n-2j-2}}{}^{\b_{n-2j-2}}h_{\a_{n-2j-1}\dots\a_n)\b_2\dots\b_{n-2j-2}\g_2}\bigg] \bigg\}=0~.\notag
\end{align}
In the last line we have used the fact that the second and third terms vanish for $j=\lfloor n/2 \rfloor$ and the first and last terms vanish for $j=0$ to shift the summation variable. This shows that in any conformally flat space \eqref{3.7} is the unique tensor satisfying the properties listed at the beginning of this section.

\subsection{CHS actions}
 
For every conformally flat spacetime, the  tensor \eqref{3.7}
has the following properties:
\begin{enumerate}

\item $\mathfrak{C}_{\a(n)}$ is conserved,
 \begin{align} \label{3.99}
\nabla^{\b(2)}\mathfrak{C}_{\a(n-2)\b(2)}=0~.
 \end{align}
 
 \item $\mathfrak{C}_{\a(n)}$ is 
 invariant under the gauge transformations \eqref{33},
 \begin{align}\label{3.9}
 \delta_{\xi}h_{\a(n)}=\nabla_{(\a_1\a_2}\xi_{\a_3\dots\a_n)}
 \quad \implies
 \quad  \delta_{\xi}\mathfrak{C}_{\a(n)}=0~. \
 \end{align}

\end{enumerate}

 In a general curved space, $\mathfrak{C}_{\a(n)}$ must reduce to the expression \eqref{3.7} in the conformally flat limit. Therefore, in such spaces the right hand side of eq. \eqref{3.7} constitutes the skeleton of $\mathfrak{C}_{\a(n)}$. It immediately follows that its divergence and 
 gauge variation under \eqref{33} are  proportional to terms involving $W_{\a(4)}$ and its covariant derivatives,
 \begin{align}
 \nabla^{\b(2)}\mathfrak{C}_{\a(n-2)\b(2)}=\mathcal{O}(W_{\a(4)})~,\qquad \delta_{\xi}\mathfrak{C}_{\a(n)}=\mathcal{O}(W_{\a(4)})~. \label{3.144}
 \end{align}
 
The properties listed in eq. \eqref{3.444} ensure that the linearised conformal higher-spin action  
\begin{align}\label{3.5}
S_{\rm{CS}}^{(n)} [ {h}] 
=-\frac{\text{i}^n}{2^{\left \lfloor{n/2}\right \rfloor }
} \int \rd^3 x \, e\, { h}^{\a(n)} 
{\mathfrak{C}}_{\a(n) } (h)~,
\end{align}
is invariant under the conformal gauge group $\mathcal{G}$. 
Furthermore, by virtue of \eqref{3.99} and \eqref{3.9}, in any conformally flat space \eqref{3.5} is invariant under the gauge transformations \eqref{33},
\begin{align}
W_{ab}=0 \quad \implies \quad \delta_{\xi}S_{\text{CS}}^{(n)}=0~.
\end{align}

Upon degauging and setting $n=4$, the model \eqref{3.5} coincides with the action for linearised conformal gravity given by eq. \eqref{CG3}. 

We would like to point out the following interesting realisation of $\mathfrak{C}_{\a(n)}$ in terms of the projection operators
\begin{align}
\Pi^{(\pm)}_{~\a}{}^{\b}=\frac{1}{2}\bigg(\delta_{\a}{}^{\b}\pm\frac{\nabla_{\a}{}^{\b}}{\sqrt{\Box_c}}\bigg)~,\qquad \Pi^{(\pm n)}_{~\a(n)}{}^{\b(n)}=\Pi^{(\pm )}_{~(\a_1}{}^{\b_1}\dots \Pi^{(\pm)}_{~\a_n)}{}^{\b_n}~.
\end{align}
which are obtained by extending the flat-space results of  \cite{BKLFP}.
Then one can express 
the higher-spin Cotton tensor as
\bea \label{3.166}
\mathfrak{C}_{\a(n)}&=(\Box_c)^{(n-1)/2}\bigg(\Pi^{(n)}-(-1)^n\Pi^{(-n)}\bigg)h_{\a(n)}~.
\eea
We may use the expressions \eqref{3.166} to rewrite the action \eqref{3.5} in terms of the projection operators. 

\subsection{Generalised CHS models}

As an extension of the previous constructions, we now consider  a conformal higher-spin gauge field $h^{(l)}_{\a(n)}$ which is primary and defined modulo gauge transformations of depth $l$,\footnote{Such gauge transformations occur in the 
description of partially massless fields in diverse dimensions
\cite{DeserN,Higuchi,BuchbinderG,DeserW1,DeserW2,DeserW3,DeserW4,Zinoviev,DNW,DeserW5,DeserW6,Brust}.} 
\begin{align}
\delta_{\xi}h^{(l)}_{\a(n)}=\nabla_{(\a_1\a_2}\cdots\nabla_{\a_{2l-1}\a_{2l}}\xi_{\a_{2l+1}\dots\a_n)}~, \label{50.39}
\end{align}
where $l$ is some integer $1\leq l \leq \lfloor \frac n2 \rfloor$. We also require that the gauge parameter $\x_{\a(n-2l)}$ be primary, after which one can show, using the identity 
\begin{align}
&\Lambda(K)^{\g(2)}\big[K_{\g(2)},\nabla_{(\a_1\a_2}\dots\nabla_{\a_{2l-1}\a_{2l}}\big]\x_{\a_{2l+1}\dots\a_n)}\phantom{LOTS OF BLANK SPACE HERE} \notag\\
&=\Lambda(K)^{\g(2)}\bigg\{4l\ve_{(\a_1|\g_1}\ve_{\g_2|\a_2}\nabla_{\a_3\a_4}\dots\nabla_{\a_{2l-1}\a_{2l}}(\mathbb{D}+l-1)\notag\\
&\phantom{=\Lambda(K)^{\g(2)}\bigg\{}-4l\nabla_{(\a_1\a_2}\dots\nabla_{\a_{2l-1}\a_{2l-2}}\ve_{\a_{2l-1}|\g_1}M_{\g_2|\a_{2l}}\bigg\}\x_{\a_{2l+1}\dots\a_n)}~,
\end{align}
 that the dimension of $h_{\a(n)}$ is fixed to $(l+1)-\frac n2$. The conformal properties of $h^{(l)}_{\a(n)}$ may then be summarised by 
\begin{align}
K_{\b(2)}h^{(l)}_{\a(n)}&=0~,\qquad 
\mathbb{D}h^{(l)}_{\a(n)}=\Big(l+1-\frac{n}{2}\Big)h^{(l)}_{\a(n)}~.\label{50.41}
\end{align}
As was done earlier in the case $l=1$, from $h^{(l)}_{\a(n)}$ we may construct a generalised higher-spin Cotton tensor $\mathfrak{C}^{(l)}_{\a(n)}(h)$ that possesses the following conformal properties,
\begin{align}
K_{\b(2)}\mathfrak{C}^{(l)}_{\a(n)}&=0~,\qquad 
\mathbb{D}\mathfrak{C}^{(l)}_{\a(n)}=\Big(2-l+\frac{n}{2}\Big)\mathfrak{C}^{(l)}_{\a(n)}~. \label{50.42}
\end{align}
In any conformally flat space, the properties \eqref{50.42} determine $\mathfrak{C}^{(l)}_{\a(n)}$ uniquely, up to an overall normalisation, to be 
 \begin{align}\label{50.43}
 \mathfrak{C}^{(l)}_{\a(n)}&=\frac{1}{2^{n-2l+1}}\sum_{j=l-1}^{\lceil n/2\rceil -1}\binom{n}{2j+1}\binom{j}{l-1}(\Box_c)^{j-l+1}\notag\\
 &\phantom{SPACE SPACE SPACE} \times \nabla_{(\a_1}{}^{\b_1}\dots\nabla_{\a_{n-2j-1}}{}^{\b_{n-2j-1}}h^{(l)}_{\a_{n-2j}\dots\a_n)\b_1\dots\b_{n-2j-1}}~.
 \end{align} 
To derive \eqref{50.43} we have made use of the identities \eqref{40.40}. 
The properties \eqref{50.41} and \eqref{50.42} mean that the generalised higher-spin Chern-Simons action,
\begin{align}\label{50.44}
S_{\rm{CS}}^{(n,l)} [ {h}] 
=-\frac{\text{i}^n}{2^{\left \lfloor{n/2}\right \rfloor }
} \int \rd^3 x \, e\, { h}_{(l)}^{\a(n)} 
{\mathfrak{C}}^{(l)}_{\a(n) } (h)~,
\end{align}
is invariant under the conformal gauge group $\mathcal{G}$. Moreover, in any 
conformally flat space the generalised higher-spin Cotton tensor possesses the following important properties:
\begin{enumerate}
\begin{subequations}\label{50.45}
\item $\mathfrak{C}^{(l)}_{\a(n)}$ is partially conserved,
\begin{align}
\nabla^{\b_1\b_2}\cdots \nabla^{\b_{2l-1}\b_{2l}}\mathfrak{C}^{(l)}_{\a(n-2l)\b(2l)}=0~.\label{50.45a}
\end{align}
\item $\mathfrak{C}^{(l)}_{\a(n)}$ is gauge invariant,
\begin{align}
\delta_{\xi}h^{(l)}_{\a(n)}=\nabla_{(\a_1\a_2}\cdots\nabla_{\a_{2l-1}\a_{2l}}\xi_{\a_{2l+1}\dots\a_n)}\quad \implies\quad  \delta_{\xi}\mathfrak{C}^{(l)}_{\a(n)}=0~.\label{50.45b}
\end{align}
\end{subequations}
\end{enumerate}
As a consequence, the action \eqref{50.44} is also gauge invariant,
\begin{align}
W_{ab}=0 \quad \implies \quad
\delta_{\xi}S_{\rm{CS}}^{(n,l)} =0~.\label{50.46}
\end{align}
The proofs for the properties \eqref{50.45} are non-trivial and are given in appendix \ref{AppendixC}.\footnote{It would be of interest to apply the methods of 
\cite{HHL,HLLMP} to demonstrate that \eqref{50.43} is the most general solution 
of the $l$-folded conservation equation \eqref{50.45a} 
in the case of Minkowski spacetime.}

An interesting question to ask is the following. For a given spin, which values of $l$ yield first and second-order Lagrangians in the action \eqref{50.44}? To answer this question, we observe that the number of covariant derivatives in \eqref{50.44} is $(n-2l+1)$ so that $l=\frac 12 n$ and $l=\frac 12 (n-1)$, respectively. Since $l$ must be an integer it immediately follows that first-order conformal models exist only for bosonic spin whilst second-order models must be fermionic. 
These models are said to have `maximal depth' since $l$ assumes its maximal value of $l=\lfloor \frac n2 \rfloor$.  Our conclusions regarding second-order models are in agreement with those drawn long ago in \cite{Sachs}. 


\subsection{Degauging}
In the gauge \eqref{degauge}, the spinor covariant derivative assumes the form 
 \begin{align}\label{3.13}
 \nabla_{\a(2)}=\mathcal{D}_{\a(2)}-\frac{1}{4}P_{\a(2),}{}^{\b(2)}K_{\b(2)}~.
 \end{align}
We may decompose $P_{\a(2),\b(2)}$ into irreducible components as
\begin{align}
P_{\a(2),\b(2)}=R_{\a(2)\b(2)}+\frac{1}{6}\ve_{\a_1(\b_1}\ve_{\b_2)\a_2}R~,
\end{align} 
where $R_{\a(4)}$ is the background traceless Ricci tensor. The goal is then to replace all occurrences of $\nabla_{\a(2)}$ with \eqref{3.13} and use the algebra \eqref{confal4} to eliminate $K_a$. In general, this is a difficult technical problem, particularly for higher-derivative tensors such as \eqref{3.7}. 

As an example, in AdS$_3$ the conformal covariant derivative is 
\begin{align} \label{ads}
\nabla_{\a(2)}=\mathcal{D}_{\a(2)}-\mathcal{S}^2K_{\a(2)}
\end{align}
 whilst the conformal d'Alembertian is
\begin{align} 
\Box_c=\square-6\mathcal{S}^2\mathbb{D}+\mathcal{S}^2\mathcal{D}^{\a(2)}K_{\a(2)}-\frac{1}{2}\mathcal{S}^4K^{\a(2)}K_{\a(2)}~.
\end{align}
 Here and in \eqref{ads}, the parameter $\mathcal{S}$ is related to the AdS scalar curvature through $R=-24\mathcal{S}^2$. 
 Making use of the above relations, one may show that the degauged version of \eqref{3.7}, for small $n$, coincides with the ones given in \cite{SK-MP} (up to conventions). However, for general $n$ we were not able to obtain a closed form expression.

It should be pointed out that various aspects of the bosonic higher-spin 
Cotton tensors $C_{\a(n)}$, with $n $ even, were studied in \cite{LN,BBB}.


\section{Conformal higher-spin models in four dimensions} \label{section4}

In this section we work in four dimensions, $D=4$, and make use of  the two-component spinor notation
and conventions in  \cite{BK}.
It is convenient to replace the Lorentz generators $M_{ab} =- M_{ba}$ 
with operators carrying spinor indices, $M_{\a\b} =M_{\b\a}$ 
and $\bar M_{\ad\bd} =\bar M_{\bd\ad}$, which are defined as 
\begin{subequations}
\bea
&M_{\alpha \beta} = \frac{1}{2} (\sigma^{ab})_{\alpha \beta} M_{ab}~, \qquad
\bar M_{\ad \bd} = -\frac{1}{2} (\tilde\sigma^{ab})_{\ad \bd} M_{ab}~, \quad \\
&M^{ab} = (\sigma^{ab})_{\alpha \beta} M^{\alpha \beta} - (\tilde\sigma^{ab})_{\ad \bd} \bar M^{\ad \bd}~,
\eea
\end{subequations}
and act only on undotted and dotted indices, respectively.

In the two-component spinor notation eq. \eqref{2.61}  
takes the form
\begin{subequations}\label{3.17}
\bea
\big[\nabla_{\a\ad},\nabla_{\b\bd} \big]&=&-\big(\ve_{\ad\bd}C_{\a\b\g\d}M^{\g\d}+\ve_{\a\b}\bar{C}_{\ad\bd\gd\dd}\bar{M}^{\gd\dd}\big) \notag\\
&&
-\frac{1}{4}\big(\ve_{\ad\bd}\nabla^{\d\gd}C_{\a\b\d}{}^{\g}+\ve_{\a\b}\nabla^{\g\dd}\bar{C}_{\ad\bd\dd}{}^{\gd}\big)K_{\g\gd}~,\label{3.17c}
\eea
whilst the commutation relations of the remaining generators with 
$\nabla_{\b\bd} $ 
are given by
\bea
\big[M_{\a\g},\nabla_{\b\bd} \big]&=&
\ve_{\b(\a}\nabla_{\g)\bd}~,\qquad 
\big[ \bar{M}_{\ad\gd},\nabla_{\b\bd} \big]=\ve_{\bd(\ad}\nabla_{\b\gd)}~,~~~~~\\
&&~~\big[\mathbb{D},\nabla_{\b\bd} \big]=\nabla_{\b\bd}~,\\
\big[K_{\a\ad},\nabla_{\b\bd}\big] &=& 4\ve_{\ad\bd}M_{\a\b}+4\ve_{\a\b}\bar{M}_{\ad\bd}-4\ve_{\a\b}\ve_{\ad\bd}\mathbb{D}~.
\eea
\end{subequations}
In eq. \eqref{3.17c} the self-dual and anti self-dual Weyl tensors  $C_{\a\b\g\d}$ and
$\bar{C}_{\ad\bd\gd\dd}$ are related to $C_{abcd}$
as follows
\begin{subequations}\label{56.98}
\begin{align}
C_{\a\b\g\d}&=\frac{1}{2}(\s^{ab})_{\a\b}(\s^{cd})_{\g\d}C_{abcd}=C_{(\a\b\g\d)}~,\\
\bar{C}_{\ad\bd\gd\dd}&=\frac{1}{2}(\ts^{ab})_{\ad\bd}(\ts^{cd})_{\gd\dd}C_{abcd}=\bar{C}_{(\ad\bd\gd\dd)}~,\\
C_{\a\b\g\d\ad\bd\gd\dd}=(\s^a)_{\a\ad}(\s^b&)_{\b\bd}(\s^c)_{\g\gd}(\s^d)_{\d\dd}C_{abcd}=2\ve_{\ad\bd}\ve_{\gd\dd}C_{\a\b\g\d}+2\ve_{\a\b}\ve_{\g\d}\bar{C}_{\ad\bd\gd\dd}~.
\end{align}
 \end{subequations}

\subsection{CHS prepotentials and field strengths}
We introduce conformal higher-spin gauge fields by generalising the constructions in 
\cite{KMT} and earlier works
 \cite{FL-4D,FL}. Given two positive integers $m$ and $n$, a conformal
higher-spin gauge prepotential $\phi_{\a(m)\ad(n)}$ is 
a primary field defined modulo gauge transformations 
\begin{align}\label{3.26}
\delta_{\lambda}\phi_{\a(m)\ad(n)}=\nabla_{(\a_1(\ad_1}\lambda_{\a_2\dots\a_{m})\ad_2\dots\ad_n)}~,
\end{align} 
 where the gauge parameter $\lambda_{\a(m-1)\ad(n-1)}$ is also assumed to be primary.
This gauge freedom uniquely fixes the conformal dimension of the gauge field, 
 \begin{align}\label{3.19}
 K_{\b\bd}\phi_{\a(m)\ad(n)}=0~,\qquad 
 \mathbb{D}\phi_{\a(m)\ad(n)}=\Big(2-\frac{1}{2}(m+n)\Big)\phi_{\a(m)\ad(n)}~.
 \end{align}
In the $m\neq n$ case, the gauge prepotential $\phi_{\a(m)\ad(n)}$ 
and its gauge parameter $\lambda_{\a(m-1)\ad(n-1)}$ are complex.

From $\phi_{\a(m)\ad(n)}$ one may construct two descendants (and their conjugates) 
to which we refer as higher-spin Weyl tensors and denote by $\hat{\mathfrak{C}}_{\a(m+n)}$ and $\check{\mathfrak{C}}_{\a(m+n)}$. They possess the following key properties:
\begin{enumerate}
\item $\hat{\mathfrak{C}}_{\a(m+n)}$ and $\check{\mathfrak{C}}_{\a(m+n)}$ are of the form $\hat{\mathcal{A}}\phi_{\a(m)\ad(n)}$ and $\check{\mathcal{A}}\bar{\phi}_{\a(n)\ad(m)}$, respectively. Here $\hat{\mathcal{A}}$ and $\check{\mathcal{A}}$ are linear differential operators involving the covariant derivatives, the Weyl tensor $C_{abcd}~$, and its covariant derivatives.
\item Both $\hat{\mathfrak{C}}_{\a(m+n)}$ and $\check{\mathfrak{C}}_{\a(m+n)}$ are primary fields 
of dimension $2-\frac{1}{2}(m-n)$  and $2-\frac{1}{2}(n-m)$, respectively,
\begin{subequations}\label{4.555}
\begin{align}
K_{\b\bd}\hat{\mathfrak{C}}_{\a(m+n)}&=0~,\qquad 
\mathbb{D}\hat{\mathfrak{C}}_{\a(m+n)}=\Big(2-\frac{1}{2}(m-n)\Big)\hat{\mathfrak{C}}_{\a(m+n)}~,\\
K_{\b\bd}\check{\mathfrak{C}}_{\a(m+n)}&=0~,\qquad \mathbb{D}\check{\mathfrak{C}}_{\a(m+n)}=\Big(2-\frac{1}{2}(n-m)\Big)\check{\mathfrak{C}}_{\a(m+n)}~.
\end{align}
\end{subequations}
\end{enumerate}

Strictly speaking, we should use a more detailed notation for 
$\hat{\mathfrak{C}}_{\a(m+n)}$ and $\check{\mathfrak{C}}_{\a(m+n)}$ 
that would explicitly indicate the values of $m$ and $n$,
say ${\mathfrak{C}}^{(m,n)}_{\a(m+n)}$ instead of $\hat{\mathfrak{C}}_{\a(m+n)}$. This is 
because there are several choices for $m$ and $n$ such that $m+n ={\rm const}$.
However, in the  hope that no confusion will arise we do not use such a cumbersome  notation.

In some respect the $4D$ case is simpler than its $3D$ counterpart. 
For instance, in a general curved space the higher-spin Weyl tensors take the form
\begin{subequations}\label{3.222}
\begin{align}
\hat{\mathfrak{C}}_{\a(m+n)}&=\nabla_{(\a_1}{}^{\bd_1}\dots\nabla_{\a_n}{}^{\bd_n}\phi_{\a_{n+1}\dots\a_{n+m})\bd_1\dots\bd_n}~,\label{3.22a}\\
\check{\mathfrak{C}}_{\a(m+n)}&=\nabla_{(\a_1}{}^{\bd_1}\dots\nabla_{\a_m}{}^{\bd_m}\bar{\phi}_{\a_{m+1}\dots\a_{m+n})\bd_1\dots\bd_m}~.\label{3.22b}
\end{align}
\end{subequations}
It is clear that both \eqref{3.22a} and \eqref{3.22b} are of the form specified in property one, and also that they have the correct Weyl weights as prescribed in property two. We now show that they are primary. 
 
Using the algebra \eqref{3.17} it is possible to prove, via induction on $j$, that the following identity holds
\begin{align}
&\big[K_{\g\gd},\nabla_{(\a_1}{}^{\bd_1}\dots\nabla_{\a_j}{}^{\bd_j}\big]\z_{\a_{j+1}\dots\a_{j+i})\bd_1\dots\bd_{j}} \phantom{LOTS OF BLANK SPACE HERE} \notag\\
&=-\bigg\{4j\nabla_{(\a_1}{}^{\bd_1}\dots\nabla_{\a_{j-1}}{}^{\bd_{j-1}}\ve_{\a_j|\g}\delta_{\gd|}{}^{\bd_j}\mathbb{D}+4j(j-1)\nabla_{(\a_1}{}^{\bd_1}\dots\nabla_{\a_{j-1}}{}^{\bd_{j-1}}\ve_{\a_j|\g}\delta_{\gd|}{}^{\bd_j}\notag\\
&\phantom{~~~~~~}+4j\nabla_{(\a_1}{}^{\bd_1}\dots\nabla_{\a_{j-1}}{}^{\bd_{j-1}}M_{\a_j|\g}\delta_{\gd|}{}^{\bd_j} \notag\\
&\phantom{~~~~~~}+4j\nabla_{(\a_1}{}^{\bd_1}\dots\nabla_{\a_{j-1}}{}^{\bd_{j-1}}\ve_{\a_j|\g}\bar{M}_{\gd|}{}^{\bd_j}\bigg\}\z_{\a_{j+1}\dots\a_{j+i})\bd_1\dots\bd_j}~, \label{3.24}
\end{align}
 where $\z_{\a(i)\ad(j)}$ is an arbitrary primary field. When the field in \eqref{3.24} is restricted to carry the Weyl weight specified in \eqref{3.19}, upon setting $j=n$ and $i=m$ and evaluating, one finds that the right hand side vanishes.  This demonstrates that $\hat{\mathfrak{C}}_{\a(m+n)}$ is primary. A similar argument holds for $\check{\mathfrak{C}}_{\a(m+n)}$.
  
In a general curved space, one may construct the following primary descendants from the higher-spin Weyl tensors,
\begin{subequations}\label{4.122}
\begin{align}
\hat{\mathfrak{B}}_{\a(n)\bd(m)}&=\nabla_{(\bd_1}{}^{\g_1}\cdots\nabla_{\bd_m)}{}^{\g_m}\hat{\mathfrak{C}}_{\a_1\dots\a_n\g_1\dots\g_m}\label{4.122a}~,\\\check{\mathfrak{B}}_{\a(m)\bd(n)}&=\nabla_{(\bd_1}{}^{\g_1}\cdots\nabla_{\bd_n)}{}^{\g_n}\check{\mathfrak{C}}_{\a_1\dots\a_m\g_1\dots\g_n}~. \label{4.122b}
\end{align} 
\end{subequations}
Both \eqref{4.122a} and \eqref{4.122b} have Weyl weight $2+\frac 12 (m+n)$. The proof that they are primary is similar to that of the higher-spin Weyl tensors and makes use of the identity \eqref{3.24} and the properties \eqref{4.555}.
The primary fields \eqref{4.122a} and \eqref{4.122b} originate from
two alternative expressions for one and the same conformal invariant 
  \begin{align} \label{4.999}
\int\text{d}^4x\, e \, \hat{\mathfrak{C}}^{\a(m+n)}\check{\mathfrak{C}}_{\a(m+n)}  = \int\text{d}^4x\, e \, \phi^{\a(m)\bd(n)}\check{\mathfrak{B}}_{\a(m)\bd(n)} = \int\text{d}^4x\, e \, \bar{\phi}^{\a(n)\bd(m)}\hat{\mathfrak{B}}_{\a(n)\bd(m)}~.
\end{align}
  The derivation of \eqref{4.999} 
is given in appendix \ref{appendixE}. 
We will call $\hat{\mathfrak{B}}_{\a(n)\bd(m)}$
  and $\check{\mathfrak{B}}_{\a(m)\bd(n)}$ (linearised)
  higher-spin Bach tensors.
  
  
  \subsection{CHS actions}
  
 The expressions \eqref{3.22a} and \eqref{3.22b} are determined uniquely, modulo terms involving the background Weyl tensor $C_{abcd}$, by the key properties listed earlier. However,  when the background spacetime is conformally flat, \eqref{3.22a} and \eqref{3.22b} are the unique higher-spin Weyl tensors. 
 By virtue of the commutator \eqref{3.17c} they are also invariant under gauge transformations \eqref{3.26},
 \be
 C_{abcd}=0 \quad \implies \quad  \delta_{\lambda}\hat{\mathfrak{C}}_{\a(m+n)}= \delta_{\lambda}\check{\mathfrak{C}}_{\a(m+n)}=0~. \label{4.888}
 \ee
 Since the commutator of covariant derivatives is proportional to the Weyl tensor, it follows that the gauge variation of \eqref{3.22a} and \eqref{3.22b} under \eqref{3.26} in an arbitrarily curved space is proportional to the Weyl tensor and its covariant derivatives,
 \begin{align}\label{4.88}
 \delta_{\lambda}\hat{\mathfrak{C}}_{\a(m+n)}=\mathcal{O}(C_{abcd})~,\qquad \delta_{\lambda}\check{\mathfrak{C}}_{\a(m+n)}=\mathcal{O}(C_{abcd})~.
 \end{align}
 
As a consequence of the properties \eqref{4.555}, the linearised conformal higher-spin action 
\begin{align}\label{3.21}
S^{(m,n)}_{\text{CHS}} [ \phi, \bar{\phi}] 
= \text{i}^{m+n}\int \rd^4 x \, e\,  \hat{\mathfrak{C}}^{\a(m+n)}\check{\mathfrak{C}}_{\a(m+n)} 
+{\rm c.c.}
\end{align}
is invariant under the gauge group $\mathcal{G}$. Additionally, by virtue of \eqref{4.888}, in any conformally flat space it is also invariant under the gauge transformations \eqref{3.26},
 \begin{align}
 C_{abcd}=0 \quad \implies \quad \delta_{\lambda}S_{\text{CHS}}^{(m,n)} =0~.
 \end{align}
In any conformally flat background the two terms in the right-hand side of  \eqref{3.21} coincide
 because of the identity
\begin{align}
\text{i}^{m+n+1}\int \rd^4 x \, e\,  \hat{\mathfrak{C}}^{\a(m+n)}\check{\mathfrak{C}}_{\a(m+n)} 
+{\rm c.c.}\approx 0~.
\end{align}
In appendices \ref{appendix-gravitino} and \ref{appendix-graviton} we discuss how the action \eqref{3.21} 
can be deformed to make it gauge invariant in Bach-flat backgrounds for 
low spin values.

 When spacetime is conformally flat, the tensors \eqref{4.122} possess the following properties:
\begin{enumerate}
\item $\hat{\mathfrak{B}}_{\a(n)\bd(m)}$ and $\check{\mathfrak{B}}_{\a(m)\bd(n)}$ are invariant under the gauge transformations \eqref{3.26}, 
\begin{subequations}\label{4.133}
\be 
\delta_{\lambda}\hat{\mathfrak{B}}_{\a(n)\bd(m)}=\delta_{\lambda}\check{\mathfrak{B}}_{\a(m)\bd(n)}=0~. \label{4.133a}
\ee 
\item $\hat{\mathfrak{B}}_{\a(n)\bd(m)}$ and $\check{\mathfrak{B}}_{\a(m)\bd(n)}$ are transverse, 
\be 
\nabla^{\g\gd}\hat{\mathfrak{B}}_{\g\a(n-1)\gd\bd(m-1)}=\nabla^{\g\gd}\check{\mathfrak{B}}_{\g\a(m-1)\gd\bd(n-1)}=0~.\label{4.133b}
\ee 
\item The complex conjugates of $\hat{\mathfrak{B}}_{\a(n)\bd(m)}$ and $\check{\mathfrak{B}}_{\a(m)\bd(n)}$ satisfy
\be 
\overline{\hat{\mathfrak{B}}}_{\a(m)\bd(n)}=\check{\mathfrak{B}}_{\a(m)\bd(n)}~,\qquad \overline{\check{\mathfrak{B}}}_{\a(n)\bd(m)}=\hat{\mathfrak{B}}_{\a(n)\bd(m)} ~.\label{4.133c}
\ee
\end{subequations}
\end{enumerate}

The first two properties are obvious. The third property contains non-trivial information and when written out in its entirety reads
\begin{subequations}\label{9.1}
\begin{align}
\nabla_{(\a_1}{}^{\gd_1}\cdots\nabla_{\a_m)}{}^{\gd_m}\overline{\hat{\mathfrak{C}}}_{\bd_1\dots\bd_n\gd_1\dots\gd_m}&=\nabla_{(\bd_1}{}^{\g_1}\cdots\nabla_{\bd_n)}{}^{\g_n}\check{\mathfrak{C}}_{\a_1\dots\a_m\g_1\dots\g_n}~,\label{9.1a}\\
\nabla_{(\a_1}{}^{\gd_1}\cdots\nabla_{\a_n)}{}^{\gd_n}\overline{\check{\mathfrak{C}}}_{\bd_1\dots\bd_m\gd_1\dots\gd_n}&=\nabla_{(\bd_1}{}^{\g_1}\cdots\nabla_{\bd_m)}{}^{\g_m}\hat{\mathfrak{C}}_{\a_1\dots\a_n\g_1\dots\g_m}~.\label{9.1b}
\end{align}
\end{subequations}
To prove \eqref{9.1a} one can assume, without loss of generality, that $m\geq n$. It may then be shown that both sides of the equality evaluate to
\begin{align}
\frac{1}{(m+n)!}\sum_{j=0}^{n}&\binom{m}{j}\binom{n}{j}m!n!(\Box_c)^j\nabla_{(\a_1}{}^{\gd_1}\cdots\nabla_{\a_{m-j}}{}^{\gd_{m-j}}\nabla_{(\bd_1}{}^{\g_1}\cdots\nabla_{\bd_{n-j}}{}^{\g_{n-j}}~~~~~~~~~~~~~~~~~~\notag\\
&\phantom{WASTING SOME SPACE}\times \bar{\phi}_{\a_{m-j+1}\dots\a_m)\g_1\dots\g_{n-j}\bd_{n-j+1}\dots\bd_n)\gd_1\dots\gd_{m-j}}~. \notag
\end{align}
The proof for \eqref{9.1b} is similar. 

It is instructive to introduce the spin projection operators $\Pi^{(m,n)}$ and $\Pi^{[m,n]}$ which are defined by their action on tensor fields, 
\begin{subequations}\label{4.144}
\begin{align}
\Pi^{(m,n)}\phi_{\a(m)\bd(n)}&=\Delta_{\bd_1}{}^{\g_1}\cdots\Delta_{\bd_n}{}^{\g_n}\Delta_{(\g_1}{}^{\gd_1}\cdots\Delta_{\g_n}{}^{\gd_n}\phi_{\a_1\dots\a_m)\gd_1\dots\gd_n} ~,\label{4.144a}\\
\Pi^{[m,n]}\phi_{\a(m)\bd(n)}&=\Delta_{\a_1}{}^{\gd_1}\cdots\Delta_{\a_m}{}^{\gd_m}\Delta_{(\gd_1}{}^{\g_1}\cdots\Delta_{\gd_m}{}^{\g_m}\phi_{\g_1\dots\g_m\bd_1\dots\bd_n)}\label{4.144b} ~.
\end{align}
\end{subequations}
Here we have made use of the involutive operator\footnote{The operator $\D_{\a\bd}$ 
is a generalisation of the flat-space one used in \cite{SG,GGRS}
to construct (super)projectors. For two special cases, $m=n$ and $m=n+1$,
the projection operators defined by \eqref{4.144} are equivalent to 
the  Behrends-Fronsdal projection operators  \cite{BF,Fronsdal58}.
} 
\begin{align}
\Delta_{\a}{}^{\bd}=\frac{\nabla_{\a}{}^{\bd}}{\sqrt{\Box_c}}~,\qquad \Delta_{\a}{}^{\bd}\Delta_{\bd}{}^{\b}=\delta_{\a}{}^{\b}~,\qquad \Delta_{\ad}{}^{\b}\Delta_{\b}{}^{\bd}=\delta_{\ad}{}^{\bd}~.
\end{align}
In a fashion similar to the proof of \eqref{9.1}, it may be shown that both projection operators are equal to one another,
\begin{align}
&\Pi^{(m,n)}\phi_{\a(m)\bd(n)}=\Pi^{[m,n]}\phi_{\a(m)\bd(n)}~.
\end{align}
In addition, they satisfy the following properties
\begin{subequations}\label{4.199}
\begin{align}
\Pi^{(m,n)}\Pi^{(m,n)}=\Pi^{(m,n)}~&,\qquad \Pi^{[m,n]}\Pi^{[m,n]}=\Pi^{[m,n]}~,\label{4.199a}\\
\nabla^{\g\gd}\Pi^{(m,n)}\phi_{\g\a(m-1)\gd\bd(n-1)}=0~&,\qquad \nabla^{\g\gd}\Pi^{[m,n]}\phi_{\g\a(m-1)\gd\bd(n-1)}=0~\label{4.199b}.
\end{align} 
\end{subequations}
Using \eqref{4.144a}, one can express the higher-spin Weyl tensors as
\begin{subequations}\label{4.155}
\begin{align}
\hat{\mathfrak{C}}_{\a(m+n)}&=(\Box_c)^{\frac n2}\Delta_{\a_1}{}^{\bd_1}\cdots\Delta_{\a_n}{}^{\bd_n}\Pi^{(m,n)}\phi_{\a_{n+1}\dots\a_{n+m}\bd_1\dots\bd_n}~,\label{4.155a}\\
\check{\mathfrak{C}}_{\a(m+n)}&=(\Box_c)^{\frac m2}\Delta_{\a_1}{}^{\bd_1}\cdots\Delta_{\a_m}{}^{\bd_m}\Pi^{(n,m)}\bar{\phi}_{\a_{m+1}\dots\a_{m+n}\bd_1\dots\bd_m}~\label{4.155b}.
\end{align}
\end{subequations}
We note that due to \eqref{4.199b}, both $\hat{\mathfrak{C}}_{\a(m+n)}$ and $\check{\mathfrak{C}}_{\a(m+n)}$ as written above are totally symmetric.
 
If we once again assume that $m\geq n$, we may use the descendants \eqref{4.122} and the projectors \eqref{4.144} to give two new realisations of the CHS action\footnote{See appendix \ref{appendixE} for a discussion on the technical issue of integration by parts.} \eqref{3.21},
\begin{subequations} \label{4.222}
\begin{align}
S_{\text{CHS}}^{(m,n)}[\phi,\bar{\phi}]&=\text{i}^{m+n}\int\text{d}^4x\,e\,\bar{\phi}^{\a(n)\bd(m)}\hat{\mathfrak{B}}_{\a(n)\bd(m)}+\text{c.c.}  \label{4.222a}\\
&=\text{i}^{m+n}\int\text{d}^4x\,e\, \bar{\phi}^{\a(n)\bd(m)}(\Box_c)^n \nabla_{\bd_{n+1}}{}^{\a_{n+1}}\cdots\nabla_{\bd_m}{}^{\a_m}\Pi^{(m,n)}\phi_{\a(m)\bd(n)}+\text{c.c.}~
\label{4.222b}
\end{align}
\end{subequations}

When $m=n=s$, the prepotential may be chosen to be real, \begin{align}
 h_{\a(s)\ad(s)}:=\phi_{\a(s)\ad(s)}=\bar{h}_{\a(s)\ad(s)}~.
 \end{align}
In this case there is only one higher-spin Weyl tensor 
$\hat{\mathfrak{C}}_{\a(2s)}=\check{\mathfrak{C}}_{\a(2s)}=\mathfrak{C}_{\a(2s)}$, and one higher-spin Bach tensor $\hat{\mathfrak{B}}_{\a(s)\bd(s)}=\check{\mathfrak{B}}_{\a(s)\bd(s)}=\mathfrak{B}_{\a(s)\bd(s)}=\bar{ \mathfrak{B}}_{\a(s)\bd(s)}$,
\begin{align}
\mathfrak{C}_{\a(2s)}&=\nabla_{(\a_1}{}^{\bd_1}\dots\nabla_{\a_s}{}^{\bd_s}h_{\a_{s+1}\dots\a_{2s})\bd_1\dots\bd_s}~, \label{3.29}\\
\mathfrak{B}_{\a(s)\bd(s)}&=\nabla_{(\bd_1}{}^{\g_1}\cdots\nabla_{\bd_s)}{}^{\g_s}\mathfrak{C}_{\a_1\dots\a_s\g_1\dots\g_s}~,
\end{align}
and the action \eqref{4.222} assumes the simple form 
\begin{align}
S_{\text{CHS}}^{(s,s)} [h]
&= 2(-1)^{s}\int \rd^4 x \, e\, h^{\a(s)\bd(s)}(\Box_c)^{s}\Pi^{(s,s)}
h_{\a(s)\bd(s)}
\label{4.277}
~.
\end{align}

Finally, when $m-1=n=s$ the action \eqref{4.222} becomes
\begin{align}
S_{\text{CHS}}^{(s+1,s)}
&= (-1)^{s}\text{i}\int \rd^4 x \, e\,\bar{\phi}^{\a(s)\bd(s+1)}\nabla_{\bd_{s+1}}{}^{\a_{s+1}}(\Box_c)^{s}\Pi^{(s+1,s)}\phi_{\a(s+1)\bd(s)}+\text{c.c.}~
\label{4.288}
\end{align}
In the case of Minkowski space, the actions \eqref{4.277} and \eqref{4.288} coinicde
with those proposed by Fradkin and Tseytlin \cite{FT}.


 \subsection{Generalised CHS models}

As a simple extension of the previous constructions, we now consider  
a generalised 
gauge field $\phi^{(l)}_{\a(m)\ad(n)}$ which is primary and defined modulo gauge transformations of depth $l$,\footnote{Such gauge transformations occur in the 
description of partially massless fields in diverse dimensions
\cite{DeserN,Higuchi,BuchbinderG,DeserW1,DeserW2,DeserW3,DeserW4,Zinoviev,DNW,DeserW5,DeserW6,Brust}.
Special families of  generalised CHS models were studied in
\cite{Sachs,EO,Vasiliev2009,BG2013,BT2015,GrigorievH}.
}  
\begin{align}
\delta_{\lambda}\phi^{(l)}_{\a(m)\ad(n)}=\nabla_{(\a_1(\ad_1}\cdots\nabla_{\a_l\ad_l}\lambda_{\a_{l+1}\dots\a_m)\ad_{l+1}\dots\ad_n)}~,
\label{gauge28}
\end{align}
with $l$ a positive  integer,  $1\leq l \leq \text{min}(m,n)$. Using an identity 
similar to \eqref{3.24}, one may show that by requiring the gauge parameter $\lambda_{\a(m-l)\ad(n-l)}$ to also be primary, the dimension of $\phi^{(l)}_{\a(m)\ad(n)}$ is fixed to $(l+1)-\frac 12 (m+n)$. The conformal properties of $\phi^{(l)}_{\a(m)\ad(n)}$  may then be summarised by 
\begin{align}
K_{\b\bd}\phi^{(l)}_{\a(m)\ad(n)}&=0~,\qquad 
\mathbb{D}\phi^{(l)}_{\a(m)\ad(n)}=\Big((l+1)-\frac{1}{2}(m+n)\Big)\phi^{(l)}_{\a(m)\ad(n)}~.\label{10.40}
\end{align}

As was done earlier in the case $l=1$, from $\phi^{(l)}_{\a(m)\ad(n)}$ we may construct generalised higher-spin Weyl tensors $\hat{\mathfrak{C}}^{(l)}_{\a(m+n-l+1)\ad(l-1)}(\phi)$ and $\check{\mathfrak{C}}^{(l)}_{\a(m+n-l+1)\ad(l-1)}(\bar{\phi})$ possessing the following conformal properties,
\begin{subequations}\label{10.41}
\begin{align}
K_{\b\bd}\hat{\mathfrak{C}}^{(l)}_{\a(m+n-l+1)\ad(l-1)}&=0~,\qquad 
\mathbb{D}\hat{\mathfrak{C}}^{(l)}_{\a(m+n-l+1)\ad(l-1)}=\Big(2-\frac{1}{2}(m-n)\Big)\hat{\mathfrak{C}}^{(l)}_{\a(m+n-l+1)\ad(l-1)}~,\\
K_{\b\bd}\check{\mathfrak{C}}^{(l)}_{\a(m+n-l+1)\ad(l-1)}&=0~,\qquad \mathbb{D}\check{\mathfrak{C}}^{(l)}_{\a(m+n-l+1)\ad(l-1)}=\Big(2-\frac{1}{2}(n-m)\Big)\check{\mathfrak{C}}^{(l)}_{\a(m+n-l+1)\ad(l-1)}~.
\end{align}
\end{subequations}

In a general curved space one may show, using the identity \eqref{3.24}, 
that these properties are satisfied by the following expressions:
\begin{subequations}\label{10.42}
\begin{align}
\hat{\mathfrak{C}}^{(l)}_{\a(m+n-l+1)\ad(l-1)}&= \nabla_{(\a_1}{}^{\bd_1}\cdots\nabla_{\a_{n-l+1}}{}^{\bd_{n-l+1}}\phi^{(l)}_{\a_{n-l+2}\dots\a_{m+n-l+1})\bd_1\dots\bd_{n-l+1}\ad_1\dots\ad_{l-1}}~,\\
\check{\mathfrak{C}}^{(l)}_{\a(m+n-l+1)\ad(l-1)}&=\nabla_{(\a_1}{}^{\bd_1}\cdots\nabla_{\a_{m-l+1}}{}^{\bd_{m-l+1}}\bar{\phi}^{(l)}_{\a_{m-l+2}\dots\a_{m+n-l+1})\bd_1\dots\bd_{m-l+1}\ad_1\dots\ad_{l-1}}~.
\end{align}
\end{subequations}
Associated with these generalised higher-spin Weyl tensors is the action
\begin{align}
S_{\text{CHS}}^{(m,n,l)}={\rm i}^{m+n}\int \text{d}^4x \, e \,\hat{\mathfrak{C}}_{(l)}^{\a(m+n-l+1)\ad(l-1)}\check{\mathfrak{C}}^{(l)}_{\a(m+n-l+1)\ad(l-1)} +\text{c.c.}~,\label{10.43}
\end{align}
which is invariant under the conformal gauge group $\mathcal{G}$.

In general, if the background Weyl tensor $ C_{abcd}$ is non-vanishing, 
the primary descendants \eqref{10.42} are not invariant 
 under the gauge transformations \eqref{gauge28}.
However, in any conformally flat space the generalised higher-spin Weyl tensors 
\eqref{10.42} prove to be
gauge invariant 
\begin{align}
C_{abcd}=0\quad\implies\quad \delta_{\lambda}\hat{\mathfrak{C}}^{(l)}_{\a(m+n-l+1)\ad(l-1)}=\delta_{\lambda}\check{\mathfrak{C}}^{(l)}_{\a(m+n-l+1)\ad(l-1)} =0~,
\end{align}
and hence so too is the action \eqref{10.43},
\begin{align}
C_{abcd}=0\quad\implies\quad \delta_{\lambda}S_{\text{CHS}}^{(m,n,l)}=0~.
\end{align}

 The equations of motion that follow from \eqref{10.43} are the vanishing of the generalised higher-spin Bach tensors,
\begin{subequations}\label{10.60}
\begin{align}
\hat{\mathfrak{B}}^{(l)}_{\a(n)\ad(m)}&=\nabla_{(\ad_1}{}^{\b_1}\cdots\nabla_{\ad_{m-l+1}}{}^{\b_{m-l+1}}\hat{\mathfrak{C}}^{(l)}_{\b_1\dots\b_{m-l+1}\a_1\dots\a_n\ad_{m-l+2}\dots\ad_{m})}~,\label{10.60a}\\ 
\check{\mathfrak{B}}^{(l)}_{\a(m)\ad(n)}&=\nabla_{(\ad_1}{}^{\b_1}\cdots\nabla_{\ad_{n-l+1}}{}^{\b_{n-l+1}}\check{\mathfrak{C}}^{(l)}_{\b_1\dots\b_{n-l+1}\a_1\dots\a_m\ad_{n-l+2}\dots\ad_{n})}~.
\label{10.60b}
\end{align}
\end{subequations}
Both \eqref{10.60a} and \eqref{10.60b} are primary in any curved spacetime. In a conformally flat spacetime, they satisfy the $l$-extended versions of the properties \eqref{4.133}, namely: 
\begin{enumerate}
\item $\hat{\mathfrak{B}}^{(l)}_{\a(n)\ad(m)}$ and $\check{\mathfrak{B}}^{(l)}_{\a(m)\ad(n)}$ are invariant under the gauge transformations \eqref{gauge28}, 
\begin{subequations}\label{10.133}
\be 
\delta_{\lambda}\hat{\mathfrak{B}}^{(l)}_{\a(n)\ad(m)}=\delta_{\lambda}\check{\mathfrak{B}}^{(l)}_{\a(m)\ad(n)}=0~. \label{10.133a}
\ee 
\item $\hat{\mathfrak{B}}^{(l)}_{\a(n)\ad(m)}$ and $\check{\mathfrak{B}}^{(l)}_{\a(m)\ad(n)}$ are partially conserved, 
\be 
\nabla^{\b_1\bd_1}\cdots\nabla^{\b_l\bd_l}\hat{\mathfrak{B}}^{(l)}_{\b(l)\a(n-l)\bd(l)\ad(m-l)}=\nabla^{\b_1\bd_1}\cdots\nabla^{\b_l\bd_l}\check{\mathfrak{B}}^{(l)}_{\b(l)\a(m-l)\bd(l)\ad(n-l)}=0~.\label{10.133b}
\ee 
\item The complex conjugates of $\hat{\mathfrak{B}}^{(l)}_{\a(n)\bd(m)}$ and $\check{\mathfrak{B}}^{(l)}_{\a(m)\bd(n)}$ satisfy
\be 
\overline{\hat{\mathfrak{B}}}^{~(l)}_{\a(m)\ad(n)}=\check{\mathfrak{B}}^{(l)}_{\a(m)\ad(n)}~,\qquad \overline{\check{\mathfrak{B}}}^{~(l)}_{\a(n)\ad(m)}=\hat{\mathfrak{B}}^{(l)}_{\a(n)\ad(m)} ~.\label{10.133c}
\ee
\end{subequations}
\end{enumerate}

As was done in the 3D case, we can once again ask which values of $l$ yield 
second-order Lagrangians in the action \eqref{10.43}.\footnote{First-order models in this context are not well defined.} A similar analysis reveals that second order models exist only for bosonic spin. These models are of maximal depth and have $l=m=n=s$ with $s=2, 3,\dots$. They were first described in \cite{Sachs,EO}.

It is of interest to provide a more detailed analysis of the maximal depth spin-2 model in a general curved space. 
Setting $l=m=n=2$ and denoting $ \mathfrak{C}_{\a(3)\ad}:=\hat{\mathfrak{C}}^{(2)}_{\a(3)\ad}=\check{\mathfrak{C}}^{(2)}_{\a(3)\ad}$ and $h_{\a(2)\ad(2)}:=\phi^{(2)}_{\a(2)\ad(2)}=\bar{h}_{\a(2)\ad(2)}$, the action \eqref{10.43} takes the form
\begin{align}
S_{\text{CHS}}^{(2,2,2)}=\int \text{d}^4x \, e \, \mathfrak{C}^{\a(3)\ad}\mathfrak{C}_{\a(3)\ad}+\text{c.c.}~,\qquad \mathfrak{C}_{\a(3)\ad}=\nabla_{(\a_1}{}^{\bd}h_{\a_2\a_3)\ad\bd}~.
\end{align}
We can still add a non-minimal term to this action whilst respecting  its $\mathcal{G}$-invariance, 
\begin{subequations} \label{57.56}
\begin{align}
\widetilde{S}^{(2,2,2)}_{\text{CHS}}=&S^{(2,2,2)}_{\text{CHS}}+\omega S_{\text{NM}}^{(2,2,2)}~,\label{57.56a}\\
S_{\text{NM}}^{(2,2,2)}=\int\text{d}^4x &\, e \, C^{\a(2)\b(2)}h_{\a(2)}{}^{\ad(2)}h_{\b(2)\ad(2)} +{\rm c.c.}~,
\end{align}
\end{subequations}
where $\omega$ is some constant and $C^{\a(2)\b(2)}$ is the self-dual part of the  Weyl tensor, eq.  \eqref{56.98}. We find that under the gauge transformations 
\begin{align} 
\delta_{\lambda}h_{\a(2)\ad(2)}=\nabla_{(\a_1(\ad_1}\nabla_{\a_2)\ad_2)}\lambda~, \label{49.10}
\end{align}
the deformed action \eqref{57.56a} varies as
\begin{align}
\delta_{\lambda}\widetilde{S}^{(2,2,2)}_{\text{CHS}}=-2\int\text{d}^4x \, e \, \lambda\bigg\{&(1+\omega)C_{\a(3)\delta}\nabla_{\ad}{}^{\delta}\mathfrak{C}^{\a(3)\ad}+2(1+\omega)\mathfrak{C}^{\a(3)\ad}\nabla_{\ad}{}^{\d}C_{\a(3)\d}\notag\\
&-\omega B^{\a(2)\ad(2)}h_{\a(2)\ad(2)}\bigg\}
+{\rm c.c.}
~,
\end{align}
where $B_{\a(2)\ad(2)}$ is the Bach tensor,
\begin{align}
 B_{\a(2) \ad(2)}=\nabla^{\b_1}{}_{(\ad_1}\nabla^{\b_2}{}_{\ad_2)}
  C_{\a(2) \b(2)}
  =\nabla_{(\a_1}{}^{\bd_1}  \nabla_{\a_2)}{}^{\bd_2}
 \bar{C}_{\ad(2) \bd(2) }=\bar{B}_{\a(2) \ad(2)}~.
\label{C.7}
\end{align}
Therefore, if we choose $\omega=-1$ then it follows that \eqref{57.56a} is gauge invariant in any Bach-flat spacetime,
\begin{align}
B_{\a(2)\ad(2)}=0\quad \implies \quad \delta_{\lambda}\widetilde{S}^{(2,2,2)}_{\text{CHS}}\bigg |_{\omega=-1}=0~.
\end{align}

This model was discussed  in Ref. \cite{BT2015}, where the authors  concluded that gauge invariance could be extended to any Einstein space. Although this statement is true, as we have just shown, it may be extended even further to Bach-flat 
backgrounds. Perhaps the reason they did not arrive at this conclusion was because they demanded gauge transformations of the type
\begin{align}\label{4.43}
\delta_{\lambda}h_{ab}=\mathcal{D}_a\mathcal{D}_b\lambda-\frac{1}{4}\eta_{ab}\Box\lambda~.
\end{align}
However, the right-hand side does not preserve its form under Weyl transformations.
The correct gauge transformation 
in a general curved background
is the degauged version of \eqref{49.10}, which in vector notation reads
\begin{align}
\delta_{\lambda}h_{ab}=\big(\mathcal{D}_a\mathcal{D}_b-\frac{1}{2}R_{ab}\big)\lambda-\frac{1}{4}\eta_{ab}\big(\Box-\frac 12 R\big)\lambda~.
\end{align} 
The two expressions coincide in the case of Einstein spaces.


 \subsection{Degauging}
 
 To conclude this section, we discuss the $4D$ degauging procedure, which turns out to be much more tractable than the $3D$ one. 
In the gauge \eqref{degauge}, the conformal covariant derivative reads
\begin{align}
\nabla_{\a\ad}=\mathcal{D}_{\a\ad}-\frac{1}{4}P_{\a\ad,}{}^{\b\bd}K_{\b\bd}~.
\end{align}
 The Schouten tensor may be decomposed into irreducible components as
 \begin{align}
P_{\a\ad,\b\bd}=\frac{1}{2}R_{\a\b\ad\bd}-\frac{1}{12}\ve_{\a\b}\ve_{\ad\bd}R
 \end{align}
 where $R_{\a\b\ad\bd}=(\s^a)_{\a\ad}(\s^b)_{\b\bd}\big(R_{ab}-\frac{1}{4}\eta_{ab}R\big)=R_{(\a\b)(\ad\bd)}$ is the traceless Ricci tensor. The aim is then to express
 all descendants
 in terms of the torsion-free Lorentz covariant derivative, the curvature and the prepotential.
 
 Using the identity \eqref{3.24} and the degauged covariant derivative, it is possible to show that the following identity holds true
 \begin{align}
 &\mathcal{D}_{(\a_1}{}^{\bd_1}\dots\mathcal{D}_{\a_{j-1}}{}^{\bd_{j-1}}\nabla_{\a_{j}}{}^{\bd_{j}}\dots\nabla_{\a_n}{}^{\bd_n}\phi_{\a_{n+1}\dots\a_{n+m})\bd_1\dots\bd_{n}} \notag\\
 &=\mathcal{D}_{(\a_1}{}^{\bd_1}\dots\mathcal{D}_{\a_{j-1}}{}^{\bd_{j-1}}\bigg\{\mathcal{D}_{\a_{j}}{}^{\bd_{j}}\nabla_{\a_{j+1}}{}^{\bd_{j+1}}\dots\nabla_{\a_n}{}^{\bd_n} \notag\\
 &\phantom{INSERT BLANK s}-\frac{1}{2}j(n-j)R_{\a_j\a_{j+1}}{}^{\bd_j\bd_{j+1}}\nabla_{\a_{j+2}}{}^{\bd_{j+2}}\dots\nabla_{\a_n}{}^{\bd_n}
 \bigg\}\phi_{\a_{n+1}\dots\a_{n+m})\bd_1\dots\bd_{n}}~.\label{4.300}
 \end{align}
 Therefore, in any background spacetime with a vanishing traceless Ricci tensor, or in other words an Einstein space, 
 \bea
 R_{\a\b\ad\bd}=0 \quad \Longleftrightarrow \quad
R_{ab} = \l \eta_{ab}~,
 \eea
 the degauging procedure is trivial and we obtain\footnote{An identity similar to \eqref{4.300} holds with $\phi_{\a(m)\ad(n)}$ replaced with $\hat{\mathfrak{C}}_{\a(m+n)}$. }
 \begin{subequations}
\begin{align}
\hat{\mathfrak{C}}_{\a(m+n)}&=\mathcal{D}_{(\a_1}{}^{\bd_1}\dots\mathcal{D}_{\a_n}{}^{\bd_n}\phi_{\a_{n+1}\dots\a_{n+m})\bd_1\dots\bd_n}~,\label{3.33a}\\
\check{\mathfrak{C}}_{\a(m+n)}&=\mathcal{D}_{(\a_1}{}^{\bd_1}\dots\mathcal{D}_{\a_m}{}^{\bd_m}\bar{\phi}_{\a_{m+1}\dots\a_{m+n})\bd_1\dots\bd_m}~,\label{3.33b}\\
\hat{\mathfrak{B}}_{\a(n)\bd(m)}&=\mathcal{D}_{(\bd_1}{}^{\g_1}\cdots\mathcal{D}_{\bd_m)}{}^{\g_m}\hat{\mathfrak{C}}_{\a_1\dots\a_n\g_1\dots\g_m}\label{3.33c}~,\\\check{\mathfrak{B}}_{\a(m)\bd(n)}&=\mathcal{D}_{(\bd_1}{}^{\g_1}\cdots\mathcal{D}_{\bd_n)}{}^{\g_n}\check{\mathfrak{C}}_{\a_1\dots\a_m\g_1\dots\g_n}~. \label{3.33d}
\end{align}
 \end{subequations}
  
 In the case where $R_{\a(2)\ad(2)}$ does not vanish we have not yet been able to obtain a closed-form expression for the degauged version of any of the above tensors, for arbitrary $m$ and $n$. 
 However, the  expressions for $\mathfrak{C}_{\a(2s)}$ with
  $s=2,3,4,5$ 
  are as follows:
\begin{subequations} 
 \begin{align}
 \mathfrak{C}_{\a(4)}&=\mathcal{D}_{(\a_1}{}^{\bd_1}\mathcal{D}_{\a_2}{}^{\bd_2}h_{\a_3\a_4)\bd_1\bd_2} -\frac{1}{2}R_{(\a_1\a_2}{}^{\bd_1\bd_2}h_{\a_3\a_4)\bd_1\bd_2}~,\phantom{HELLO THERE EASTE}\label{3.34}\\
 \mathfrak{C}_{\a(6)}&=\mathcal{D}_{(\a_1}{}^{\bd_1}\mathcal{D}_{\a_2}{}^{\bd_2}\mathcal{D}_{\a_3}{}^{\bd_3}h_{\a_4\a_5\a_6)\bd_1\bd_2\bd_3}\notag\\
  &\phantom{C~}-\big(\mathcal{D}_{(\a_1}{}^{\bd_1}R_{\a_2\a_3}{}^{\bd_2\bd_3}\big)h_{\a_4\a_5\a_6)\bd_1\bd_2\bd_3}\notag\\
 &\phantom{C~}-2R_{(\a_1\a_2}{}^{\bd_1\bd_2}\mathcal{D}_{\a_3}{}^{\bd_3}h_{\a_4\a_5\a_6)\bd_1\bd_2\bd_3}~,\\
  \mathfrak{C}_{\a(8)}&=\mathcal{D}_{(\a_1}{}^{\bd_1}\mathcal{D}_{\a_2}{}^{\bd_2}\mathcal{D}_{\a_3}{}^{\bd_3}\mathcal{D}_{\a_4}{}^{\bd_4}h_{\a_5\a_6\a_7\a_8)\bd_1\bd_2\bd_3\bd_4} \notag\\
  &\phantom{C~} -\frac{3}{2}\big(\mathcal{D}_{(\a_1}{}^{\bd_1}\mathcal{D}_{\a_2}{}^{\bd_2}R_{\a_3\a_4}{}^{\bd_3\bd_4}\big)h_{\a_5\a_6\a_7\a_8)\bd_1\bd_2\bd_3\bd_4}\notag\\
 &\phantom{C~}-5\big(\mathcal{D}_{(\a_1}{}^{\bd_1}R_{\a_2\a_3}{}^{\bd_2\bd_3}\big)\mathcal{D}_{\a_4}{}^{\bd_4}h_{\a_5\a_6\a_7\a_8)\bd_1\bd_2\bd_3\bd_4} \notag\\
 &\phantom{C~}-5R_{(\a_1\a_2}{}^{\bd_1\bd_2}\mathcal{D}_{\a_3}{}^{\bd_3}\mathcal{D}_{\a_4}{}^{\bd_4}h_{\a_5\a_6\a_7\a_8)\bd_1\bd_2\bd_3\bd_4} \notag\\
 &\phantom{C~}+\frac{9}{4}R_{(\a_1\a_2}{}^{\bd_1\bd_2}R_{\a_3\a_4}{}^{\bd_3\bd_4}h_{\a_5\a_6\a_7\a_8)\bd_1\bd_2\bd_3\bd_4} ~,\\
 \mathfrak{C}_{\a(10)}&=\mathcal{D}_{(\a_1}{}^{\bd_1}\dots \mathcal{D}_{\a_5}{}^{\bd_5}h_{\a_6\dots\a_{10})\bd_1\dots\bd_5} \notag\\
&\phantom{C~} -2\big(\mathcal{D}_{(\a_1}{}^{\bd_1}\mathcal{D}_{\a_2}{}^{\bd_2}\mathcal{D}_{\a_3}{}^{\bd_3}R_{\a_4\a_5}{}^{\bd_4\bd_5}\big)h_{\a_6\dots \a_{10})\bd_1 \dots \bd_5}\notag\\
 &\phantom{C~}-9\big(\mathcal{D}_{(\a_1}{}^{\bd_1}\mathcal{D}_{\a_2}{}^{\bd_2}R_{\a_3\a_4}{}^{\bd_3\bd_4}\big)\mathcal{D}_{\a_5}{}^{\bd_5}h_{\a_6\dots \a_{10})\bd_1\dots \bd_5}\notag\\
 &\phantom{C~}-15\big(\mathcal{D}_{(\a_1}{}^{\bd_1}R_{\a_2\a_3}{}^{\bd_2\bd_3}\big)\mathcal{D}_{\a_4}{}^{\bd_4}\mathcal{D}_{\a_5}{}^{\bd_5}h_{\a_6\dots\a_{10})\bd_1\dots\bd_5} \notag\\
 &\phantom{C~}-10R_{(\a_1\a_2}{}^{\bd_1\bd_2}\mathcal{D}_{\a_3}{}^{\bd_3}\mathcal{D}_{\a_4}{}^{\bd_4}\mathcal{D}_{\a_5}{}^{\bd_5}h_{\a_6\dots\a_{10})\bd_1\dots \bd_{5}} \notag\\
 &\phantom{C~}+16R_{(\a_1\a_2}{}^{\bd_1\bd_2}\big(\mathcal{D}_{\a_3}{}^{\bd_3}R_{\a_4\a_5}{}^{\bd_4\bd_5}\big)h_{\a_6\dots\a_{10})\bd_1\dots\bd_5} \notag\\
 &\phantom{C~}+16R_{(\a_1\a_2}{}^{\bd_1\bd_2}R_{\a_3\a_4}{}^{\bd_3\bd_4}\mathcal{D}_{\a_5}{}^{\bd_5}h_{\a_6\dots \a_{10})\bd_1\dots \bd_5}~.
 \end{align}
 \end{subequations}
 Modulo terms involving the background Weyl tensor, eq. \eqref{3.34} proves to coincide with the linearised self-dual  Weyl tensor $C_{\a(4)}$.
 
The expressions for $\hat{\mathfrak{C}}_{\a(m+n)}$ with $m-1=n=s$ for $s=1,2,3,4$ are as follows:
\begin{subequations} 
 \begin{align}
 \hat{\mathfrak{C}}_{\a(3)}&=\mathcal{D}_{(\a_1}{}^{\bd_1}\phi_{\a_2\a_3)\bd_1} ~,\\
 \hat{\mathfrak{C}}_{\a(5)}&=\mathcal{D}_{(\a_1}{}^{\bd_1}\mathcal{D}_{\a_2}{}^{\bd_2}\phi_{\a_3\a_4\a_5)\bd_1\bd_2}-\frac 12 R_{(\a_1\a_2}{}^{\bd_1\bd_2}\phi_{\a_3\a_4\a_5)\bd_1\bd_2}~,\phantom{HELLO THERE EASTE}\\
  \hat{\mathfrak{C}}_{\a(7)}&=\mathcal{D}_{(\a_1}{}^{\bd_1}\mathcal{D}_{\a_2}{}^{\bd_2}\mathcal{D}_{\a_3}{}^{\bd_3}\phi_{\a_4\a_5\a_6\a_7)\bd_1\bd_2\bd_3}\notag\\
  &\phantom{C~} -\big(\mathcal{D}_{(\a_1}{}^{\bd_1}R_{\a_2\a_3}{}^{\bd_2\bd_3}\big)\phi_{\a_4\a_5\a_6\a_7)\bd_1\bd_2\bd_3}\notag\\
  &\phantom{C~} -2R_{(\a_1\a_2}{}^{\bd_1\bd_2}\mathcal{D}_{\a_3}{}^{\bd_3}\phi_{\a_4\a_5\a_6\a_7)\bd_1\bd_2\bd_3}~,\\
 \hat{\mathfrak{C}}_{\a(9)}&=\mathcal{D}_{(\a_1}{}^{\bd_1}\mathcal{D}_{\a_2}{}^{\bd_2}\mathcal{D}_{\a_3}{}^{\bd_3} \mathcal{D}_{\a_4}{}^{\bd_4}\phi_{\a_5\dots\a_{9})\bd_1\dots\bd_4} \notag\\
&\phantom{C~} -\frac 32 \big(\mathcal{D}_{(\a_1}{}^{\bd_1}\mathcal{D}_{\a_2}{}^{\bd_2}R_{\a_3\a_4}{}^{\bd_3\bd_4}\big)\phi_{\a_5\dots \a_{9})\bd_1 \dots \bd_4}\notag\\
 &\phantom{C~}-5\big(\mathcal{D}_{(\a_1}{}^{\bd_1}R_{\a_2\a_3}{}^{\bd_2\bd_3}\big)\mathcal{D}_{\a_4}{}^{\bd_4}\phi_{\a_5\dots \a_{9})\bd_1\dots \bd_4}\notag\\
 &\phantom{C~}-5R_{(\a_1\a_2}{}^{\bd_1\bd_2}\mathcal{D}_{\a_3}{}^{\bd_3}\mathcal{D}_{\a_4}{}^{\bd_4}\phi_{\a_5\dots\a_{9})\bd_1\dots\bd_4} \notag\\
 &\phantom{C~}+\frac 94 R_{(\a_1\a_2}{}^{\bd_1\bd_2}R_{\a_3\a_4}{}^{\bd_3\bd_4}\phi_{\a_5\dots \a_{9})\bd_1\dots \bd_4}~.
 \end{align}
 \end{subequations}
 Finally, the expressions for $\check{\mathfrak{C}}_{\a(m+n)}$ with $m-1=n=s$ for $s=1,2,3,4$ are as follows:
\begin{subequations} 
 \begin{align}
 \check{\mathfrak{C}}_{\a(3)}&=\mathcal{D}_{(\a_1}{}^{\bd_1}\mathcal{D}_{\a_2}{}^{\bd_2}\bar{\phi}_{\a_1)\bd_1\bd_2} -\frac 12 R_{(\a_1\a_2}{}^{\bd_1\bd_2}\bar{\phi}_{\a_3)\bd_1\bd_2}~,\phantom{HELLO THERE EASTER EGG}\\
  \check{\mathfrak{C}}_{\a(5)}&=\mathcal{D}_{(\a_1}{}^{\bd_1}\mathcal{D}_{\a_2}{}^{\bd_2}\mathcal{D}_{\a_3}{}^{\bd_3}\bar{\phi}_{\a_4\a_5)\bd_1\bd_2\bd_3} \notag\\
 &\phantom{C~} -\big(\mathcal{D}_{(\a_1}{}^{\bd_1}R_{\a_2\a_3}{}^{\bd_2\bd_3}\big)\bar{\phi}_{\a_4\a_5)\bd_1\bd_2\bd_3}\notag\\
  &\phantom{C~} -2R_{(\a_1\a_2}{}^{\bd_1\bd_2}\mathcal{D}_{\a_3}{}^{\bd_3}\bar{\phi}_{\a_4\a_5)\bd_1\bd_2\bd_3}~,\\
\check{ \mathfrak{C}}_{\a(7)}&=\mathcal{D}_{(\a_1}{}^{\bd_1}\mathcal{D}_{\a_2}{}^{\bd_2}\mathcal{D}_{\a_3}{}^{\bd_3} \mathcal{D}_{\a_4}{}^{\bd_4}\bar{\phi}_{\a_5\a_6\a_7)\bd_1\bd_2\bd_3\bd_4} \notag\\
&\phantom{C~} -\frac 32 \big(\mathcal{D}_{(\a_1}{}^{\bd_1}\mathcal{D}_{\a_2}{}^{\bd_2}R_{\a_3\a_4}{}^{\bd_3\bd_4}\big)\bar{\phi}_{\a_5\a_6\a_7)\bd_1 \bd_2\bd_3 \bd_4}\notag\\
 &\phantom{C~}-5\big(\mathcal{D}_{(\a_1}{}^{\bd_1}R_{\a_2\a_3}{}^{\bd_2\bd_3}\big)\mathcal{D}_{\a_4}{}^{\bd_4}\bar{\phi}_{\a_5\a_6 \a_7)\bd_1\bd_2\bd_3 \bd_4}\notag\\
 &\phantom{C~}-5R_{(\a_1\a_2}{}^{\bd_1\bd_2}\mathcal{D}_{\a_3}{}^{\bd_3}\mathcal{D}_{\a_4}{}^{\bd_4}\bar{\phi}_{\a_5\a_6\a_7)\bd_1\bd_2\bd_3\bd_4} \notag\\
 &\phantom{C~}+\frac 94 R_{(\a_1\a_2}{}^{\bd_1\bd_2}R_{\a_3\a_4}{}^{\bd_3\bd_4}\bar{\phi}_{\a_5\a_6\a_7)\bd_1\bd_2\bd_3\bd_4}~,\\
 \check{\mathfrak{C}}_{\a(9)}&=\mathcal{D}_{(\a_1}{}^{\bd_1}\dots \mathcal{D}_{\a_5}{}^{\bd_5}\bar{\phi}_{\a_6\dots\a_{9})\bd_1\dots\bd_5} \notag\\
&\phantom{C~} -2\big(\mathcal{D}_{(\a_1}{}^{\bd_1}\mathcal{D}_{\a_2}{}^{\bd_2}\mathcal{D}_{\a_3}{}^{\bd_3}R_{\a_4\a_5}{}^{\bd_4\bd_5}\big)\bar{\phi}_{\a_6\dots \a_{9})\bd_1 \dots \bd_5}\notag\\
 &\phantom{C~}-9\big(\mathcal{D}_{(\a_1}{}^{\bd_1}\mathcal{D}_{\a_2}{}^{\bd_2}R_{\a_3\a_4}{}^{\bd_3\bd_4}\big)\mathcal{D}_{\a_5}{}^{\bd_5}\bar{\phi}_{\a_6\dots \a_9)\bd_1\dots \bd_5}\notag\\
 &\phantom{C~}-15\big(\mathcal{D}_{(\a_1}{}^{\bd_1}R_{\a_2\a_3}{}^{\bd_2\bd_3}\big)\mathcal{D}_{\a_4}{}^{\bd_4}\mathcal{D}_{\a_5}{}^{\bd_5}\bar{\phi}_{\a_6\dots\a_{9})\bd_1\dots\bd_5} \notag\\
 &\phantom{C~}-10R_{(\a_1\a_2}{}^{\bd_1\bd_2}\mathcal{D}_{\a_3}{}^{\bd_3}\mathcal{D}_{\a_4}{}^{\bd_4}\mathcal{D}_{\a_5}{}^{\bd_5}\bar{\phi}_{\a_6\dots\a_{9})\bd_1\dots \bd_{5}} \notag\\
 &\phantom{C~}+16R_{(\a_1\a_2}{}^{\bd_1\bd_2}\big(\mathcal{D}_{\a_3}{}^{\bd_3}R_{\a_4\a_5}{}^{\bd_4\bd_5}\big)\bar{\phi}_{\a_6\dots\a_{9})\bd_1\dots\bd_5} \notag\\
 &\phantom{C~}+16R_{(\a_1\a_2}{}^{\bd_1\bd_2}R_{\a_3\a_4}{}^{\bd_3\bd_4}\mathcal{D}_{\a_5}{}^{\bd_5}\bar{\phi}_{\a_6\dots \a_{9})\bd_1\dots \bd_5}~.
 \end{align}
 \end{subequations}

Here we do not attempt to degauge the generalised higher-spin Weyl and Bach tensors introduced in the previous subsection. However, we do note that they will also degauge trivially in any Einstein space.  


\section{SCHS theories in three dimensions}\label{section5}

In three dimensions,  $\cN$-extended conformal supergravity was formulated in superspace
as the gauge theory of the superconformal group  in \cite{BKNT-M1}. Upon degauging, 
this formulation reduces to the conventional one, sketched in \cite{HIPT} and fully developed in \cite{KLT-M11}, with the local structure group $\rm SL(2 , \dsR) \times SO(\cN)$.
The former formulation is known as $\cN$-extended conformal superspace, while
the latter  is often referred to as SO$(\cN)$ superspace. 
In this section we only make use of the conformal 
superspace formulations for  $\cN=1,2$ and 3.
To start with, we recall the main facts about the $3D$ $\cN$-extended superconformal algebra
and primary superfields in conformal superspace following \cite{BKNT-M1}.

The $3D$ $\cN$-extended superconformal algebra,
 ${\mathfrak{osp}}(\cN|4, {\mathbb R})$, 
 contains bosonic and fermionic generators. Its even part
 ${\mathfrak{so}}(3,2) \oplus {\mathfrak{so}}(\cN)$
 includes the generators 
of $\mathfrak{so}(\cN)$, $N_{KL}=- N_{LK}$, where 
$K,L=1,\dots, \cN$, in addition to the generators of the conformal group
described in section \ref{subsection2.1}.
Their commutation relations are:
\begin{subequations} \label{SCA}
\begin{gather}
[N_{KL} , N^{IJ}] = 2 \d^I_{[K} N_{L]}{}^J - 2 \d^J_{[K} N_{L]}{}^I \ . \label{SCA.e}
\end{gather}
The odd part of  ${\mathfrak{osp}}(\cN|4, {\mathbb R})$ is spanned by 
 the $Q$-supersymmetry ($Q_\a^I$) and
 $S$-supersymmetry ($S_\a^I$) generators.
 In accordance with \cite{BKNT-M1}, the fermionic operators $Q_\a^I$ obey the
algebra
\begin{gather}
\{ Q_\a^I \ , Q_\b^J \} = 2 \ri \d^{IJ} (\g^c)_{\a\b} P_c 
\ , 
\quad [Q_\a^I , P_b ] = 0 \ , \label{5.1b}\\
[M_{\a\b} , Q_\g^I] = \ve_{\g(\a} Q_{\b)}^I \ , \quad [\mathbb D, Q_\a^I] = \hf Q_\a^I \ , \quad [N_{KL} , Q_\a^I] = 2 \d^I_{[K} Q_{\a L]} \ ,
\end{gather}
while the operators $S_\a^I$ obey the
algebra
\begin{gather}
\{ S_\a^I , S_\b^J \} = 2 \ri \d^{IJ} (\g^c)_{\a\b} K_c \ , \quad [S_\a^I , K_b] = 0 \ , \label{5.1d} \\
[M_{\a\b} , S_\g^I] = \ve_{\g(\a} S_{\b)}^I \ , \quad [\mathbb D, S_\a^I] = - \hf S_\a^I \ , \quad [N_{KL} , S_\a^I] = 2 \d^I_{[K} S_{\a L]} \ .
\end{gather}
In the supersymmetric case, the translation $(P_a)$ and special conformal
$(K_a)$ generators are extended to $P_A = (P_a , Q_\a^I)$ and 
$K_A = (K_a , S_\a^I)$, respectively.
The remainder of the algebra of $K_A$ with $P_A$ is given by
\begin{gather}
[K_a , Q_\a^I ] = - \ri (\g_a)_\a{}^\b S_\b^I \ , \quad [S_\a^I , P_a] = \ri (\g_a)_\a{}^{\b} Q_{\b}^I \ , \\
\{ S_\a^I , Q_\b^J \} = 2 \ve_{\a\b} \d^{I J} \mathbb D - 2 \d^{I J} M_{\a\b} - 2 \ve_{\a\b} N^{IJ} \ .
\end{gather}
\end{subequations}
 
The superspace geometry of $\cN$-extended conformal supergravity
is formulated in terms of the covariant derivatives of the form
\bea
\nabla_A  =(\nabla_a, \nabla_\a^I)
= E_A {}^M\pa_M- \hf \Omega_A{}^{bc} M_{bc} - \hf \Phi_A{}^{PQ} N_{PQ} - B_A \mathbb D - \mathfrak{F}_A{}^B K_B \ .
\eea
Here 
$\Omega_A{}^{bc}$  
is the Lorentz connection,  $\Phi_A{}^{PQ}$  the  ${\rm SO }(\cN)$ connection, $B_A$
 the dilatation connection,   and $\mathfrak F_A{}^B$ the special
superconformal connection.
The graded commutation relations of $\nabla_A$ with the generators 
$M_{bc}$, $N_{PQ} $, $\mathbb D $ and $K_B$ are obtained from \eqref{SCA}
by the replacement $P_A \to \nabla_A$. However the relations \eqref{5.1b} turn into 
\begin{align}
[ \nabla_A , \nabla_B \}
	&= -\cT_{AB}{}^C \nabla_C
	- \frac{1}{2} \cR(M)_{AB}{}^{cd} M_{cd}
	- \frac{1}{2} \cR(N)_{AB}{}^{PQ} N_{PQ}
	\non \\ & \quad
	- \cR({\mathbb D})_{AB} \mathbb D
	- \cR(S)_{AB}{}^\g_K S_\g^K
	- \cR(K)_{AB}{}^c K_c~.
\end{align}
To describe the off-shell conformal supergravity multiplet, the torsion and curvature 
tensors should obey certain $\cN$-dependent covariant constraints 
given in \cite{BKNT-M1}.
The complete solutions to the constraints are derived  in \cite{BKNT-M1}. We will reproduce the 
$\cN=1$ and $\cN=2$ solutions below.

The  generators $K_A = (K_a, S_\alpha^I)$ are used to define conformal {\it primary} superfields:
\be 
K_A \Phi = 0 \ .
\ee
In accordance with \eqref{5.1d}, 
 if a superfield is annihilated by the $S$-supersymmetry generator,
then it is necessarily primary,
\bea
S_\a^I \F =0 \quad \implies \quad K_a \F =0~.
\eea

\subsection{$\cN=1$ SCHS theories}

The algebra of $\cN=1$ conformal covariant derivatives \cite{BKNT-M1} is 
\begin{subequations} \label{5.1}
\begin{align} \{ \nabla_\a , \nabla_\b \} &= 2 \ri \nabla_{\a\b} \ , \\
[ \nabla_a , \nabla_\b ] &= \frac{1}{4} (\g_a)_\b{}^\g W_{ \g\d \s} K^{\d\s}  ~, \\
[\nabla_a , \nabla_b] &= - \frac{\ri}{8} \ve_{abc} (\g^c)^{\a\b} \nabla_\a W_{\b\g\d} K^{\g\d} - \frac{1}{4} \ve_{abc} (\g^c)^{\a\b} W_{\a\b\g} S^\g \label{N=1Algebra.3}\ .
\end{align}
\end{subequations}
It is written in terms of 
 the $\cN = 1$ super Cotton tensor $W_{\a\b\g}$ 
which is a primary superfield of dimension 5/2,
\bea 
 S_\d W_{\a\b\g} = 0 \ ,\quad
\mathbb D W_{\a\b\g} = \frac{5}{2} W_{\a\b\g} 
\ , 
\eea
 obeying the Bianchi identity 
\bea
\nabla^\a W_{\a \b\g} = 0 \ . 
\eea
The super Cotton tensor $W_{\a\b\g}$ was originally introduced in \cite{KT-M12}.

Consider a real primary superfield $L$ of dimension $+2$,
\bea
S_\a L =0~, \qquad {\mathbb D} L  =2L~.
\eea
Then the functional 
\bea
I =   \int \rd^{3|2}z \, E\, L ~, \qquad E^{-1} = {\rm Ber} (E_A{}^M)
\eea
is locally superconformal. We will use this action principle to construct $\cN=1$ locally superconformal higher-spin actions.

We now introduce SCHS gauge prepotentials by extending the definitions given 
in \cite{K16,KT,SK-MP} to conformal superspace. 
Given a positive integer $n>0$,
a real tensor superfield $ H_{\a(n) } $ is said to be a SCHS
gauge prepotential if (i)  it is  primary and of dimension $(1-{n}/{2})$, 
\bea
S_\b H_{\a(n) } =0 ~, \qquad {\mathbb D} H_{\a(n)} = \left(1 -\frac{n}{2} \right) H_{\a(n)}~;
\label{5.8}
\eea
and (ii)  it is defined modulo gauge transformations of the form
\bea
\d_\L { H}_{\a(n) } =\ri^n \nabla_{(\a_1} \L_{\a_2 \dots \a_n) }~,
\label{5.9}
\eea
with the  gauge parameter $\L_{\a(n-1)}$
being real but otherwise unconstrained. The dimension of $H_{\a(n) }$ is 
uniquely fixed by requiring  $\L_{\a(n-1)}$ 
and the right-hand side of \eqref{5.9} to be primary.

Let us first discuss the case $n=1$ corresponding to a superconformal 
vector multiplet. Associated with the prepotential $H_\a$ 
is the real spinor descendant
\bea
{\mathfrak W}_\a (H)=-\frac{\ri}{2} \nabla^\b \nabla_\a { H}_\b~,
\label{5.12}
\eea
which proves to be gauge invariant, 
\bea
\d_\L  {\mathfrak W}_\a =0~,
\label{5.13}
\eea
and  primary, 
\bea
S_\b {\mathfrak W}_\a = 0~, \qquad {\mathbb D} {\mathfrak W}_\a =\frac 32
{\mathfrak W}_\a~.
\label{5.14}
\eea
The field strength \eqref{5.12} obeys the Bianchi identity 
\bea
\nabla^\a {\mathfrak W}_\a =0~.
\label{5.15}
\eea
In general, this conservation equation is superconformal, for some primary 
spinor  $ {\mathfrak W}_\a $, if the dimension of 
${\mathfrak W}_\a $ is equal to 3/2.
The Chern-Simons action 
\bea
S_{\rm SCS}[H]
= - \frac{\ri}{2}
   \int \rd^{3|2}z \, E\,
 { H}^{\a} 
{\mathfrak W}_{\a}( H) 
\eea
has the following basic properties: (i) it
is locally superconformal;  and (ii) it is invariant under the gauge transformations 
\eqref{5.9} with $n=1$.

It turns out that some of the properties of the conformal vector supermultiplet ($n=1$), given by eqs. 
\eqref{5.13}--\eqref{5.15}, cannot  be extended  to $n>1$ in the case
of an  arbitrary curved background. So
let us first consider a conformally flat superspace, 
\bea
W_{\a\b\g}=0~.
\eea
Then it follows from \eqref{5.1} that the conformally covariant derivatives
$\nabla_A =(\nabla_a, \nabla_\a)$ obey the same graded commutation relations 
as the flat-superspace covariant derivatives. This allows us to use the flat-superspace 
results of \cite{K16} provided local
superconformal invariance can be kept under control.
We associate with the gauge prepotential $H_{\a(n)}$ 
the following linearised higher-spin super Cotton tensor
\bea
{\mathfrak W}_{\a_1 \dots \a_n} 
&:=& \frac{1}{2^{n}} 
\sum\limits_{j=0}^{\left \lfloor{n/2}\right \rfloor}
\bigg\{
\binom{n}{2j}  (\Box_c)^{j}\nabla_{(\a_{1}}{}^{\b_{1}}
\dots
\nabla_{\a_{n-2j}}{}^{\b_{n-2j}}H_{\a_{n-2j+1}\dots\a_{n})\b_1 \dots\b_{n-2j}}~~~~
\nonumber \\
&&
-\frac{\ri}{2} 
\binom{n}{2j+1}\nabla^{2}(\Box_c)^{j}\nabla_{(\a_{1}}{}^{\b_{1}}
\dots\nabla_{\a_{n-2j -1}}{}^{\b_{n-2j -1}}H_{\a_{n-2j}\dots\a_{n})
\b_1 \dots \b_{n-2j -1} }\bigg\}~,~~~~~
\eea
where we have denoted $\nabla^2= \nabla^\a \nabla_\a$.
Making use of \eqref{SCA} it may be shown that ${\mathfrak W}_{\a(n)} (H)$ 
has the following properties: 
\begin{enumerate}
\item It is 
primary, 
\bea
S_\b {\mathfrak W}_{\a(n)} =0~, \qquad  {\mathbb D}{\mathfrak W}_{\a(n)} 
= \left(1+\frac{n}{2}\right){\mathfrak W}_{\a(n)}~.
\label{5.19}
\eea  
\item It is conserved, 
\bea
\nabla^\b {\mathfrak W}_{\b \a(n-1)} =0~.
\label{5.20}
\eea
\item It is invariant under 
the gauge transformations \eqref{5.9}, 
\bea
\d_\L {\mathfrak W}_{\a(n)} =0~.
\label{5.21}
\eea
\end{enumerate}
These properties imply that the Chern-Simons-type action
\bea
{S}_{\rm{SCS}}^{(n)} [ {H}] 
= - \frac{\ri^n}{2^{\left \lfloor{n/2}\right \rfloor +1}}
   \int \rd^{3|2}z \, E\,
 {H}^{\a(n)} 
{\mathfrak W}_{\a(n) }( H) 
\eea
has the following fundamental properties: (i) it
is locally superconformal;  and (ii) it is invariant under the gauge transformations 
\eqref{5.9}.
It is worth pointing out the existence of an 
alternative representation for ${\mathfrak W}_{\a (n)}$ inspired by the flat-superspace
construction of \cite{K16}. It is given by  
\bea
{\mathfrak W}_{\a_1 \dots \a_n} = 
\Big( -\frac{\ri}{2}\Big)^n
\nabla^{\b_1} \nabla_{\a_1} \dots \nabla^{\b_n} \nabla_{\a_n} H_{\b_1 \dots \b_n}
= {\mathfrak W}_{(\a_1 \dots \a_n )} ~.
\eea

To conclude our $\cN=1$ discussion, we remark that the off-shell formulations for 
massless and massive higher-spin $\cN=1$ supermultiplets in Minkowski and anti-de Sitter 
backgrounds were constructed in \cite{KT,SK-MP}. These theories are realised in terms of 
the conformal gauge prepotentials $H_{\a (n)}$ in conjunction with certain compensating 
supermultiplets.


\subsection{$\cN=2$ SCHS theories}

In the $\cN=2$ case it is convenient to replace the real spinor covariant derivatives $\nabla_\a^I$ with 
complex ones,
\bea \nabla_\a = \frac{1}{\sqrt{2}} (\nabla_\a^{1} - \ri \nabla_\a^{2}) \ , \quad \bar{\nabla}_\a = - \frac{1}{\sqrt{2}} (\nabla_\a^{1} + \ri \nabla_\a^{2}) \ ,
\eea
which are eigenvectors, 
\bea
 [J , \nabla_\a] = \nabla_\a \ , \quad [J , \bar{\nabla}_\a] = - \bar{\nabla}_\a ~,
\eea
of the U(1) generator $J$ defined by
\bea 
J := - \frac{\ri}{2} \ve^{KL} N_{KL} \ .
\eea
It is also useful to introduce the operators
\bea 
S_\a := \frac{1}{\sqrt{2}} (S_\a^{1} + \ri S_\a^{2}) \ , \quad 
\bar{S}_\a := \frac{1}{\sqrt{2}} (S_\a^{1} - \ri S_\a^{2}) \ ,
\eea
which have the properties
\bea
[J , \bar{S}_\a] = \bar{S}_\a \ , \quad [J , S_\a] = - S_\a \ .
\eea
The graded commutation relations specific to the new basis are
\begin{subequations} \label{gensCB}
\begin{align}
\{ S_\a , S_\b \} = 0 \ , \quad \{ \bar{S}_\a , \bar{S}_\b \}& = 0 \ , 
\quad \{ S_\a , \bar{S}_\b \} = 2 \ri K_{\a\b} \ , \\
[K_a , \nabla_\a ] = - \ri (\g_a)_\a{}^\b \bar{S}_\b \ ,& \quad [K_a , \bar{\nabla}_\a ] = \ri (\g_a)_\a{}^\b S_\b \ , \\
[\bar{S}_\a , \nabla_a] = \ri (\g_a)_\a{}^\b \nabla_{\b} \ , \quad &[S_\a , \nabla_a] = - \ri (\g_a)_\a{}^\b \bar{\nabla}_{\b} \ , \\
\{ \bar{S}_\a , \nabla_\b \} = 0 \ , &\quad \{ S_\a , \bar{\nabla}_\b \} = 0 \ , \\
\{ \bar{S}_\a , \bar{\nabla}_\b \} = - 2 \ve_{\a\b} {\mathbb D} + 2 M_{\a\b} - 2 \ve_{\a\b} J \ ,& 
\quad \{ S_\a , \nabla_\b \} = 2 \ve_{\a\b} {\mathbb D} - 2 M_{\a\b} - 2 \ve_{\a\b} J \ . 
\end{align}
\end{subequations}

In the complex basis,
the algebra of $\cN=2$  covariant derivatives \cite{BKNT-M1} is  
\begin{subequations} 
\begin{align} 
\{ \nabla_\a , \nabla_\b \} &= 0 \ , \quad \{ \bar{\nabla}_\a , \bar{\nabla}_\b \} = 0 \ , \\
\{ \nabla_\a , \bar{\nabla}_\b \} &= - 2 \ri \nabla_{\a\b} - \ve_{\a\b} W_{\g\d} K^{\g\d}\ , \\
[\nabla_a , \nabla_\b] = \frac{\ri}{2} (&\g_a)_\b{}^\g \nabla_\g W^{\a\d} K_{\a\d} - (\g_a)_{\b\g} W^{\g\d} \bar{S}_\d \ , \\
[\nabla_a , \nabla_b] = - \frac{\ri}{8} \ve_{abc} (\g^c)^{\g\d} \Big( \ri [\nabla_\g &, \bar{\nabla}_\d] W_{\a\b} K^{\a\b} + 4 \bar{\nabla}_\g W_{\d\b} \bar{S}^\b + 4 \nabla_\g W_{\d \b} S^\b 
- 8 W_{\g\d} J \Big) \ , 
\end{align}
\end{subequations}
where 
 the $\cN = 2$ super Cotton tensor $W_{\a\b}$ is a primary real superfield,
 \bea
 S_\g W_{\a\b} = 0 \quad \Longleftrightarrow \quad \bar S_\g W_{\a\b} = 0 \ ,\quad
 \mathbb D W_{\a\b} = 2 W_{\a\b} \ , 
\eea
with the fundamental property 
\bea
\nabla^\a W_{\a\b} = 0 ~. 
\eea
In SO(2) superspace \cite{KLT-M11}, the super Cotton tensor $W_{\a\b}$ was introduced
originally  in \cite{Kuzenko12}. 

Given an integer $n>0$, a real  tensor superfield ${H}_{\a(n) } $ 
is said to be a superconformal gauge prepotential
if (i)  it is primary and of dimension $(-{n}/{2})$, 
\bea
S_\b H_{\a(n) } =0  \quad \Longleftrightarrow \quad \bar S_\b H_{\a(n) }=0 ~,\quad
\quad {\mathbb D} H_{\a(n)}  =  -\frac{n}{2}H_{\a(n)}~;
\eea
and (ii) it is defined modulo gauge transformations of the form
\bea
\d_\L { H}_{\a(n) } =\bar  \nabla_{(\a_1} \L_{\a_2 \dots \a_n) }
-(-1)^n\nabla_{(\a_1} \bar \L_{\a_2 \dots \a_n) }~,
\label{55.28}
\eea
where the  gauge parameter $\L_{\a(n-1)}$
is a primary  complex superfield of U(1) charge $+1$, that is,
$J \L_{\a(n-1) } = \L_{\a(n-1)}$. The dimension of the gauge prepotential  is uniquely fixed by requiring 
 $H_{\a(n)}$ and  $\L_{\a(n-1)}$ to be primary.

In the remainder of this section we assume 
that the background curved superspace $\cM^{3|4}$
is conformally flat, 
\bea
W_{\a \b}=0~.
\eea

Associated with the gauge prepotential $H_{\a(n)} $ 
is the following real descendant
\bea
&&{\mathfrak W}_{\a (n)}  (H)
= \frac{1}{2^{n-1}} 
\sum\limits_{j=0}^{\left \lfloor{n/2}\right \rfloor}
\bigg\{
\binom{n}{2j} 
\Delta ( \Box_c)^{j}\nabla_{(\a_{1}}{}^{\b_{1}}
\dots
\nabla_{\a_{n-2j}}{}^{\b_{n-2j}}H_{\a_{n-2j+1}\dots\a_{n})\b_1 \dots\b_{n-2j}}~~~~
\nonumber \\
&&\qquad \qquad +
\binom{n}{2j+1}\Delta^{2}(\Box_c)^{j}\nabla_{(\a_{1}}{}^{\b_{1}}
\dots\nabla_{\a_{n-2j -1}}{}^{\b_{n-2j -1}}H_{\a_{n-2j}\dots\a_{n})
\b_1 \dots \b_{n-2j -1} }\bigg\}~,~~~~~
\label{eq:HSFSUniversal}
\eea
where 
$
\D = \frac{\ri}{2} \nabla^\a \bar \nabla_\a$.
This descendant proves to be primary, 
\bea
S_\b {\mathfrak W}_{\a (n)} =0 \quad \Longleftrightarrow \quad 
\bar S_\b {\mathfrak W}_{\a (n)} =0~,
\quad {\mathbb D} {\mathfrak W}_{\a (n)}  
=  \left(1+\frac{n}{2} \right)  {\mathfrak W}_{\a (n)}~,
\eea
and gauge invariant,
\bea
\d_\L {\mathfrak W}_{\a(n)}&=&0~.
\eea
Moreover, it obeys the conservation equation
\bea
\nabla^\b {\mathfrak W}_{\b \a_1 \dots \a_{n-1}} =0 
\quad \Longleftrightarrow \quad
 \bar \nabla^\b {\mathfrak W}_{\b \a_1 \dots \a_{n-1}} =0 ~.
\eea
These properties  imply
that the action
\bea
{S}^{(n)}_{\rm{SCS}}
[ {H}] 
= - \frac{\ri^n}{2^{\left \lfloor{n/2}\right \rfloor +1}}
   \int \rd^{3|4}z \, E\,
 {H}^{\a(n)} 
{\mathfrak W}_{\a(n) }( { H}) 
\eea
is superconformal and invariant under the gauge transformations 
\eqref{55.28}.

To conclude our $\cN=2$ analysis, we remark that the off-shell formulations for 
massless and massive higher-spin $\cN=2$ supermultiplets in Minkowski superspace, 
as well as in the (1,1) and (2,0) anti-de Sitter 
backgrounds were constructed in \cite{KO,HKO,Hutomo:2018iqo}. 
These theories are realised in terms of 
the conformal gauge prepotentials $H_{\a (n)}$ in conjunction with certain compensating 
supermultiplets.


\subsection{$\cN=3$ SCHS gauge prepotentials}

We introduce $\cN=3$ SCHS prepotentials $H_{\a(n)}$, with $n$ a positive integer, 
with the following properties: (i) it is primary and of dimension $ -(1+{n}/{2})$,
\bea
S^J_\b H_{\a(n) } =0 ~, \qquad {\mathbb D} H_{\a(n)} = -\left( 1+\frac{n}{2} \right) H_{\a(n)}~;
\eea
and (ii)  it is defined modulo gauge transformations of the form
\bea
\d_\L H_{\a(n) } =\ri^n \nabla^I_{(\a_1} \L^I_{\a_2 \dots \a_n) }~,
\label{5.43}
\eea
with the  primary gauge parameter $\L^I_{\a(n-1)}$
being real but otherwise unconstrained. In the right-hand side of \eqref{5.43},
summation over $I$ is assumed. 
The  prepotential $H_\a$ corresponds to linearised $\cN=3$ conformal supergravity
\cite{BKNT-M1}.

The $\cN=3$ story is still incomplete since higher-spin super Cotton tensors 
are not yet known. 


\section{SCHS theories in four dimensions}\label{section6}

In $\cN=1$ conformal superspace \cite{ButterN=1} in four dimensions, 
the covariant derivatives $\nabla_A = (\nabla_a, \nabla_\alpha, \bar\nabla^\ad)$
have the form
\begin{align}
\nabla_A &= E_A{}^M \pa_M - \hf \Omega_A{}^{bc} M_{bc} - \ri \Phi_A Y
	- B_A \mathbb{D} - \mathfrak{F}_{A}{}^B K_B \non  \\
	&= E_A{}^M \pa_M - \Omega_A{}^{\b\g} M_{\b\g} 
	- \bar{\Omega}_A{}^{\bd\gd} \bar{M}_{\bd\gd}
	- \ri \Phi_A Y - B_A \mathbb{D} - \mathfrak{F}_{A}{}^B K_B ~.
\label{6.1}
\end{align}
Here  
$\Omega_A{}^{bc}$  
is the Lorentz connection,  $\Phi_A$  the  $\rm U(1)_R$ connection, $B_A$
 the dilatation connection, and  $\mathfrak F_A{}^B$ the special
superconformal connection.
Below we list the graded commutation relations for the $\cN=1$ superconformal 
algebra $\mathfrak{su}(2,2|1)$  following the conventions
adopted in \cite{ButterN,BKN}, keeping in mind that (i) the translation generators 
$P_A = (P_a, Q_\a ,\bar Q^\ad)$ are replaced with $\nabla_A$; and (ii) the graded commutator 
$[\nabla_A , \nabla_B\}$ differs from that obtained from $[P_A , P_B\}$ by torsion and curvature dependent terms, 
\bea
[\nabla_A, \nabla_B\} 
         = -\cT_{AB}{}^C \nabla_C - \hf \cR_{AB}{}^{cd} (M)M_{cd} 
	- \ri \cR_{AB}(Y) Y - \cR_{AB} (\mathbb{D}) \mathbb{D} - \cR_{AB}{}^C (K)K_C 
	~.~
	\label{6.2}
\eea

The Lorentz generators $M_{ab}$ act on the covariant derivatives as 
\begin{align}
[M_{ab}, \nabla_c ] &= 2 \eta_{c [a} \nabla_{b]}~, \quad
[M_{ab}, \nabla_\g] = (\s_{ab})_\g{}^\d \nabla_\d ~,\quad 
[M_{ab}, \bar\nabla^\gd] = (\tilde{\s}_{ab})^\gd{}_\dd \bar\nabla^\dd~.
\end{align}
The Lorentz generators with spinor indices act on the spinor covariant derivatives
\begin{subequations}
\begin{align}
[M_{\alpha \beta}, \nabla_\gamma] = \ve_{\gamma (\alpha} \nabla_{\beta)}~, \qquad
[\bar M_{\ad \bd}, \bar\nabla_{\gd }] = \ve_{\gd (\ad} \bar\nabla_{\bd) }~.
\end{align}
The  $\rm U(1)_R$ and dilatation generators obey
\begin{align}
[Y, \nabla_\a] &= \nabla_\a ~,\quad [Y, \bar\nabla^\ad] = - \bar\nabla^\ad~,   \\
[\mathbb{D}, \nabla_a] = \nabla_a ~, \quad
&[\mathbb{D}, \nabla_\a] = \hf \nabla_\a ~, \quad
[\mathbb{D}, \bar\nabla^\ad ] = \hf \bar\nabla^\ad ~.
\end{align}
The special superconformal generators $K^A = (K^a, S^\alpha, \bar S_\ad)$
transform in the obvious way under the Lorentz group,
\begin{align}
[M_{ab}, K_c] &= 2 \eta_{c [a} K_{b]} ~, \quad
[M_{ab} , S^\g] = - (\s_{ab})_\b{}^\g S^\b ~, \quad
[M_{ab} , \bar S_\gd] = - (\s_{ab})^\bd{}_\gd \bar S_\bd~, 
\end{align}
and carry opposite $\rm U(1)_R$ and dilatation weight to $\nabla_A$:
\begin{align}
[Y, S^\a] &= - S^\a ~, \quad
[Y, \bar{S}_\ad] = \bar{S}_\ad~,  \\
[\mathbb{D}, K_a] = - K_a ~, \quad
&[\mathbb{D}, S^\a] = - \hf S^\a~, \quad
[\mathbb{D}, \bar{S}_\ad ] = - \hf \bar{S}_\ad ~.
\end{align}
Among themselves, the generators $K^A$ obey the algebra
\begin{align}
\{ S^\a , \bar{S}_\ad \} &= 2 \ri  (\s^a)^\a{}_\ad K_a~,
\end{align}
with all the other (anti-)commutators vanishing. Finally, the algebra of $K^A$ with $\nabla_B$ is given by
\begin{align}
[K^a, \nabla_b] &= 2 \delta^a_b \mathbb{D} + 2 M^{a}{}_b ~, \\
\{ S^\a , \nabla_\b \} &= 2 \d^\a_\b \mathbb{D} - 4  M^\a{}_\b 
- 3 \d^\a_\b Y ~, \\
\{ \bar{S}_\ad , \bar{\nabla}^\bd \} &= 2  \d^\bd_\ad \mathbb{D} 
+ 4  \bar{M}_\ad{}^\bd +  3\d_\ad^\bd Y  ~, \\
[K^a, \nabla_\b] &= -\ri (\s^a)_\b{}^\bd \bar{S}_\bd \ , \qquad \qquad \qquad[K^a, \bar{\nabla}^\bd] = 
-\ri ({\s}^a)^\bd{}_\b S^\b ~,  \\
[S^\a , \nabla_b] &= \ri (\s_b)^\a{}_\bd \bar{\nabla}^\bd \ , \qquad \qquad \quad \qquad[\bar{S}_\ad , \nabla_b] = 
\ri ({\s}_b)_\ad{}^\b \nabla_\b \ ,
\end{align}
\end{subequations}
where all other graded commutators vanish.

In conformal superspace \cite{ButterN=1}, 
the torsion and  curvature tensors in \eqref{6.2} are subject to covariant constraints such that 
$[\nabla_A, \nabla_B\}$ is expressed in terms of the super Weyl tensor
$W_{\alpha \beta \gamma}= W_{(\a\b\g)}$, its conjugate $\bar W_{\ad \bd \gd}$ and their covariant derivatives. The solutions to the constraints are given by
\begin{subequations}\label{eq_csgAlg}
\begin{align}
\{\nabla_\alpha, \nabla_\beta\} = \{\bar\nabla_\ad, \bar\nabla_\bd\} = 0 ~,&\qquad
\nabla_{\alpha \ad} := \frac{\ri}{2} \{\nabla_\alpha, \bar\nabla_\ad\}~, 
\label{eq_csgAlg1} \\
[\nabla_\beta, \nabla_{\alpha \ad}] =
	2\ri \ve_{\beta \alpha} \bar W_{\ad \bd \gd} \bar M^{\bd \gd}
	- &R(\bar S)_{\beta\, \alpha \ad\,}{}_\gd \bar S^\gd
	- R(K)_{\beta\, \alpha \ad}{}^c K_c~, \label{eq_csgAlg2} 
\end{align}
\end{subequations}
where $R(\bar S)_{\beta\, \alpha \ad\,}{}_\gd$ and $R(K)_{\beta\, \alpha \ad}{}^c$
involve derivatives of the superfield $\bar W_{\ad \bd \gd}$. Their precise
expressions will not be necessary for our discussion; they can be found in the original 
publication \cite{ButterN=1}. 

Consider a primary superfield $\J$ (with suppressed indices),
$K_B \J =0$. Its dimension $\D$ and $\rm U(1)_R$ charge $q$ are defined 
as ${\mathbb D} \J = \D \J$ and $Y \J = q \J$. As is well known, for every 
primary covariantly chiral superfield $\f_{\a(n)}$, 
its $\rm U(1)_R$ charge is determined in terms of its dimension, 
\bea
K_B \f_{\a(n)} =0~, \quad \bar \nabla_\bd \f_{\a(n)} =0 \quad
\implies \quad q = -\frac 23  \D~.
\eea

The super Weyl tensor $W_{\a\b\g}$ is a primary chiral 
superfield of dimension 3/2, 
\bea
K_B W_{\a \b \g} =0~, \quad \bar \nabla_\bd W_{\a\b\g}=0 ~, \quad 
{\mathbb D} W_{\a\b\g} = \frac 32 W_{\a\b\g}~.
\eea
It obeys the Bianchi identity  
\bea
B_{\a\ad} :=  \ri \nabla^\b{}_{\ad} \nabla^\g W_{\a\b\g}
=\ri \nabla_{\a}{}^{ \bd} \bar \nabla^\gd \bar W_{\ad\bd\gd}
= \bar B_{\a\ad}~.
\label{6.99}
\eea
Upon degauging (see \cite{ButterN=1} for the technical details of the degauging procedure) $B_{\a\ad}$
takes the form given in 
\cite{BK,BK88} (see also \cite{KMT}). 
It is clear that
$B_{\a\ad}$   is
the $\cN=1$ supersymmetric generalisation of the Bach tensor \eqref{C.7}.
One may check that $B_{\a\ad}$  is  primary, 
\bea
K_B B_{\a\ad} &=&0~, 
\qquad {\mathbb D} B_{\a\ad} = 3 B_{\a\ad} ~,
\eea
and obeys  the conservation equation 
\bea
\nabla^\a B_{\a\ad}=0 \quad & \Longleftrightarrow &\quad  
\bar \nabla^\ad B_{\a\ad} =0~.
\label{616}
\eea

The super-Bach tensor defined by eq. \eqref{6.99}  naturally originates
(see \cite{BK,BK88} for the technical details)
as a functional derivative of the conformal 
supergravity action\footnote{In Minkowski superspace, the linearised action for conformal supergravity  was constructed by Ferrara and Zumino \cite{FZ2}.}
  \cite{Siegel78,Zumino},
\bea 
I_{\rm CSG} =  \int \rd^4x\, \rd^2\q\, \cE\,  W^{\a\b \g}W_{\a\b\g} 
+{\rm c.c.} ~,
\label{6.17}
\eea
with respect to the gravitational superfield $H^{\a\ad}$ \cite{Siegel78}, specifically
\bea
\d  \int \rd^4x \rd^2 \q \, \cE\, W^{\a \b \g}W_{\a\b\g } =
 \int \rd^4x \rd^2 \q  \rd^2 \bar \q \, E\, \D H^{\a\ad} B_{\a\ad}~,
 \eea
where $\cE$ denotes the chiral integration measure, and 
$\D H^{\a\ad} $ the covariant variation of the gravitational superfield
defined in \cite{GrisaruSiegel}. The  conservation equation \eqref{616}
expresses the gauge invariance of the conformal supergravity action. 

We introduce SCHS gauge prepotentials
by generalising the construction of \cite{KMT}.
Given two positive integers $m$ and $n$, 
a SCHS gauge prepotential
 $\U_{\a(m)\ad(n)} $  
is a primary superfield, $K_B \U_{\a(m)\ad(n)}  =0$, defined modulo
gauge transformations \cite{KMT}
\bea
 \d_{ \L, \z} \U_{\a_1 \dots \a_m \ad_1 \dots \ad_{n}} 
 =  \bar \nabla_{(\ad_1} \L_{\a_1 \dots \a_m \ad_2 \dots \ad_{n} )}
+ \nabla_{(\a_1}\z_{\a_2 \dots \a_m)\ad_1 \dots \ad_{n}}  \ ,
\label{6.6}
\eea
with unconstrained  primary gauge parameters $ \L_{\a (m) \ad (n-1)} $ 
and $\z_{\a(m-1)\ad(n)}$. The conditions that $\U_{\a(m)\ad(n)} $, 
$ \L_{\a(m)\ad(n-1)} $  
and $\z_{\a(m-1)\ad(n)} $ be primary superfields uniquely fix the dimension 
and $\rm U(1)_R$ charge of $\U_{\a(m)\ad(n)} $, 
\bea
{\mathbb D} \U_{\a(m)\ad(n)}  = -\hf (m+n) \U_{\a(m)\ad(n)} ~,
\qquad Y \U_{\a(m)\ad(n)} =\frac 13 (m-n) \U_{\a(m)\ad(n)} ~.
\eea

Associated with $\U_{\a(m)\ad(n)} $ and its conjugate $\bar \U_{\a(n)\ad(m)} $
are higher-derivative
descendants
\begin{subequations} \label{super-Weyl}
\bea
\hat{\mathfrak W}_{ \a (m+n+1)} &:=& -\frac{1}{4}\bar \nabla^2 
\nabla_{(\a_1}{}^{\bd_1} \cdots 
\nabla_{\a_{n}}{}^{\bd_{n}}
\nabla_{\a_{n+1}} \U_{\a_{n+2} \dots \a_{m+n+1} )\bd_1 \dots \bd_{n}} ~,
\\
\check{\mathfrak W}_{\a (m+n+1)} &:=& 
-\frac{1}{4}\bar \nabla^2 \nabla_{(\a_1}{}^{\bd_1} 
\cdots  \nabla_{\a_{m}}{}^{\bd_{m}}
\nabla_{\a_{m+1}} \bar \U_{\a_{m+2} \dots \a_{m+n+1} )\bd_1 \dots \bd_{m}}~.
\eea
\end{subequations}
By construction they  are obviously covariantly chiral, 
\bea
\bar \nabla_\bd \hat{\mathfrak W}_{ \a (m+n+1)} =0~,
\qquad \bar \nabla_\bd \check{\mathfrak W}_{ \a (m+n+1)} =0~.
\eea
What is less trivial is the fact that  they are primary, 
\begin{subequations}
\bea
K_B \hat{\mathfrak W}_{ \a (m+n+1)} &=&0~, 
\qquad {\mathbb D} \hat{\mathfrak W}_{ \a (m+n+1)} 
= \hf (3+n-m) \hat{\mathfrak W}_{ \a (m+n+1)}~,\\
K_B \check{\mathfrak W}_{ \a (m+n+1)} &=&0~, 
\qquad {\mathbb D} \check{\mathfrak W}_{ \a (m+n+1)} 
= \hf (3+m-n) \check{\mathfrak W}_{ \a (m+n+1)}~.
\eea
\end{subequations}
These properties imply that the following action
\bea
S^{(m,n)}_{\rm SCHS}=\ri^{m+n}  \int \rd^4x \rd^2 \q \, \cE\, \hat{\mathfrak W}^{\a_1 \dots \a_{m+n+1}}
\check{\mathfrak W}_{\a_1 \dots \a_{m+n+1}} +{\rm c.c.} 
\label{6.11}
\eea
is locally superconformal. 

Consider a conformally flat background superspace, 
\bea
W_{\a\b\g}=0~.
\eea
Then it may be shown that the chiral descendants \eqref{super-Weyl}
are invariant under the gauge transformations \eqref{6.6}, 
\bea
 \d_{ \L, \z} \hat{\mathfrak W}_{ \a (m+n+1)} =0~,
 \qquad 
  \d_{ \L,  \z} \check{\mathfrak W}_{ \a (m+n+1)} =0~.
  \eea
  As a result, the  higher-spin actions \eqref{6.11} are gauge invariant.  
It is clear that these actions
are modelled on the conformal supergravity action.

There are several special cases that require a separate discussion. 
Firstly, choosing $m=n=s>0$ allows us to impose the reality condition
\bea
\U_{\a(s)\ad(s)} = \bar \U_{\a(s)\ad(s)} \equiv H_{\a(s) \ad(s)}~,
\eea
and then \eqref{6.6} turns into the transformation law \cite{KMT}
\bea
 \d_{ \L} H_{\a_1 \dots \a_s \ad_1 \dots \ad_{s}} 
 =  \bar \nabla_{(\ad_1} \L_{\a_1 \dots \a_s \ad_2 \dots \ad_{s} )}
- \nabla_{(\a_1}\bar \L_{\a_2 \dots \a_s)\ad_1 \dots \ad_{s}}  \ ,
\eea
which is a curved-superspace extension of \eqref{1.8}. The gauge prepotential $H_{\a(s) \ad(s)}$
 describes the conformal superspin-$(s+\hf) $ multiplet, with the lowest choice $s=1$ 
corresponding to linearised conformal supergravity.
It is one of the dynamical 
variables in terms of which the off-shell massless superspin-$(s+\hf)$ multiplets
in Minkowski and anti-de Sitter backgrounds  are formulated \cite{KSP,KS94}.

The second special case corresponds to $m= n+1 =s>1$.
The gauge prepotential $\U_{\a(s) \ad(s-1)}$ and its conjugate,
along with certain compensating supermultiplets, 
are used to describe the off-shell massless superspin-$s$ multiplet
in Minkowski and anti-de Sitter backgrounds, originally proposed in  \cite{KS,KS94}
and recently  reformulated in \cite{HK2,BHK}.
  
Thirdly, the case $m=1$ and $n=0$, which corresponds the superconformal gravitino multiplet, 
has been excluded from the previous consideration since the transformation 
law \eqref{6.6} is not defined.  This supermultiplet is characterised by the gauge 
freedom \cite{KMT}
\bea
\d \U_\a = \nabla_\a \z + \l_\a ~, \qquad \bar \nabla_\bd \l_\a =0~, 
\eea
which is a curved superspace extension of the transformation 
law given by Gates and Siegel \cite{GS} who studied an off-shell formulation 
for the  massless gravitino supermultiplet in Minkowski superspace. 


 
 \section{Concluding comments} \label{section7}
 
There exist two modern approaches to formulate conformal  geometry.
One of them was developed by mathematicians and  is often referred to as 
 tractor calculus \cite{BEG,Gover}, with its roots going back to the work of Thomas
\cite{Thomas}. The other formalism was created
by supergravity practitioners \cite{KTvN1}.  It describes conformal gravity 
as the gauge theory of the conformal group, which was reviewed in section 
\ref{section2}. It may be shown that the former approach is obtained from the latter 
by imposing the gauge condition \eqref{degauge}, which makes transparent the fact that 
the $\mathfrak{so}(D,2)$ connection \eqref{2.177} encodes  
the tractor connection of \cite{BEG,Gover}.
This means that the two approaches to conformal geometry are essentially equivalent
and complementary.

Tractor calculus has been used to construct families of conformal differential operators. 
Moreover, there have appeared 
interesting applications of this formalism in physics,  see
\cite{Gover:2008sw,Gover:2008pt,Bonezzi:2010jr,Grigoriev:2011gp,Joung:2013doa} 
and references therein. 
At the same time, tractor calculus is not practical if one is interested in constructing  
superconformal field theories, and alternative ideas are required. Fortunately, 
the work of Butter in four dimensions \cite{ButterN=1,ButterN=2}
and its extensions to three, five and six dimensions 
\cite{BKNT-M1,BKNT-M2,BKNT-M5D,BKNT} have provided powerful tools 
to describe conformal supergravity and its higher-order invariants 
in superspace.\footnote{The conformal superspace approach is 
at the heart of the construction of all $\cN=4$ conformal supergravity theories
in four dimensions \cite{Butter:2016mtk}.}
In this paper we have demonstrated the power of conformal (super)space
to generate (super) CHS theories.  

In four dimensions, all of the higher-spin models which have been constructed in this paper are (super) Weyl-invariant on general curved backgrounds. However, their higher-spin gauge invariance is present, in general, 
only for conformally flat backgrounds.
In principle, extensions to more general (Bach-flat) backgrounds are possible by deforming the action with 
non-minimal {\it primary} terms containing factors of the (super) Weyl tensor 
and its covariant derivatives.
Instructive examples are provided by  (i) the model \eqref{57.56} with $\o=-1$; 
(ii) the conformal gravitino model considered in
appendix \ref{appendix-gravitino}; (iii) the conformal graviton model considered in Appendix \ref{appendix-graviton}; and (iv) the superconformal gravitino model 
studied in \cite{KMT}. Conformal (super)space is an ideal formalism for constructing 
such deformations since the algebra of covariant derivatives is determined by 
the (super) Weyl tensor and its covariant derivatives. 

The structure of the (super) CHS actions presented in this paper indicate
that there should exist a generating formalism to formulate all of these models 
in terms of a single hyper-action. Recently 
there has been much interest in the so-called tensorial or hyperspace 
approach to the description of massless higher-spin (super)fields 
\cite{HS1,HS2,HS3,HS4,HS5,HS6,HS7,HS8,HS9,HS10,HS11,HS12,HS13,HS14,HS15}, see also \cite{Sorokin:2017irs} for a pedagogical review.
 It would be interesting to study whether the conformal (super)space methods
 can be extended to hyperspace. 
 \\


 \noindent
{\bf Acknowledgements:}\\
SMK acknowledges email correspondence with Daniel Butter, Dmitri Sorokin
and especially Arkady Tseytlin.
MP is grateful to Gabriele Tartaglino-Mazzucchelli for illuminating discussions.
The work of SMK is supported in part by the Australian 
Research Council, project No. DP160103633.
The work of MP is supported by the Hackett Postgraduate Scholarship UWA,
under the Australian Government Research Training Program. 

 
\appendix

\section{3D notation and conventions}\label{appendixA}


In $3D$ we follow the notation and conventions adopted in
\cite{KLT-M11}.
In particular, the Minkowski metric is
$\eta_{ab}=\mbox{diag}(-1,1,1)$.
The spinor indices are  raised and lowered using
the $\rm SL(2,{\mathbb R})$ invariant tensors
\bea
\ve_{\a\b}=\left(\begin{array}{cc}0~&-1\\1~&0\end{array}\right)~,\qquad
\ve^{\a\b}=\left(\begin{array}{cc}0~&1\\-1~&0\end{array}\right)~,\qquad
\ve^{\a\g}\ve_{\g\b}=\d^\a_\b
\eea
by the standard rule:
\bea
\psi^{\a}=\ve^{\a\b}\psi_\b~, \qquad \psi_{\a}=\ve_{\a\b}\psi^\b~.
\label{A2}
\eea

We make use of real gamma-matrices,  $\g_a := \big( (\g_a)_\a{}^\b \big)$, 
which obey the algebra
\be
\gamma_a \gamma_b=\eta_{ab}{\mathbbm 1} + \varepsilon_{abc}
\gamma^c~,
\label{A3}
\ee
where the Levi-Civita tensor is normalised as
$\varepsilon^{012}=-\varepsilon_{012}=1$. The completeness
relation for the gamma-matrices reads
\be
(\gamma^a)_{\alpha\beta}(\gamma_a)^{\rho\sigma}
=-(\delta_\alpha^\rho\delta_\beta^\sigma
+\delta_\alpha^\sigma\delta_\beta^\rho)~.
\label{A4}
\ee
Here the symmetric matrices 
$(\gamma_a)^{\alpha\beta}$ and $(\gamma_a)_{\alpha\beta}$
are obtained from $(\g_a)_\a{}^{\b}$ by the rules (\ref{A2}).
Some useful relations involving $\g$-matrices are 
\bsubeq
\bea
\ve_{abc}(\g^b)_{\a\b}(\g^c)_{\g\d}&=&
\ve_{\g(\a}(\g_a)_{\b)\d}
+\ve_{\d(\a}(\g_a)_{\b)\g}
~,
\\
\tr[\g_a\g_b\g_{c}\g_d]&=&
2\eta_{ab}\eta_{cd}
-2\eta_{ac}\eta_{db}
+2\eta_{ad}\eta_{bc}
~.
\eea
\esubeq
Given a three-vector $x_a$,
it  can be equivalently described by a symmetric second-rank spinor $x_{\a\b}$
defined as
\bea
x_{\a\b}:=(\g^a)_{\a\b}x_a=x_{\b\a}~,\qquad
x_a=-\hf(\g_a)^{\a\b}x_{\a\b}~.
\eea
In the $3D$ case,  an
antisymmetric tensor $F_{ab}=-F_{ba}$ is Hodge-dual to a three-vector $F_a$, 
specifically
\bea
F_a=\hf\ve_{abc}F^{bc}~,\qquad
F_{ab}=-\ve_{abc}F^c~.
\label{hodge-1}
\eea
Then, the symmetric spinor $F_{\a\b} =F_{\b\a}$, which is associated with $F_a$, can 
equivalently be defined in terms of  $F_{ab}$: 
\bea
F_{\a\b}:=(\g^a)_{\a\b}F_a=\hf(\g^a)_{\a\b}\ve_{abc}F^{bc}
~.
\label{hodge-2}
\eea
These three algebraic objects, $F_a$, $F_{ab}$ and $F_{\a \b}$, 
are in one-to-one correspondence with each other, 
$F_a \leftrightarrow F_{ab} \leftrightarrow F_{\a\b}$.
The corresponding inner products are related to each other as follows:
\bea
-F^aG_a=
\hf F^{ab}G_{ab}=\hf F^{\a\b}G_{\a\b}
~.
\eea

The Lorentz generators with two vector indices ($M_{ab} =-M_{ba}$),  one vector index ($M_a$)
and two spinor indices ($M_{\a\b} =M_{\b\a}$) are related to each other by the rules:
$M_a=\hf \ve_{abc}M^{bc}$ and $M_{\a\b}=(\g^a)_{\a\b}M_a$.
These generators 
act on a vector $V_c$ 
and a spinor $\J_\g$ 
as follows:
\bea
M_{ab}V_c=2\eta_{c[a}V_{b]}~, ~~~~~~
M_{\a\b}\J_{\g}
=\ve_{\g(\a}\J_{\b)}~.
\label{generators}
\eea
The $D=3$ conformal algebra in spinor notation is
\begin{subequations}\label{confal5}
\begin{align}
[M_{\a\b},M_{\g\d}]&=\ve_{\g(\a}M_{\b)\d}+\ve_{\d(\a}M_{\b)\g}~, \phantom{inserting blank space inserting}\\
[M_{\a\b},P_{\g\d}]&=\ve_{\g(\a}P_{\b)\d}+\ve_{\d(\a}P_{\b)\g}~, \qquad \qquad \qquad \qquad ~ ~[\mathbb{D},P_{\a\b}]=P_{\a\b}~,\\
[M_{\a\b},K_{\g\d}]&=\ve_{\g(\a}K_{\b)\d}+\ve_{\d(\a}K_{\b)\g}~, \qquad \qquad \qquad \qquad [\mathbb{D},K_{\a\b}]=-K_{\a\b}~,\\
[K_{\a\b},P_{\g\d}]&=4\ve_{\g(\a}\ve_{\b)\d}\mathbb{D}-4\ve_{(\g(\a}M_{\b)\d)}~,
\end{align}
\end{subequations}
where $M_{\a\b}=(\g^a)_{\a\b}M_a, P_{\a\b}=(\g^a)_{\a\b}P_a$ and $K_{\a\b}=(\g^a)_{\a\b}K_a$.

To describe conformal gravity in three dimensions we made use of the symmetric Cartan-Killing metric on $\mathfrak{so}(3,2)$, $\Gamma_{\tilde{a}\tilde{b}}=f_{\tilde{a}\tilde{d}}{}^{\tilde{c}}f_{\tilde{b}\tilde{c}}{}^{\tilde{d}}$, see \cite{BKNT-M2} for the technical details. 
In accordance with \eqref{strcon1}, 
the non-vanishing components of $\Gamma_{\tilde{a}\tilde{b}}$ are
\begin{align}
\Gamma_{M_{ab},M_{cd}}=-12\eta_{a[c}\eta_{d]b}~,\qquad \Gamma_{K_a,P_b}=-12\eta_{ab}~,\qquad \Gamma_{\mathbb{D}\mathbb{D}}=6~.
\end{align}



\section{Converting between conventions}\label{appendixB}

In several papers such as Ref. \cite{SK-MP}, a different set of conventions were used. The purpose of this appendix is to provide a summary of how to easily convert between conventions.

Suppose that a field $\Phi$ transforms under some tensor representation of the gauge group $\mathcal{G}$. We may write the action of the generators $X_{\tilde{a}}$ on $\Phi$ as
\begin{align}
X_{\tilde{a}}\Phi=t_{\tilde{a}}\Phi
\end{align}
 for some matrix  $t_{\tilde{a}}$.
  Since the operator $X_{\tilde{a}}$ passes through $t_{\tilde{b}}$, we find that the commutator is
 \begin{align}
 [X_{\tilde{a}},X_{\tilde{b}}]\Phi=-[t_{\tilde{a}},t_{\tilde{b}}]\Phi~.
 \end{align}
The $X_{\tilde{a}}$ and $t_{\tilde{a}}$ therefore satisfy
\begin{align}
[X_{\tilde{a}},X_{\tilde{b}}]=-f_{\tilde{a}\tilde{b}}{}^{\tilde{c}}X_{\tilde{c}}~,\qquad [t_{\tilde{a}},t_{\tilde{b}}]=+f_{\tilde{a}\tilde{b}}{}^{\tilde{c}}t_{\tilde{c}}~.
\end{align}
 In particular, the action of the Lorentz generators is $M_{ab}\Phi=m_{ab}\Phi$. The Lorentz generators of Ref. \cite{SK-MP} correspond to $m_{ab}$, i.e. no change is necessary here.
 
In section \ref{section2}, the connection one-forms, torsion and curvature tensors were defined through
\begin{align}
\nabla_a=e_a-\omega_a{}^{\un{b}}X_{\un{b}}~,\qquad[\nabla_a,\nabla_b]=-\mathcal{T}_{ab}{}^{c}\nabla_c-\mathcal{R}_{ab}{}^{\un{c}}X_{\un{c}}~,
\end{align} 
 which differs to the definitions in Refs. \cite{SK-MP} and \cite{BK} by a minus sign. Thus, to flow between conventions, one must impose the gauge $\mathfrak{b}_a=0$ and make the following replacements
 \begin{align}
 T_{ab}{}^{c}&\longmapsto -T_{ab}{}^{c}~,\\
 R_{abcd}&\longmapsto -R_{abcd}~,\\
 \omega_{abc}&\longmapsto -\omega_{abc}~.
 \end{align}
 Additionally, since the Cotton, Weyl and Schouten tensors, given by \eqref{2.25}, \eqref{Weyl} and \eqref{2.97} respectively, are defined in terms of $R_{abcd}$, we must also rescale each of them by $-1$. This accounts for the sign discrepancy between the second and third terms of \eqref{lincot} and the first two terms of \eqref{3.6c}.
 
 
\section{Properties of the generalised HS Cotton tensor}\label{AppendixC} 
In this section we present the main steps that are needed in order to prove the two properties \eqref{50.45} of the generalised higher-spin Cotton tensor $\mathfrak{C}^{(l)}_{\a(n)}$. Namely, that in any conformally flat spacetime it is partially conserved and gauge invariant.

It may be shown that the $l^{\text{th}}$ divergence of $\mathfrak{C}^{(l)}_{\a(n)}$ is given by
\begin{align}
&2^{n-2l+1}\nabla^{\b_1\b_2}\cdots\nabla^{\b_{2l-1}\b_{2l}}\mathfrak{C}^{(l)}_{\a(n-2l)\b(2l)}\notag\\
&=\sum_{j=l-1}^{\lceil \frac n2 \rceil -1}\binom{n}{2j+1}\binom{j}{l-1}(\Box_c)^{j-l+1}\nabla^{\b_1\b_2}\cdots\nabla^{\b_{2l-1}\b_{2l}}\notag\\
&\phantom{Spa}\times\frac{1}{n!}\sum_{k=0}^{l}\binom{l}{k}\binom{n-2j-1}{2k}(2k)!\binom{n-2l}{n-2j-2k-1}(n-2j-2k-1)!(2j+1)! \notag\\
&\phantom{Spa}\times\nabla_{\b_1}{}^{\g_1}\cdots\nabla_{\b_{2k}}{}^{\g_{2k}}\nabla_{(\a_1}{}^{\g_{2k+1}}\cdots\nabla_{\a_{n-2j-2k-1}}{}^{\g_{n-2j-1}}h^{(l)}_{\a_{n-2j-2k}\dots\a_{n-2l})\b_{2k+1}\dots\b_{2l}\g_1\dots\g_{n-2j-1}} \notag\\
&=\sum_{k=0}^{l}\sum_{j=l-k}^{\lceil \frac n2 \rceil -k-1}(-1)^k\binom{n}{2j+1}\binom{j}{l-1}\binom{l}{k}\binom{n-2j-1}{2k}\binom{n-2l}{n-2j-2k-1}\frac{(2k)!(2j+1)!}{n!}\notag\\
&\phantom{Spa}\times(n-2j-2k-1)!(\Box_c)^{j+k-l+1}\nabla^{\g_1\g_2}\cdots\nabla^{\g_{2l-1}\g_{2l}}\nabla_{(\a_1}{}^{\g_{2l+1}}\cdots\nabla_{\a_{n-2j-2k-1}}{}^{\g_{n-2j-2k+2l-1}}\notag\\
&\phantom{Spa}\times h^{(l)}_{\a_{n-2j-2k}\dots\a_{n-2l})\g_1\dots\g_{n-2j-2k+2l-1}} \notag\\
&=\sum_{j=l}^{\lceil \frac n2 \rceil -1}\binom{n-2l}{n-2j-1}(\Box_c)^{j-l+1}\nabla^{\g_1\g_2}\cdots\nabla^{\g_{2l-1}\g_{2l}}\nabla_{(\a_1}{}^{\g_{2l+1}}\cdots\nabla_{\a_{n-2j-1}}{}^{\g_{n-2j+2l-1}}\notag\\
&\phantom{Spa}\times h^{(l)}_{\a_{n-2j}\dots\a_{n-2l})\g_1\dots\g_{n-2j+2l-1}}\bigg\{\sum_{k=0}^{l}(-1)^k\binom{j-k}{l-1}\binom{l}{k}\bigg\} ~.\notag
\end{align}
Making use of the combinatoric identity
 \begin{align}
 \sum_{k=0}^{l}(-1)^k \binom{j-k}{l-1}\binom{l}{k}=0 \qquad \forall ~j\geq l~,\label{G.1}
 \end{align}
 which may be proved by induction on $l$, it follows that the last line in the above is equal to zero. In the second line we have used the combinatoric factors to shift the upper and lower bounds of the summation over $j$. Then, in the third line we have shifted the dummy variable $j\mapsto j-k$.

 Under the gauge transformations \eqref{50.39}, it may be shown that $\mathfrak{C}^{(l)}_{\a(n)}$ transforms as 
 \begin{align}
 \delta_{\x}&\bigg(2^{n-2l+1}\mathfrak{C}^{(l)}_{\a(n)}\bigg)\notag\\
 =&\sum_{j=l-1}^{\lceil \frac n2 \rceil-1}\binom{n}{2j+1}\binom{j}{l-1}(\Box_c)^{j-l+1}\nabla_{(\a_1}{}^{\b_1}\cdots\nabla_{\a_{n-2j-1}}{}^{\b_{n-2j-1}}\notag\\
 &\phantom{Spa}\times\frac{1}{n!}\sum_{k=0}^{l}\binom{n-2j-1}{2k}\binom{l}{k}(2k)!\binom{2j+1}{2l-2k}(2l-2k)!(n-2l)! \notag\\
 &\phantom{Spa}\times\nabla_{|\b_1\b_2}\cdots\nabla_{\b_{2k-1}\b_{2k}|} \nabla_{\a_{n-2j}\a_{n-2j+1}}\cdots\nabla_{\a_{n-2j-2k+2l-2}\a_{n-2j-2k+2l-1}}\notag\\
 &\phantom{Spa}\times\x_{\a_{n-2j-2k+2l}\dots\a_n)\b_{2k+1}\dots\b_{n-2j-1}} \notag\\
 =&\sum_{k=0}^{l}\sum_{j=l-k}^{\lceil \frac n2 \rceil-k-1}(-1)^k\binom{n}{2j+1}\binom{j}{l-1}\binom{n-2j-1}{2k}\binom{l}{k}\binom{2j+1}{2l-2k}\frac{(2k)!(2l-2k)!(n-2l)!}{n!} \notag\\
 &\phantom{Spa}\times (\Box_c)^{j+k-l+1}\nabla_{(\a_1\a_2}\cdots\nabla_{\a_{2l-1}\a_{2l}}\nabla_{\a_{2l+1}}{}^{\b_{2l+1}}\cdots\nabla_{\a_{n-2j-2k+2l+1}}{}^{\b_{n-2j-2k+2l+1}}\notag\\
 &\phantom{Spa}\times\x_{\a_{n-2j-2k+2l+2}\dots\a_n)\b_{2l+1}\dots\b_{n-2j-2k+2l+1}} \notag\\
 =&\sum_{j=l}^{\lceil \frac n2 \rceil-1}\binom{n-2l}{2j-2l+1}(\Box_c)^{j-l+1}\nabla_{(\a_1\a_2}\cdots\nabla_{\a_{2l-1}\a_{2l}}\nabla_{\a_{2l+1}}{}^{\b_{2l+1}}\cdots\nabla_{\a_{n-2j+2l+1}}{}^{\b_{n-2j+2l+1}} \notag\\
 &\phantom{Spa}\times\x_{\a_{n-2j+2l+2}\dots\a_n)\b_{2l+1}\dots\b_{n-2j+2l+1}}\bigg\{\sum_{k=0}^{l}(-1)^k \binom{j-k}{l-1}\binom{l}{k}\bigg\}\notag.
 \end{align}
It follows that the gauge variation vanishes after making use of \eqref{G.1} once more. In the second line of the above we have used the combinatoric factors to shift the upper and lower bounds of the summation over $j$. Then, in the third line we have shifted the dummy variable $j\mapsto j-k$.

 
\section{Integration by parts}\label{appendixE}

In this section we discuss integration by parts in conformal space. To demonstrate the technique we give a detailed analysis of how it works in the case of the $4D$ CHS action \eqref{3.21} in a general curved space. However, before we begin the analysis let us give some general remarks.

 To integrate by parts in $D$ dimensions, we must impose the gauge\footnote{In fact, since most Lagrangians we consider are primary, all dependence on $\mathfrak{b}_a$ drops out and we needn't choose the gauge $\mathfrak{b}_a=0$. However, the two are equivalent because in both cases the K-symmetry is exhausted.} \eqref{degauge}. As discussed earlier, the conformal covariant derivative then takes the form 
\begin{align}
\mathfrak{b}_a=0\quad \implies\quad \nabla_{a}=\mathcal{D}_{a}+\frac 12 P_{a}{}^{b}K_{b}~. 
\end{align}
Consider some vector $V^{a}$, we obtain the following identity regarding total conformal derivatives
\begin{align} \label{D.2}
\int\text{d}^Dx \, e \, \nabla_{a}V^{a}=\frac 12 \int\text{d}^Dx \, e \, P_{ab}K^{a}V^{b}~. 
\end{align}
In the usual way we have integrated out the total derivative arising from the torsion-free Lorentz covariant derivative. One then uses the conformal algebra  and the conformal properties of the physical fields which comprise $V_a$ to eliminate the generator $K_a$.

Consider an integral of the form
\begin{align}
I=\int\text{d}^Dx\, e \, \mathcal{L}~,\quad \mathbb{D}\mathcal{L}=D\mathcal{L}~,\quad K_a\mathcal{L}=0~.
\end{align}
This means that $I$ is invariant under the gauge group $\mathcal{G}$ and since $\mathcal{L}$ is primary all dependence on $\mathfrak{b}_a$ drops out.  Let us further suppose that $\mathcal{L}$ takes the form
\begin{align}
\mathcal{L}=g^J\mathcal{A}h_J \label{Z.0}
\end{align}
where $g_J$ and $h_J$ are primary fields with abstract index structure and $\mathcal{A}$ is a linear differential operator such that $\mathcal{A}h_J\equiv \mathcal{A}_J{}^{I}h_I$ is also primary.  We define the transpose of the operator $\mathcal{A}$ by
\begin{align}
\int\text{d}^Dx\, e \, g^J\mathcal{A}h_J=\int\text{d}^Dx\, e \, h^J\mathcal{A}^Tg_J+\int\text{d}^Dx\, e \, \Omega \label{Z.1}
\end{align}
where $\Omega$ is a total conformal derivative and may be written as $\Omega=\nabla_aV^a$ for some vector $V^a$ with Weyl weight $D-1$. The first term on the right hand side of \eqref{Z.1} is the result of integrating $I$ by parts in the usual way. 

In general we cannot conclude that the second term on the right hand side of \eqref{Z.1} vanishes. However, under the condition that $\mathcal{A}^Tg_J$ is primary then $\Omega$ must also be primary. It follows that 
\begin{align}
0=K_a\Omega=[K_a,\nabla_b]V^b+\nabla^bK_aV_b= (2\eta_{ab}\mathbb{D}+2M_{ab})V^b+\nabla^bK_aV_b=\nabla^bK_aV_b~.\label{Y.0}
\end{align}
 It is clear that the condition $\nabla^bK_aV_b=0$ is satisfied if $V_a$ is primary. What is not so clear is that any solution $V_a$ to this equation is necessarily primary. However, for all cases known to us this is true.
 Application of the rule \eqref{D.2} then allows us to conclude that the second term on right side of \eqref{Z.1} vanishes up to a total derivative,
\begin{align}
\int\text{d}^Dx\, e \, \Omega=\int\text{d}^Dx\, e \, \bigg(\mathcal{D}^aV_a+\frac 12 P^{ab}K_aV_b\bigg)\approx 0~.
\end{align} 
 Therefore, we arrive at the following rule for integration by parts:
\begin{align}
\int\text{d}^Dx\, e \, g^J\mathcal{A}h_J=\int\text{d}^Dx\, e \, h^J\mathcal{A}^Tg_J \label{Y.10}
\end{align}
if $K_ag_I=K_ah_I=K_a(\mathcal{A}h_I)=K_a(\mathcal{A}^Tg_I)=0$. We remark that most of the Lagrangians proposed in the main body of this paper are of the form \eqref{Z.0}.

 As an example, presented below are the steps one must take in order to integrate the $4D$ CHS action by parts in a general curved space. For convenience we do not include the complex conjugated part of the action. 
\begin{align}
S^{(m,n)}_{\text{CHS}}
&= \text{i}^{m+n}\int \rd^4 x \, e\, \check{\mathfrak{C}}^{\a(m+n)} \hat{\mathfrak{C}}_{\a(m+n)}  \notag\\
=&~\text{i}^{m+n}\int \rd^4 x \, e\, (-1)^n\nabla_{\a_1\ad_1}\dots\nabla_{\a_m\ad_m}\bar{\phi}_{\a_{m+1}\dots\a_{m+n}}{}^{\ad_1\dots\ad_m} \hat{\mathfrak{C}}^{\a(m+n)}  \notag \\
=&~\text{i}^{m+n}\int \rd^4 x \, e\, (-1)^n\bigg\{(-1)^m\bar{\phi}_{\a_{m+1}\dots\a_{m+n}}{}^{\ad_1\dots\ad_m} \nabla_{\a_1\ad_1}\dots\nabla_{\a_m\ad_m}\hat{\mathfrak{C}}^{\a(m+n)} \notag\\
&-\sum_{j=1}^{m}(-1)^j\nabla_{\a_1\ad_1}\bigg[\nabla_{\a_{j+1}\ad_{j+1}}\cdots\nabla_{\a_m\ad_m}\bar{\phi}_{\a_{m+1}\dots\a_{m+n}}{}^{\ad_1\dots\ad_m} \nabla_{\a_2\ad_2}\cdots\nabla_{\a_j\ad_j}\hat{\mathfrak{C}}^{\a(m+n)}\bigg] \bigg\}\notag \\
=&~\text{i}^{m+n}\int \rd^4 x \, e\,  \bar{\phi}^{\a(n)\ad(m)} \hat{\mathfrak{B}}_{\a(n)\ad(m)} -(-1)^n \text{i}^{m+n}\int \rd^4 x \, e\,\nabla_{\a_1\ad_1}V^{\a_1\ad_1} \label{Y.7}
\end{align}
where
\begin{align}
V^{\a_1\ad_1}=\sum_{j=1}^{m}(-1)^{j}\nabla_{\a_{j+1}\ad_{j+1}}\cdots\nabla_{\a_m\ad_m}\bar{\phi}_{\a_{m+1}\dots\a_{m+n}}{}^{\ad_1\dots\ad_m} \nabla_{\a_2\ad_2}\cdots\nabla_{\a_j\ad_j}\hat{\mathfrak{C}}^{\a(m+n)}~. \notag
\end{align}
In accordance with the previous discussion, since the integrand of the first term in \eqref{Y.7} is of the form $h^J\mathcal{A}^Tg_J$ and all primary conditions listed below \eqref{Y.10} are met, the total conformal derivative must vanish. However, to support our belief that  $V_a$ appearing in \eqref{Y.0} is primary, we show that this is indeed the case for the current example, 
\begin{align}
K_{\g\gd}V^{\a_1\ad_1}=&\phantom{+}\sum_{j=1}^{m-1}(-1)^jK_{\g\gd}\nabla_{\a_{j+1}\ad_{j+1}}\cdots\nabla_{\a_m\ad_m}\bar{\phi}_{\a_{m+1}\dots\a_{m+n}}{}^{\ad_1\dots\ad_m} \nabla_{\a_2\ad_2}\cdots\nabla_{\a_j\ad_j}\hat{\mathfrak{C}}^{\a(m+n)} \notag\\
&+\sum_{j=2}^{m}(-1)^j\nabla_{\a_{j+1}\ad_{j+1}}\cdots\nabla_{\a_m\ad_m}\bar{\phi}_{\a_{m+1}\dots\a_{m+n}}{}^{\ad_1\dots\ad_m} K_{\g\gd}\nabla_{\a_2\ad_2}\cdots\nabla_{\a_j\ad_j}\hat{\mathfrak{C}}^{\a(m+n)} \notag \\
=&\phantom{+}\sum_{j=2}^{m}(-1)^j \bigg( \big[K_{\g\gd},\nabla_{\a_{j}\ad_{j}}\cdots\nabla_{\a_m\ad_m}\big]\bar{\phi}_{\a_{m+1}\dots\a_{m+n}}{}^{\ad_1\dots\ad_m} \nabla_{\a_2\ad_2}\cdots\nabla_{\a_{j-1}\ad_{j-1}}\hat{\mathfrak{C}}^{\a(m+n)}\notag \\
&\phantom{\times~}-\nabla_{\a_{j+1}\ad_{j+1}}\cdots\nabla_{\a_m\ad_m}\bar{\phi}_{\a_{m+1}\dots\a_{m+n}}{}^{\ad_1\dots\ad_m} \big[K_{\g\gd}\nabla_{\a_2\ad_2}\cdots\nabla_{\a_j\ad_j}\big]\hat{\mathfrak{C}}^{\a(m+n)} \bigg) 
=0~.\notag
\end{align}
In the first line we have used the fact that the first term vanishes for $j=m$ whilst the second term vanishes for $j=1$. This allows us to translate the summation index of the first term in the second line. In going from the second to the third line, we have used the identity \eqref{3.24} twice, whereupon all terms in the round brackets cancel among themselves.


 \section{Conformal gravitino model} \label{appendix-gravitino}
 
As an application of the techniques developed earlier, we will discuss in detail the construction of a gauge-invariant model for the conformal gravitino in any $4D$ Bach-flat spacetime.
This model can be extracted from the action for $\cN=1$ conformal supergravity 
 \cite{KTvN1,KTvN2} by linearising it around a Bach-flat background. 

The conformal gravitino is described by a complex primary field $\phi_{\a(2)\ad}$ of dimension 
$+1/2$ and 
 its conjugate, which are defined modulo gauge transformations of the type
 \begin{align}\label{C.1}
 \delta_{\lambda}\phi_{\a(2)\ad}=\nabla_{(\a_1\ad}\lambda_{\a_2)}~,
 \end{align}
 where the complex gauge parameter $\lambda_{\a}$ is primary of dimension $-1/2$.
 
 Associated with the gravitino are the two field strengths 
 \begin{align}
 \hat{\mathfrak{C}}_{\a(3)}=\nabla_{(\a_1}{}^{\bd}\phi_{\a_2\a_3)\bd}~,\qquad \check{\mathfrak{C}}_{\a(3)}=\nabla_{(\a_1}{}^{\bd_1}\nabla_{\a_2}{}^{\bd_2}\bar{\phi}_{\a_3)\bd(2)}~,
 \label{C.2}
 \end{align}
 and their conjugates, which are primary fields of dimensions $+3/2$ and $+5/2$ respectively. Under the gauge transformation \eqref{C.1}, their variations are given by 
 \begin{align}
 \d_{\lambda}\hat{\mathfrak{C}}_{\a(3)}=C_{\a(3)\d}\lambda^{\d}~,\qquad \d_{\lambda}\check{\mathfrak{C}}_{\a(3)}=\frac{1}{2}C_{\a(3)\d}\nabla^{\d\dd}\bar{\lambda}_{\dd}-\bar{\lambda}_{\dd}\nabla^{\d\dd}C_{\a(3)\d}~.
 \end{align} 
 In accordance with the results from section 4, the action \eqref{3.21} with $m=2, n=1$, which we now denote by
\begin{align}\label{C.5}
S^{(3/2)}_{\text{CHS}}[\phi,\bar{\phi}]=-\text{i}\int\text{d}^4x~e~\hat{\mathfrak{C}}^{\a(3)}\check{\mathfrak{C}}_{\a(3)}+\text{c.c.}~,
\end{align} 
 is invariant under the conformal gauge group $\mathcal{G}$, but not under \eqref{C.1}. However, if \eqref{C.5} is supplemented by the non-minimal term that is linear in the Weyl tensor,
 \begin{align}
 &S_{\text{Linear}}^{(3/2)}=
 \ri\int\text{d}^4x\,e\, \phi^{\a(2)\ad}\check{\mathfrak{J}}_{\a(2)\ad}
+\text{c.c.}~,\\
\check{\mathfrak{J}}_{\a(2)\ad} &= C_{\a(2)}{}^{\b(2)}\nabla_{\b_1}{}^{\bd}\bar{\phi}_{\b_2\ad\bd}-\bar{\phi}_{\b_1\ad\bd}\nabla_{\b_2}{}^{\bd}C_{\a(2)}{}^{\b(2)} ~,\label{C.55}
 \end{align}
where $\check{\mathfrak{J}}_{\a(2)\ad}$ is a dimension $+7/2$ primary field, 
then the resulting action
 \begin{align}
 S_{\text{Gravitino}}=S_{\text{CHS}}^{(3/2)}+S_{\text{Linear}}^{(3/2)}
  \label{C.6}
 \end{align}
 is invariant under  \eqref{C.1} provided the background Bach tensor 
 \eqref{C.7}
 vanishes, 
 \bea
 B_{\a(2)\ad(2)}=0~.
 \eea

We remark that the following primary deformation of the linearised Bach tensor,  
\begin{align}
\check{\mathcal{B}}_{\a(2)\ad}=\check{\mathfrak{B}}_{\a(2)\ad}-\check{\mathfrak{J}}_{\a(2)\ad}~,
\end{align}
 which may be used to rewrite the action \eqref{C.6} as
 \begin{align}
 S_{\text{Gravitino}}=-\text{i}\int\text{d}^4x \, e \, \phi^{\a(2)\ad}\check{\mathcal{B}}_{\a(2)\ad}+\text{c.c.}~,
 \end{align}
is transverse and gauge invariant in any Bach-flat spacetime,
\begin{align}
B_{\a(2)\ad(2)}=0\quad\implies\quad \nabla^{\g\gd}\check{\mathcal{B}}_{\a\g\gd}=\d_{\l}\check{\mathcal{B}}_{\a(2)\ad}=0~.
\end{align}

To conclude, we present the degauged versions of the above results. In the gauge \eqref{degauge}, the gauge transformations \eqref{C.1} are
\begin{align}
\delta_{\lambda}\phi_{\a(2)\ad}=\mathcal{D}_{(\a_1\ad}\lambda_{\a_2)}~. \label{C.9}
\end{align}
Under \eqref{C.9}, the degauged gravitino field strengths 
\begin{align}
 \hat{\mathfrak{C}}_{\a(3)}=\mathcal{D}_{(\a_1}{}^{\bd}\phi_{\a_2\a_3)\bd}~,\qquad \check{\mathfrak{C}}_{\a(3)}=\mathcal{D}_{(\a_1}{}^{\bd_1}\mathcal{D}_{\a_2}{}^{\bd_2}\bar{\phi}_{\a_3)\bd(2)}-\frac 12 R_{(\a_1\a_2}{}^{\bd(2)}{}\bar{\phi}_{\a_3)\bd(2)}~, \label{C.10}
\end{align}
transform as 
\begin{align}
 \d_{\lambda}\hat{\mathfrak{C}}_{\a(3)}=C_{\a(3)\d}\lambda^{\d}~,\qquad \d_{\lambda}\check{\mathfrak{C}}_{\a(3)}=\frac{1}{2}C_{\a(3)\d}\mathcal{D}^{\d\dd}\bar{\lambda}_{\dd}-\bar{\lambda}_{\dd}\mathcal{D}^{\d\dd}C_{\a(3)\d}~.
 \end{align} 
 The degauged gravitino action \eqref{C.6} remains the same except with \eqref{C.2} replaced by \eqref{C.10} in $S_{\text{CHS}}^{(3/2)}$ as well as the replacement $\nabla_{\a\ad} \mapsto \mathcal{D}_{\a\ad}$ in $S_{\text{Linear}}^{(3/2)}$. Finally, $S_{\text{Gravitino}}$ is invariant under the gauge transformations \eqref{C.9} as long as the degauged Bach tensor,
 \begin{align}
  B_{\a(2)\ad(2)}=\mathcal{D}^{\b_1}{}_{(\ad_1}\mathcal{D}^{\b_2}{}_{\ad_2)}C_{\a(2) \b(2)}-\frac{1}{2} C_{\a(2)\b(2)}R^{\b(2)}{}_{\ad(2)}~,
 \end{align}
 vanishes.


 \section{Conformal graviton model} \label{appendix-graviton}

In this appendix we construct a conformally invariant model for the graviton that is gauge invariant in any Bach-flat spacetime. This model may of course be obtained by linearising the action of $4D$ conformal gravity around a Bach-flat background. However, in principle the method presented below can be applied to higher-spin models, albeit with considerable effort.   
 
In accordance with section \ref{section4}, the conformal graviton is described by a real primary field $h_{\a(2)\ad(2)}=\bar{h}_{\a(2)\ad(2)}$ with zero Weyl weight and is defined modulo the gauge transformations
\begin{align}
\delta_{\l}h_{\a(2)\ad(2)}=\nabla_{(\a_1(\ad_1}\l_{\a_2)\ad_2)}~. \label{F.1}
\end{align} 

Associated with the graviton is the linearised Weyl tensor,
\begin{align}
\mathfrak{C}_{\a(4)}=\nabla_{(\a_1}{}^{\bd_1}\nabla_{\a_2}{}^{\bd_2}h_{\a_3\a_4)\bd_1\bd_2}~, \label{F.2}
\end{align}
which is a primary field of dimension 2. Under the gauge transformation \eqref{F.1}, its variation is given by
\begin{align}
\delta_{\l}\mathfrak{C}_{\a(4)}=\frac{1}{2}C_{\a(4)}\nabla^{\b\bd}\l_{\b\bd}-\l_{\b\bd}\nabla^{\b\bd}C_{\a(4)}-2C^{\b}{}_{(\a_1\a_2\a_3}\nabla_{\a_4)}{}^{\bd}\l_{\b\bd}~.
\end{align}
The action of linearised conformal gravity is given by \eqref{3.21}, with $m=n=2$, which we now denote by $S_{\text{CHS}}^{(2)}$,
\begin{align}
S_{\text{CHS}}^{(2)}=\int\text{d}^4x\, e \, \mathfrak{C}^{\a(4)}\mathfrak{C}_{\a(4)}+\text{c.c.} \label{F.4}
\end{align}
In general \eqref{F.4} is invariant under the gauge group $\mathcal{G}$, but only in conformally flat spacetimes is it invariant under \eqref{F.1}. Indeed, upon integrating by parts, we find that under \eqref{F.1} the action \eqref{F.4} varies as
\begin{align}
\delta_{\l}S_{\text{CHS}}^{(2)}=\int\text{d}^4x \, e \, \lambda^{\a\ad}\bigg\{4\mathfrak{C}^{\b(4)}\nabla_{\b\ad}C_{\b(3)\a}&+4C_{\b(3)\a}\nabla_{\b\ad}\mathfrak{C}^{\b(4)} \notag\\
&-C_{\b(4)}\nabla_{\a\ad}\mathfrak{C}^{\b(4)}-3\mathfrak{C}^{\b(4)}\nabla_{\a\ad}C_{\b(4)}
\bigg\}+\text{c.c.}
\end{align}
  
In the spirit of the previous appendix, to extend the gauge invariance of this model we seek a weight $+4$ primary deformation of the linearised Bach tensor, denoted by $\mathfrak{J}_{\a(2)\ad(2)}$,
\begin{align}
K_{\b\bd}\mathfrak{J}_{\a(2)\ad(2)}=0~,\qquad \mathbb{D}\mathfrak{J}_{\a(2)\ad(2)}=4\mathfrak{J}_{\a(2)\ad(2)}~. \label{F.6}
\end{align}
Restricting our attention to the construction of tensors with the  properties of $\mathfrak{J}_{\a(2)\ad(2)}$ greatly lightens the workload. In fact, beginning with the most general weight $+4$ tensor with this index structure, the condition of being primary is so strong that one may show that there are only three (up to complex conjugation) such inequivalent tensors that are linear in the Weyl tensor. They are given by
\begin{subequations}\label{F.7}
\begin{align}
\mathfrak{J}^{1}_{\a(2)\ad(2)}&= B_{\a_1}{}^{\g\gd}{}_{\ad_1}h_{\a_2\g\gd\ad_2}~,\label{F.7a}\\
\mathfrak{J}^{2}_{\a(2)\ad(2)}&=-2C_{\a(2)}{}^{\b(2)}\nabla_{\ad_1}{}^{\g}\nabla_{\b_1}{}^{\gd}h_{\g\b_2\ad_2\gd}-\nabla_{\ad_1}{}^{\g}C_{\a(2)}{}^{\b(2)}\nabla_{\g}{}^{\gd}h_{\b(2)\ad_2\gd} \notag\\
&\quad+2\nabla_{\ad_1}{}^{\g}C_{\a(2)}{}^{\b(2)}\nabla_{\b_1}{}^{\gd}h_{\b_2\g\ad_2\gd} -\nabla_{\a_1\ad_1}C_{\a_2}{}^{\b(3)}\nabla_{\b_1}{}^{\gd}h_{\b_2\b_3\ad_2\gd}\notag\\
&\quad-\nabla_{\a_1}{}^{\gd}C_{\a_2}{}^{\b(3)}\nabla_{\b_1\ad_1}h_{\b_2\b_3\ad_2\gd} +3\nabla_{\b_1}{}^{\gd}C_{\a(2)}{}^{\b(2)}\nabla_{\gd}{}^{\g}h_{\g\b_2\ad(2)}\notag\\
&\quad+\frac 12\nabla^{\g\gd}C_{\a(2)}{}^{\b(2)}\nabla_{\g\gd}h_{\b(2)\ad(2)} +h_{\b(2)\ad(2)}\Box_c C_{\a(2)}{}^{\b(2)}~,\label{F.7b}\\ 
\mathfrak{J}^{3}_{\a(2)\ad(2)}&=-C_{\a_1}{}^{\b(3)}\nabla_{\b_1\ad_1}\nabla_{\b_2}{}^{\gd}h_{\b_3\a_2\ad_2\gd}+C_{\a(2)}{}^{\b(2)}\Box_c h_{\b(2)\ad(2)}\notag\\
&\quad-2\nabla_{\b_1\ad_1}C_{\a_1}{}^{\b(3)}\nabla_{\a_2}{}^{\gd}h_{\b_2\b_3\ad_2\gd}-\nabla_{\b_1}{}^{\gd}C_{\a(2)}{}^{\b(2)}\nabla_{\gd}{}^{\g}h_{\b_2\g\ad(2)}\notag\\
&\quad-\frac 12\nabla^{\g\gd}C_{\a(2)}{}^{\b(2)}\nabla_{\g\gd}h_{\b(2)\ad(2)} -h_{\g\b_1\gd\ad_1}\nabla_{\ad_2}{}^{\g}\nabla_{\b_2}{}^{\gd}C_{\a(2)}{}^{\b(2)}\notag\\
&\quad+h_{\b(2)\ad(2)}\Box_c C_{\a(2)}{}^{\b(2)}~. \label{F.7c}
\end{align}

In the above and for the remainder of this appendix we employ the convention whereby all free dotted and undotted indices appearing in any tensor are assumed to be independently symmetrised over, e.g., $\nabla_{\a_1\ad_1}C_{\a_2}{}^{\b(3)}\nabla_{\b_1}{}^{\gd}h_{\b_2\b_3\ad_2\gd}=\nabla_{(\a_1(\ad_1}C_{\a_2)}{}^{\b(3)}\nabla_{|\b_1}{}^{\gd}h_{\b_2\b_3|\ad_2)\gd}$. 

In addition to the primary fields \eqref{F.7a}, \eqref{F.7b} and \eqref{F.7c}, 
there are precisely three (up to complex conjugation) inequivalent structures that are quadratic in the Weyl tensor and which satisfy the properties \eqref{F.6}. They are given by
\begin{align}
	\mathfrak{J}^{4}_{\a(2)\ad(2)}&=C_{\a(2)}{}^{\g(2)}C_{\g(2)}{}^{\b(2)}h_{\b(2)\ad(2)} ~,\label{F.8a}\\ 
	\mathfrak{J}^{5}_{\a(2)\ad(2)}&=C_{\a_1\g(2)}{}^{\b_1}C_{\a_2}{}^{\b_2\g(2)}h_{\b(2)\ad(2)}~,\label{F.8b}\\ 
	\mathfrak{J}^{6}_{\a(2)\ad(2)}&=C_{\a(2)}{}^{\b(2)}\bar{C}_{\ad(2)}{}^{\bd(2)}h_{\b(2)\bd(2)} ~. \label{F.8c}
\end{align}
\end{subequations}

The tensors \eqref{F.7} span all primary structures of the type $\mathfrak{J}_{\a(2)\ad(2)}$ and in particular any linear combination will also satisfy \eqref{F.6}. Furthermore, if we express them in the form
\begin{align}
\mathfrak{J}^i_{\a(2)\ad(2)}=\mathcal{A}_ih_{\a(2)\ad(2)}
\end{align}
where $\mathcal{A}_i$ is a linear differential operator then, with the exception of $\mathcal{A}_2$, it may be shown that each operator is symmetric in the sense $\mathcal{A}_i=\mathcal{A}_i^T$ (see appendix \ref{appendixE} for the definition of $\mathcal{A}_i^T$). This property reduces the amount of work required to compute the gauge variation of each of the functionals associated with $\mathfrak{J}^i_{\a(2)\ad(2)}$.

Any operator $\mathcal{A}$ may be decomposed into symmetric and antisymmetric parts, $\mathcal{A}=\mathcal{A}_{\text{S}}+\mathcal{A}_{\text{A}}$ with $\mathcal{A}_{\text{S}}=\frac 12 (\mathcal{A}+\mathcal{A}^T)$ and $\mathcal{A}_{\text{A}}=\frac 12 (\mathcal{A}-\mathcal{A}^T)$. It follows that the antisymmetric part of $\mathcal{A}$ vanishes identically in any integral of the form 
\begin{align}
\int\text{d}^4 x \, e \, h^J\mathcal{A}h_J=\int\text{d}^4 x \, e \, h^J\mathcal{A}_{\text{S}}h_J~. \label{Z.3}
\end{align}

Using \eqref{Z.3} it is possible to show that at the level of actions, the following correspondence between $\mathfrak{J}^2$ and the remaining primary structures holds
\begin{align}
\int\text{d}^4x\, e \,h^{\a(2)\ad(2)}\mathfrak{J}^2_{\a(2)\ad(2)}= \int\text{d}^4x\, e \,h^{\a(2)\ad(2)}\big(2\mathcal{A}_1-\mathcal{A}_3+2\mathcal{A}_4+\mathcal{A}_5+\mathcal{A}_6\big)h_{\a(2)\ad(2)}~.
\end{align}
Additionally, the structure $\mathfrak{J}^1$ vanishes in any Bach-flat spacetime and will be of no use. Therefore, it suffices to consider only one of the primary structures that is linear in the Weyl tensor, say $\mathfrak{J}^3$, and its associated functional
\begin{align}
S_{\text{Linear}}^{(2)}=\int\text{d}^4x\, e \,h^{\a(2)\ad(2)}\mathfrak{J}^3_{\a(2)\ad(2)}+\text{c.c.}\label{F.9}
\end{align}
One can then show that under the gauge transformation \eqref{F.1} and upon integrating by parts, the action \eqref{F.9} transforms as
\begin{align}
\d_{\l}S_{\text{Linear}}^{(2)}=~&\frac{1}{2}\delta_{\l}S_{\text{CHS}}^{(2)}+\bigg(2\int\text{d}^4x\, e \, \l^{\a\ad}\bigg\{\nabla_{\d}{}^{\dd}\big[C^{\b(2)}{}_{\g(2)}C_{\a}{}^{\g(2)\d}h_{\b(2)\ad\dd}\big]\notag\\
&+\nabla_{\b_1}{}^{\dd}\big[C^{\b(2)}{}_{\g(2)}C_{\a}{}^{\g(2)\d}h_{\b_2\d\ad\dd}\big] -\nabla_{\b_1\bd_1}\big[C_{\a}{}^{\b(3)}\bar{C}_{\ad}{}^{\bd(3)}h_{\b(2)\bd(2)}\big]\bigg\}\notag\\
&-\int \text{d}^4x\, e \, \l^{\a\ad}\bigg\{\nabla^{\b_1\bd_1}B_{\a}{}^{\b_2\bd_2}{}_{\ad}h_{\b(2)\bd(2)}+B_{\a}{}^{\b_2\bd(2)}\nabla_{\ad}{}^{\b_2}h_{\b(2)\bd(2)}\bigg\} +\text{c.c.}\bigg)~.
\end{align}

To annihilate the terms quadratic in the Weyl tensor, we define the functional
\begin{align}
S_{\text{Quadratic}}^{(2)}=\int\text{d}^4x \, e \, h^{\a(2)\ad(2)}\bigg\{\mathfrak{J}^{4}_{\a(2)\ad(2)}+\mathfrak{J}^{5}_{\a(2)\ad(2)}+\mathfrak{J}^{6}_{\a(2)\ad(2)}\bigg\}+\text{c.c.}
\end{align}
It follows that the action
\begin{align}
S_{\text{Graviton}}&=S_{\text{CHS}}^{(2)}-2S_{\text{Linear}}^{(2)}+2S_{\text{Quadratic}}^{(2)} \notag\\
&=\int\text{d}^4x \, e \, h^{\a(2)\ad(2)}\bigg\{\mathfrak{C}^{\a(4)}\mathfrak{C}_{\a(4)}+2C_{\a_1}{}^{\b(3)}\nabla_{\b_1\ad_1}\nabla_{\b_2}{}^{\gd}h_{\b_3\a_2\ad_2\gd}-2C_{\a(2)}{}^{\b(2)}\Box_c h_{\b(2)\ad(2)}\notag\\
&+4\nabla_{\b_1\ad_1}C_{\a_1}{}^{\b(3)}\nabla_{\a_2}{}^{\gd}h_{\b_2\b_3\ad_2\gd}+2\nabla_{\b_1}{}^{\gd}C_{\a(2)}{}^{\b(2)}\nabla_{\gd}{}^{\g}h_{\b_2\g\ad(2)}-2h_{\b(2)\ad(2)}\Box_c C_{\a(2)}{}^{\b(2)}\notag\\
&+\nabla^{\g\gd}C_{\a(2)}{}^{\b(2)}\nabla_{\g\gd}h_{\b(2)\ad(2)} +2h_{\g\b_1\gd\ad_1}\nabla_{\ad_2}{}^{\g}\nabla_{\b_2}{}^{\gd}C_{\a(2)}{}^{\b(2)}+2C_{\a(2)}{}^{\g(2)}C_{\g(2)}{}^{\b(2)}h_{\b(2)\ad(2)}\notag\\
& +2C_{\a_1\g(2)}{}^{\b_1}C_{\a_2}{}^{\b_2\g(2)}h_{\b(2)\ad(2)}+2C_{\a(2)}{}^{\b(2)}\bar{C}_{\ad(2)}{}^{\bd(2)}h_{\b(2)\bd(2)} \bigg\}+\text{c.c.}~, \label{Graviton}
\end{align}
whose variation under \eqref{F.1} is given by
\begin{align}
\d_{\l}S_{\text{Graviton}}=2\int \text{d}^4x\, e \, \l^{\a\ad}\bigg\{\nabla^{\b_1\bd_1}B_{\a}{}^{\b_2\bd_2}{}_{\ad}h_{\b(2)\bd(2)}+B_{\a}{}^{\b_1\bd(2)}\nabla_{\ad}{}^{\b_2}h_{\b(2)\bd(2)}\bigg\}+\text{c.c.}~,
\end{align}
is the unique model describing the graviton that is both conformally invariant in a general curved spacetime and gauge invariant in any Bach-flat spacetime,
\begin{align}
B_{\a(2)\ad(2)}=0 \quad \implies \quad \d_{\l}S_{\text{Graviton}}=0~.
\end{align}

This model was analysed in \cite{Manvelyan} using a similar methodology, the main differences being that their analysis was performed in the gauge $\mathfrak{b}_a=0$ and the graviton field was not traceless. In accordance with \eqref{1.1}, this means that their model contains an extra algebraic gauge symmetry which may be used to gauge away the trace.  The authors found two inequivalent primary Lagrangians that were linear in the Weyl tensor and used both in the construction of their gauge invariant action. Upon eliminating the trace of the graviton field, one of these structures vanishes and the other must be proportional to \eqref{F.9} modulo terms involving the Bach tensor and the square of the Weyl tensor.

Finally, we remark that the following primary deformation of the linearised Bach tensor,
\begin{align}
\mathcal{B}_{\a(2)\ad(2)}=\mathfrak{B}_{\a(2)\ad(2)}-2\mathfrak{J}^{3}_{\a(2)\ad(2)}+2\mathfrak{J}^{4}_{\a(2)\ad(2)}+2\mathfrak{J}^{5}_{\a(2)\ad(2)}+2\mathfrak{J}^{6}_{\a(2)\ad(2)}~,
\end{align}
which may be used to rewrite the action \eqref{Graviton} as
\begin{align}
S_{\text{Graviton}}=\int\text{d}^4x \, e \, h^{\a(2)\ad(2)}\mathcal{B}_{\a(2)\ad(2)}~,
\end{align}
is transverse and gauge invariant in any Bach-flat spacetime,
\begin{align}
B_{\a(2)\ad(2)}=0\quad\implies\quad \nabla^{\g\gd}\mathcal{B}_{\a\g\gd\ad}=\d_{\l}\mathcal{B}_{\a(2)\ad(2)}=0~.
\end{align}

\begin{footnotesize}

\end{footnotesize}

\end{document}